\documentclass[aps,prx,twocolumn,superscriptaddress]{revtex4-2}
\usepackage{amsmath,amssymb}
\usepackage{mathrsfs}
\usepackage{graphicx}
\usepackage{setspace}
\usepackage{braket}
\usepackage{dcolumn}
\usepackage{bm}
\usepackage{hyperref}
\usepackage{cleveref}
\usepackage{physics}
\usepackage{algorithm}
\usepackage{algorithmic}
\usepackage{xcolor}
\usepackage{amsthm}

\usepackage{comment}

\usepackage{tikz}
\usepackage{quantikz}
\usepackage{multirow}
\usepackage{placeins}
\usepackage{booktabs}

\usepackage{xcolor}

\usetikzlibrary{quantikz2}

\begin{document}


\title{Quantum Advantage in Sensing and Imaging \\ through Coherent Nuisance Suppression}

\author{Maxim Sirotin}
\email{msirotin@g.harvard.edu}
\affiliation{Department of Physics, Harvard University, Cambridge, MA 02138, USA}
\author{Aleksandr Mokeev}
\affiliation{QuTech, Delft University of Technology, 2628 CJ, Delft, The Netherlands}
\author{Ishaan Kannan}
\affiliation{Department of Physics, Harvard University, Cambridge, MA 02138, USA}
\author{Julian C. Christou}
\affiliation{NSF NOIRLab, 950 N Cherry Ave., Tucson, AZ 85719, USA}
\affiliation{Steward Observatory, 933 N Cherry Ave., Tucson, AZ 85721, USA}
\author{Jayadev Rajagopal}
\affiliation{NSF NOIRLab, 950 N Cherry Ave., Tucson, AZ 85719, USA}
\author{Babak Saif}
\affiliation{NASA Goddard Space Flight Center, 8800 Greenbelt Road, Greenbelt, Maryland 20771, USA}
\author{Mikhail D. Lukin}
\affiliation{Department of Physics, Harvard University, Cambridge, MA 02138, USA}
\author{Johannes Borregaard}
\email{borregaard@fas.harvard.edu}
\affiliation{Department of Physics, Harvard University, Cambridge, MA 02138, USA}

\begin{abstract}
Many sensing and imaging problems must contend with unknown \textit{nuisance} parameters such as aberrations, background fields, and drift, and the cost of removing these noise sources can dominate the sample budget. In contrast to classical receivers, a quantum receiver can processes samples jointly to directly access observables in which common nuisance is suppressed. We introduce a relational Fourier sampling algorithm for optimal, source-independent imaging with a programmable \textit{quantum camera}, which uses quantum pixels and a quantum computer to capture and process photon states. We demonstrate its relevance for general astronomical imaging as well as super-resolution and transient-object tracking through drifting atmospheric turbulence. A quantum camera that resolves an optical wavefront across $2^q$ spatial modes using $q$ receiving qubits, with a full-rank and strongly-correlated nuisance over those modes, can yield an exponential sample-complexity reduction of \(\exp[\Omega(2^q)]\). For typical seeing conditions at leading ground-based observatories, we predict more than three orders of magnitude fewer photons to resolve point-source positions using only 15~processing qubits and a depth-9 circuit. Our work identifies coherent nuisance suppression as a novel mechanism for provable quantum advantage with direct relevance to many sensing tasks.
\end{abstract}

{
\maketitle
}

\section{Introduction}\label{sec:intro}

Physical experiments rest on extracting information from a finite number of observations~\cite{Tsang2019,Na2017,Bates1982,Petrov86}. In sensing and imaging, information about an uncharacterized system is carried by photons, and conventional experiments destructively measure each set of samples as it arrives. Recent advances in quantum learning theory have suggested a fundamentally different form of experiment, in which coherently storing and processing multiple quantum samples before measurement substantially reduces the sample complexity of learning from quantum data, with applications ranging from learning properties of quantum states and dynamics~\cite{AharonovCotlerQi2022, Huang2021,HuangLearningExperiments2022,ChenCotlerHuangLi2021} to decoding classical communications~\cite{Kelly2026}, property testing~\cite{montanaro2017learningstabilizerstatesbell, montanaro2018surveyquantumpropertytesting}, and error mitigation~\cite{Cotler_2019, Huggins_2021}. 

These concepts have recently been leveraged in quantum computational sensing~\cite{sarovar2023quantum, khan2026quantumcomputationalsensing,deshler2026}, where quantum advantages have been proposed for tasks including learning Gaussian optical states~\cite{dimitroff2026}, sensing structured or stochastic signals~\cite{allen2025, cotler2026quantum, kannan2026exponential}, and measuring gravitational-lensing time delays~\cite{liu2026}. Experiments demonstrating collective measurements and coherent storage of optical and microwave signals establish key ingredients for such architectures~\cite{HuangLearningExperiments2022,conlon2023,stas2026entanglement, kannan2026exponential}. Identifying practical sensing tasks that benefit from coherent quantum processing becomes an important challenge as platforms capable of sophisticated quantum computational sensing emerge~\cite{huang2026vast, babbush2026grand}.

In realistic sensing tasks ranging from astronomy to microscopy, the signal is embedded in a time-dependent environment of drifting backgrounds, fluctuating references, aberrations, and uncontrolled fields, broadly referred to as \emph{nuisance} degrees of freedom~\cite{Rao2024,Hunziker2018,Booth2014,Tsang2019}. Recovering the signal requires learning or averaging over the nuisance from the same samples available for imaging, a cost that can dominate the sample complexity when signals are weak, fundamentally limiting sensitivity~\cite{Tsang2019}. This obstruction has motivated quantum strategies that characterize structured noise more efficiently~\cite{wang2026entanglement} or engineer sensor networks to reject unwanted fields while retaining sensitivity to the signal~\cite{Sekatski2020,Hamann2025,Bate2025}.

\begin{figure*}[htbp]
 \centering
 \includegraphics[width=17.5cm]{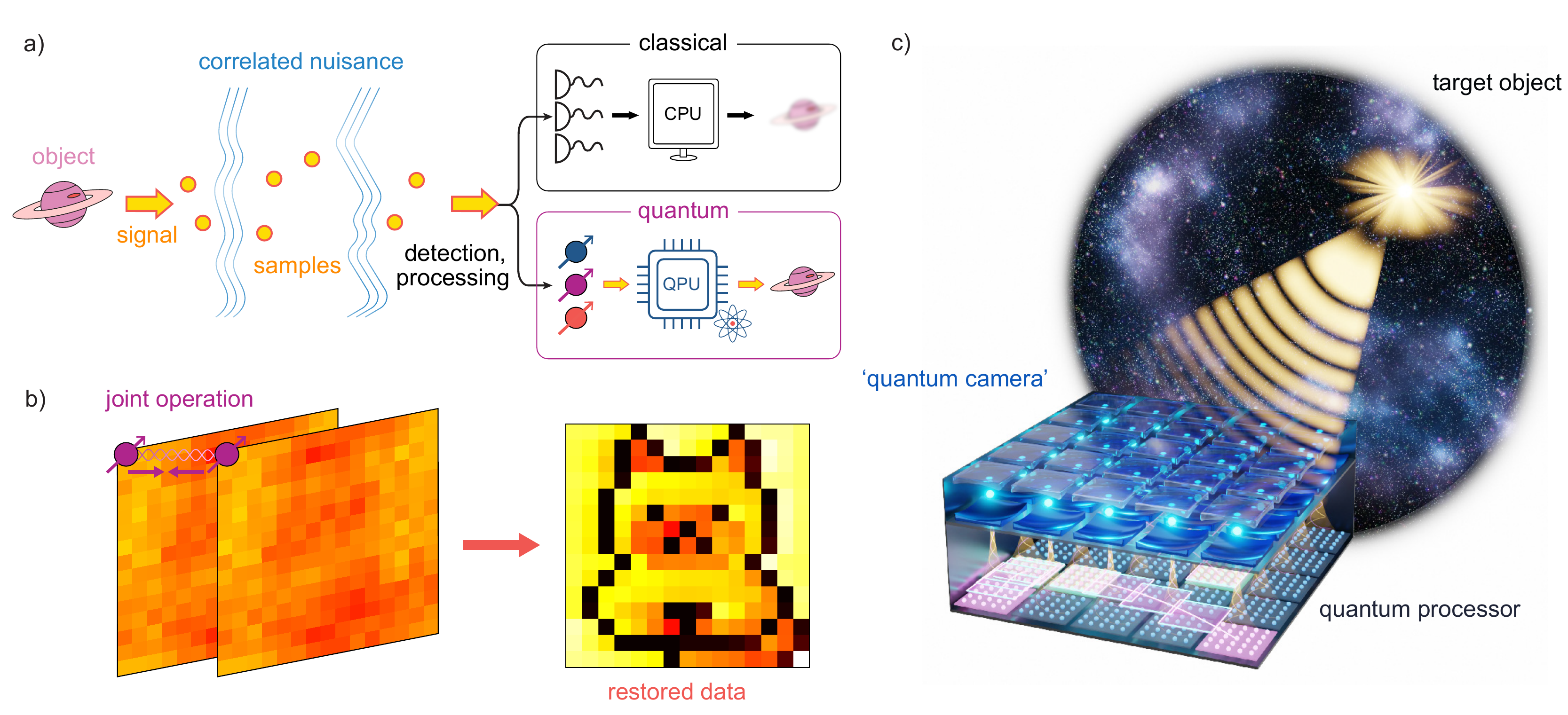}
\caption{\textbf{Quantum-enhanced learning under correlated nuisance.} (a) A stable optical signal is distorted by environmental nuisance that is correlated in time and/or space. In a conventional experiment, each sample is measured separately and only classical data are retained. In a quantum-enhanced experiment, nearby samples are stored as quantum data and processed jointly before final readout, enabling a reduction in sample complexity. (b) Numerical demonstration of the swap-based nuisance-cancellation algorithm. Conventional photon-by-photon processing sees distorted realizations of the object, while the joint quantum measurement accesses relative information in which the correlated nuisance coherently cancels, yielding a restored signal. (c) A practical example of a quantum-enhanced experiment: the conceptual architecture of a 'quantum camera' based on Refs.~\cite{stas2026entanglement, mokeev2026enhancing, khabiboulline2019optical}. The incoming starlight state is stored in an array of qubits using cavity-based spin-photon interfaces. The stored quantum data are then transferred to a quantum processor which performs image reconstruction.}
 \label{fig:concept}
\end{figure*}

In this Article, we show that by jointly processing samples that experience correlated noise, a quantum receiver can cancel the nuisance before measurement, a mechanism we refer to as \textit{coherent nuisance suppression}. Dynamical decoupling, decoherence-free encoding~\cite{lidar2014review}, and swap-test-type measurements~\cite{mokeev2026enhancing} share elements of this principle. The mechanism is distinct from the more familiar source of quantum advantage in learning, where collective measurements efficiently extract structured properties of high-dimensional states~\cite{Huang2021,HuangLearningExperiments2022,King2024,li2026exponential,cotler2026noisy, huang2026vast}; here, coherent processing instead removes a correlated disturbance before irreversible readout. Moreover, it provides a distinct mechanism for robust quantum metrological speedups that are known to degrade under noise \cite{depradenne2026restrictionsnoncliffordfaulttolerance}. This mechanism is particularly valuable in optical imaging, where correlated nuisance is the dominant bottleneck and where the advantage conferred by coherent nuisance suppression can be dramatic.

In optical imaging, photons from a stationary or slowly varying scene carry the same source information while acquiring time-dependent aberrations. When quantum processing is available, nearby photons with correlated aberrations can be measured jointly so that the common distortion is suppressed while source-dependent relations are retained~\cite{liu2026}. We introduce \textit{relational Fourier sampling} (RFS), a constant-depth quantum imaging algorithm that extracts from the incoming photons a set of Fourier observables sufficient to reconstruct an arbitrary source distribution without prior knowledge of the number, positions, or brightnesses of the objects. RFS is quantum-Fisher-information optimal for these observables in the correlated-noise models studied here, and we prove quantum advantages for estimating image parameters, such as angular separation, position, and velocity of objects to fixed additive error.

This source independence, together with the programmability of quantum processing, motivates a \textit{quantum camera} concept (Figs.~\ref{fig:concept} and \ref{fig:camera_full}): a general imaging front end in which the spatial information contained in incident photons is mapped to qubit registers and coherently processed before collapsing the photon's wavefunction via detection. Since a quantum-enabled telescope, microscope, or other sensor requires substantial optical integration, quantum hardware, and calibration, a single programmable receiver supporting many tasks is far more useful than a specialized measurement for each observable. The quantum camera proposed here would support general image reconstruction, superresolution, parameter estimation, and moving-object tracking, and other coherent primitives such as background removal~\cite{mokeev2026enhancing} can be combined with nuisance suppression. Our approach draws on the principles of classical speckle imaging techniques which use Fourier-phase retrieval techniques such as the Knox–Thompson cross-spectrum and the bispectrum~\cite{knox1974recovery, weigelt1977modified, Bates1982}, while extending them through coherent access to quantum correlations. This connection suggests that the proposed framework and algorithm could be applied to a broad range of speckle imaging problems, from astronomy to biological imaging~\cite{heeman2019clinical} and beyond, making the quantum camera a promising architecture for practical quantum advantage~\cite{babbush2026grand}.

\section{Quantum advantage \\ for correlated nuisance}
\label{sec:scalingresults}

We consider the architecture of Fig.~\ref{fig:concept}, in which each sample is coherently transduced into a quantum register so that multiple samples can be stored and jointly processed before measurement. In our approach, the information carried by weak optical wavefronts is mapped at the single-photon level onto spin-qubit registers using cavity-mediated spin–photon entangling gates, followed by a heralding measurement of the photons, as detailed in Sec.~\ref{sec:astronomy}. The stored wavefronts can then be jointly processed to extract the observable of interest. We focus on general imaging tasks in which an arbitrary image parameter must be estimated to a prescribed accuracy, rather than one known in advance. 

Nearby samples experience correlated stochastic unitary noise, and we consider two regimes. In the limiting \emph{common-phase-twirled} model, each pair of samples experiences the same unknown random phase screen across all modes, refreshed independently between pairs, so earlier pairs carry no information about the present nuisance. In the more realistic \emph{Gaussian} model, the phase screens of different samples are Gaussian correlated, so earlier observations inform the present nuisance and can be used to learn it, but the nuisance experienced by consecutive photons differs.

For the common-phase-twirled model, RFS estimates the Fourier power spectrum optimally, saturating the quantum Fisher information. We further prove a quantum advantage over all positive-partial-transpose (PPT) measurements, a class that contains separable measurements and LOCC (local operations and classical communication) protocols, and hence any strategy that measures the two samples separately with classical feedforward. For estimating all power-spectrum parameters to fixed additive error, any such measurement is limited to an $O(1/D)$ fraction of the quantum Fisher information, where $D$ is the number of resolved spatial modes. 

For the Gaussian model, classical receivers can handle the nuisance in two ways. \emph{Nuisance-learning strategies} estimate it from previous measurement statistics, as wavefront sensors commonly do in adaptive optics. Within the linear-Gaussian model considered here, Kalman filtering is the minimum-mean-square-error (MMSE) optimal causal estimator~\cite{Kalman1960}. In contrast, \emph{Nuisance-blind strategies} extract image observables from classical correlations in which the common disturbance cancels, as in speckle interferometry~\cite{Bates1982,Petrov86}. By measuring samples separately, however, they lose access to quantum coherences that remain available to a joint measurement.

In this partially correlated model, RFS still enables quantum-Fisher-information-optimal estimation of Fourier observables sufficient for full image reconstruction. Against both families of classical strategies we identify regimes of exponential sample-complexity reduction of the form
\begin{equation}
    \frac{N_{\rm cl}}{N_{\rm RFS}}
    =\Omega\left(e^{\Xi_{\rm cl}-\Xi_{\rm Q}}\right),
\end{equation}
where $N_{\rm cl}$ ($N_{\rm RFS}$) is the number of samples required by the classical (RFS) receiver and $\Xi_{\rm Q}$ captures the residual phase uncertainty caused by differences in nuisance between consecutive photons which RFS cannot remove. For nuisance-learning strategies, $\Xi_{\rm cl}=\Xi_{\rm K}$ describes the residual phase uncertainty after optimal Kalman prediction; for nuisance-blind strategies, $\Xi_{\rm cl}=\Xi_{\rm C}$ characterizes the suppression of classical information about the target observable through its least-suppressed phase combinations. 

This sample-complexity advantage depends on the observable of interest. The common-phase-twirled model gives an $\Omega(D)$ advantage against classical strategies for power-spectrum parameters, but stronger scaling is possible for observables encoded in phase combinations with a large minimum nuisance variance. The distinction matters when the task is known in advance: if the source is known to be binary with unknown separation, nuisance-learning strategies may concentrate on the relevant nuisance modes and avoid the cost of general image reconstruction, and nuisance-blind strategies may optimize their measurements directly for separation. Their measurement-nonlocality limitation nevertheless persists: we show that RFS retains an advantage even against arbitrary nuisance-blind receivers with full knowledge of the source model and nuisance statistics. 

For estimating image parameters of an unknown scene, we identify realistic observables with $\Xi_{\rm cl}-\Xi_{\rm Q}=\Omega(D)$, giving an $\exp(\Omega(D))$ advantage against both classes of strategies, which can be expressed as $\exp[\Omega(2^q)]$ with $q$ qubits needed to store a $D$-dimensional state. We model atmospheric turbulence with a frozen-flow approximation, in which a von K\'arm\'an phase screen moves across the telescope at constant wind velocity~\cite{Poyneer2009,liu2020,Jia2015}. Increasing $D$ at fixed pixel size enlarges the pupil and the available baselines, increasing the phase-difference variance over baselines below the atmospheric outer scale, while for sufficiently correlated samples the differential nuisance affecting RFS remains parametrically smaller. This yields the $\exp(\Omega(D))$ advantage in a practically relevant finite-dimensional regime. The scaling cannot continue indefinitely, because phase screens decorrelate beyond the finite outer scale, and it is not universal: for binary separation estimation with nuisance-blind classical strategies, the information is not generically suppressed by an exponent extensive in $D$ (see below and the Supplementary Information).

\section{Relational Fourier Sampling} \label{sec:nuisancesup}

\begin{figure*}[htbp]
 \centering
 \includegraphics[width=17cm]{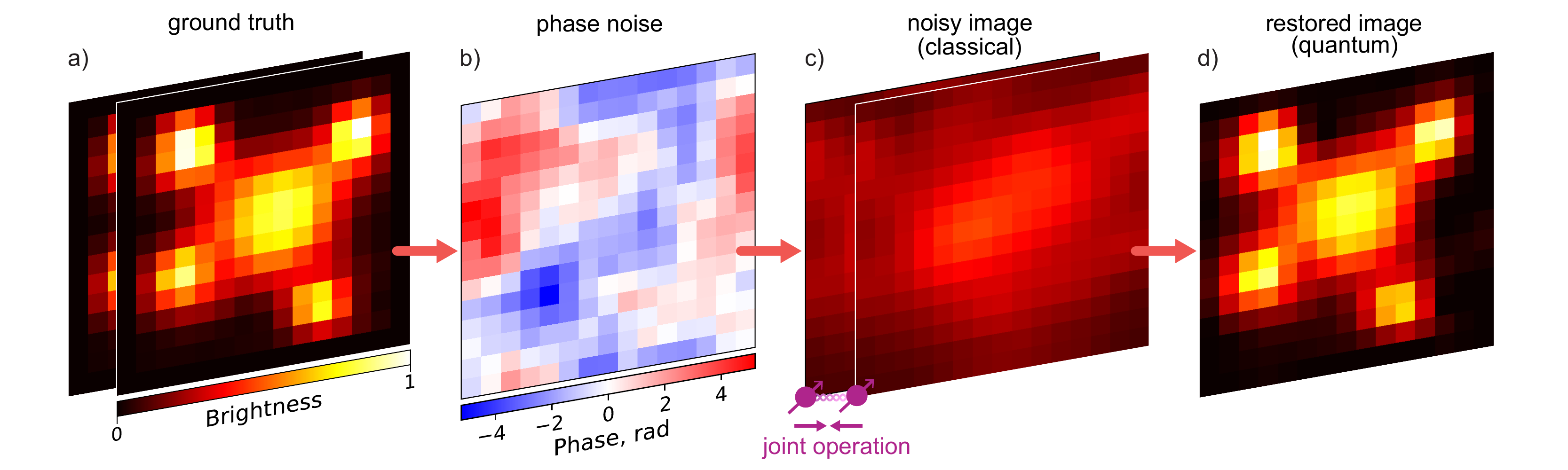}
 \caption{\textbf{Numerical realization of the full-imaging RFS algorithm under correlated phase noise.} Light from the source (a) passes through phase noise (b), producing distorted classical images under photon-by-photon detection (c). A quantum camera instead captures nearby photons as quantum data and performs relational Fourier-sampling measurements on pairs of stored samples (d). We use an illustrative Einstein-cross intensity distribution of DESI-253.2534+26.884, imaged through realistic atmospheric conditions with quantum and classical cameras.}
 \label{fig:imaging}
\end{figure*}

We consider a family of signal states
\begin{equation}
    \rho_{\bm{\phi}}
    =
    \sum_{\alpha} a_{\alpha}\,
    U(\bm{\phi}_{\alpha})
    \ket{\psi_\alpha}\bra{\psi_\alpha}
    U^\dagger(\bm{\phi}_{\alpha}),
    \label{eq:thm1}
\end{equation}
in which the unknown parameters $\{\bm{\phi}_\alpha\}$ are encoded into a statistical mixture through the unitary map $U(\bm{\phi}_\alpha)$. Each sample then undergoes an additional stochastic perturbation,
\begin{equation}
    U(\delta\bm{\phi})U(\bm{\phi}_\alpha)
    =
    U(\bm{\phi}_\alpha+\delta\bm{\phi}),
\end{equation}
with $\delta\bm{\phi}$ drawn from a correlated random process. 

Motivated by imaging and other sensing problems which probe spatial information, we focus on the coherences
\begin{equation}
    \Phi_m
    =
    \sum_{j\in\mathbb Z_D}
    \bra{j}\rho_{\bm{\phi}}\ket{j+m},
    \qquad
    m\in\mathbb Z_D,
    \label{eq:Phi_m_def}
\end{equation}
up to a convention-dependent normalization, where $\{\ket{j}\}$ spans the $D$-dimensional signal space. In the quantum camera setting discussed below, where an optical wavefront is encoded across spatial modes in the pupil plane of a telescope, $\Phi_m$ is proportional to the $m$th spatial Fourier component of the source intensity distribution. Expression $\bra{j}\rho_{\bm{\phi}}\ket{j+m}$ represents the spatial coherence of the wavefront on a pixel baseline of length $m$. The goal is to recover the information in $\{\Phi_m\}$ while remaining insensitive to the correlated perturbation. We call such quantities \textit{relational} observables, as they capture relationships in the spatial wavefunction of incident photons at different points.

To identify the information that remains accessible under nuisance, consider first the simple common-phase-twirl model, in which the same phase noise, independently random over all spatial modes, acts on both samples of a pair. Before averaging over the random noise, the two-sample state is
\begin{equation}
    \rho_{\bm{\phi}}^{(\delta,2)}
    =
    \left[U(\delta\bm{\phi})\otimes U(\delta\bm{\phi})\right]
    \left(\rho_{\bm{\phi}}\otimes\rho_{\bm{\phi}}\right)
    \left[U^\dagger(\delta\bm{\phi})\otimes U^\dagger(\delta\bm{\phi})\right].
\end{equation}
Because the common perturbation is unknown, the effective experiment is obtained by averaging over its realizations,
\begin{equation}
    \rho_{\bm{\phi}}^{\otimes 2}
    \longmapsto
    \mathcal T\!\left(\rho_{\bm{\phi}}^{\otimes 2}\right),
    \label{eq:common-twirl-main}
\end{equation}
where $\mathcal T$ denotes the common-phase twirl corresponding to averaging over all possible realizations of $U(\delta\bm{\phi})$. The twirl removes the absolute phase reference of either sample, but it preserves coherences between basis states carrying the same total nuisance charge. For each shift $m\neq0$ and basis label $j$, these surviving coherences lie in the two-dimensional exchange block
\begin{equation}
    \mathcal H_{j,m}
    =
    \mathrm{span}\!\left\{
        \ket{j,j+m},\,
        \ket{j+m,j}
    \right\},
    \label{eq:exchange-block-main}
\end{equation}
where $\ket{j,j+m}$ refers to computational basis state $\ket{j}$ of the first sample and $\ket{j+m}$ of the second. These blocks contain precisely the nuisance-invariant information needed to estimate the Fourier powers $|\Phi_m|^2$.

The Fisher-information-optimal measurement of this information projects each exchange block onto
\begin{equation}
    \ket{\pm_{j,m}}
    =
    \frac{\ket{j,j+m}\pm\ket{j+m,j}}{\sqrt 2}.
    \label{eq:rfs-basis-main}
\end{equation}
We refer to this as a block-resolved swap measurement. Each outcome yields a shift $m$ and an exchange parity $\pm$; the resolved shift identifies the observable (power spectrum component) and the parity imbalance estimates its value. Summing over $j$, the normalized probability difference directly gives the twirl-invariant Fourier power,
$|\Phi_m|^2=(p_{m,+}-p_{m,-})/(p_{m,+}+p_{m,-})$, where $p_{m,\pm}$ denotes the probability of obtaining the outcome $(m, \pm)$.

Unlike the complete twirl, spatially-correlated noise retains coherences between different shift sectors, which supply the phase information needed for full image reconstruction through the nearest-neighbor coherences $\beta_m=\Phi_m^*\Phi_{m+1}$. These are accessed with \textit{rotated} block-swap measurements: a mixing operation interferes neighboring $m,m+1$ shift sectors, so that the corresponding parity imbalance measures $\mathrm{Re}[e^{i\chi}\beta_m]$, up to a known normalization. Choosing $\chi=0$ or $\pi/2$ gives the real or imaginary quadrature (up to a known sign), respectively, see Sec.~\ref{sec:hardware} and Fig.~\ref{fig:circuits} for a specific implementation of the algorithm. Together with the Fourier powers, these phase relations finalize the RFS algorithm and reconstruct the image intensity distribution (Fig.~\ref{fig:imaging}).

\section{Quantum imaging \\ through the atmosphere}
\label{sec:astronomy}

\begin{figure*}[htbp]
 \centering
 \includegraphics[width=17cm]{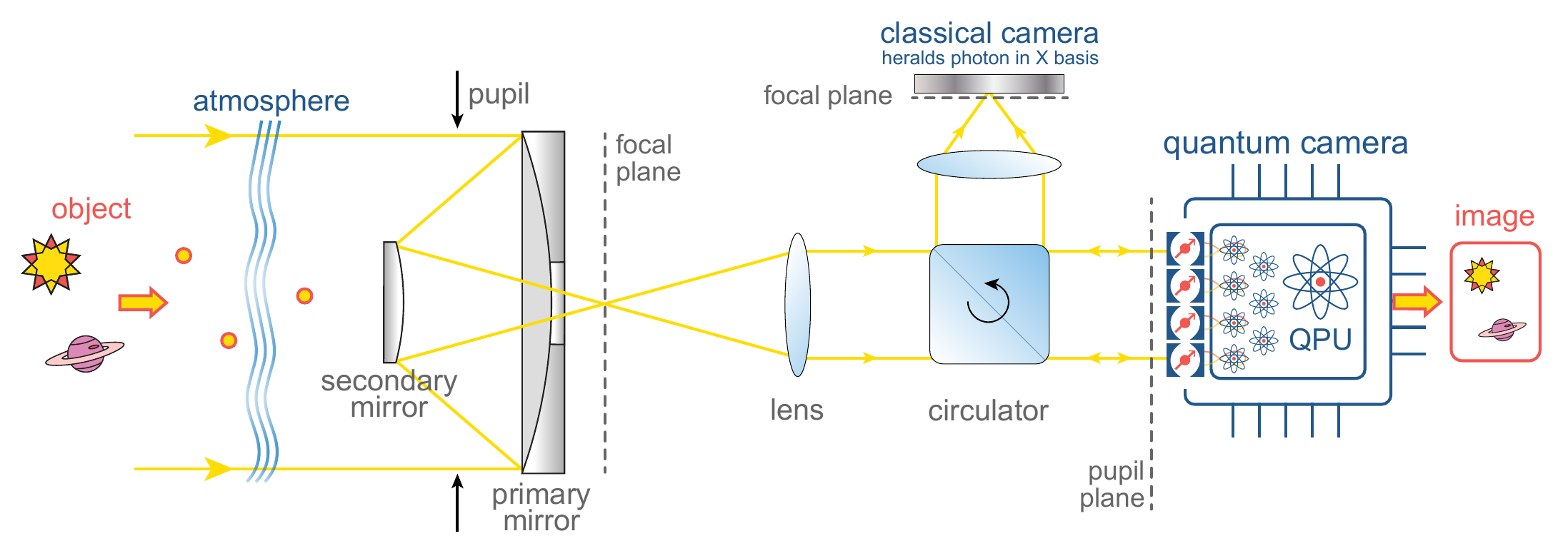}
 \caption{\textbf{Quantum camera concept for astronomical imaging.} The quantum memory array is located in the telescope pupil plane, where cavity-mediated photon--spin interactions transfer the spatial quantum state of the incident light into memory qubits. A conventional camera in the focal plane performs a measurement in the spatial Fourier basis, heralding photon arrival without revealing the original pupil mode. The detected pixel determines a known phase correction, applied through classical feed-forward. Here we illustrate an architecture based on cavity-coupled neutral atoms, in which the memory array and quantum processing unit (QPU) can be integrated inside the same vacuum chamber.}
 \label{fig:camera_full}
\end{figure*}

In ground-based optical imaging, atmospheric turbulence is an experimentally important source of nuisance (Fig.~\ref{fig:concept}c and~\ref{fig:camera_full}). Light from a distant star or other source carries a stable or predictably varying optical signal, while atmospheric propagation introduces rapidly varying wavefront distortions. Rather than detecting each photon immediately, the quantum camera transfers its spatial quantum state into a register of quantum memories, coherently processing photons arriving at different times. As a representative implementation, we consider an array of cavity-coupled quantum memories encoding the spatial modes of the optical field, following the spin--photon-interface architectures of Refs.~\cite{stas2026entanglement,mokeev2026enhancing,khabiboulline2019optical}.

Astronomical light is described by a thermal optical state containing vacuum, single-photon, and higher-photon-number components. We first consider a single point source at angular position $\theta$. In the weak-signal regime, with mean photon number $\bar n\ll1$ per detection window, the incident state can be approximated as
\begin{equation}
\rho_{\rm opt}(\theta)\simeq
(1-\bar n)|{\rm vac}\rangle\langle{\rm vac}|
+\bar n|\psi_\theta\rangle\langle\psi_\theta|
+O(\bar n^2),
\end{equation}
where the normalized single-photon spatial state is
\begin{equation}
|\psi_\theta\rangle_{\rm opt}
=
\frac{1}{\sqrt D}
\sum_{j\in\mathbb Z_D}
e^{ikx_j\theta}|1_j\rangle.
\label{eq:point_source_optical_state}
\end{equation}
Here $|1_j\rangle$ denotes a photon in pupil mode $j$, $x_j$ is its spatial coordinate, and $k=2\pi/\lambda$ is the optical wavenumber. 

Each optical mode interacts with a corresponding memory qubit initialized in $|+\rangle=(|0\rangle+|1\rangle)/\sqrt2$. A reflective phase gate flips the memory qubit state $|+\rangle\rightarrow|-\rangle$ if the photon reflects from this qubit, while $|+\rangle\rightarrow|+\rangle$ if there was no photon~\cite{reiserer2014quantum}. The single-photon component becomes entangled with the memory array,
\begin{equation}
|\psi_\theta\rangle_{\rm opt}|+\rangle^{\otimes D}
\longrightarrow
\frac{1}{\sqrt D}
\sum_{j\in\mathbb Z_D}
e^{ikx_j\theta}|1_j\rangle|e_j\rangle,
\end{equation}
where $|e_j\rangle=|-\rangle_j\bigotimes_{k\ne j}|+\rangle_k$ are orthogonal memory states corresponding to the spatial-mode basis $|j\rangle$.

The reflected photon is then detected by a conventional camera in the focal plane, implementing a measurement in the spatial Fourier basis (see Fig.~\ref{fig:camera_full}). This detection heralds the photon arrival without revealing which pupil mode it traversed, preserving the spatial coherence transferred to the memories. Conditioned on a detection at focal-plane pixel $\ell$, the normalized memory state is
\begin{equation}
|\psi_{\theta,\ell}\rangle_{\rm mem}
=
\frac{1}{\sqrt D}
\sum_{j\in\mathbb Z_D}
e^{ikx_j\theta-2\pi i j\ell/D}|e_j\rangle.
\end{equation}
The detected pixel $\ell$ and photon arrival time are recorded classically, and the known Fourier phases are removed by feed-forward. The resulting heralded memory state is
\begin{equation}
|\psi_\theta\rangle
=
\frac{1}{\sqrt D}
\sum_{j\in\mathbb Z_D}
e^{ikx_j\theta}|e_j\rangle.
\label{eq:point_source_state}
\end{equation}
Thus, conditioned on photon detection, the original single-photon spatial information is coherently transferred into the atomic register.

The description extends to arbitrary spatially incoherent sources. According to the van Cittert--Zernike theorem, the spatial coherence of light from an incoherent source is determined by the Fourier transform of its angular intensity distribution. A single point source produces a pure single-photon spatial state, whereas an extended source with angular intensity distribution $I(\theta)$ produces a statistical mixture of point-source states,
\begin{equation}
\rho_I
=
\int d\theta\,p(\theta)
|\psi_\theta\rangle\langle\psi_\theta|,
\qquad
p(\theta)=
\frac{I(\theta)}
{\int d\theta'\,I(\theta')}.
\label{eq:arbitrary_source_state}
\end{equation}

Atmospheric turbulence acts on the optical field before transduction and is represented by an additional unitary transferred to the quantum register,
\begin{equation}
    W_t
    =
    \sum_{j\in\mathbb Z_D}
    e^{i\eta_j(t)}
    \ket{j}\bra{j},
    \qquad
    \rho_I
    \longmapsto
    W_t\rho_IW_t^\dagger ,
    \label{eq:atmospheric_unitary}
\end{equation}
where $\eta_j(t)$ is the atmospheric phase distortion at pupil coordinate $x_j$ and photon arrival time $t$. The astronomical imaging problem therefore has precisely the structure considered in Sec.~\ref{sec:nuisancesup}: a stationary signal encoded in a family of quantum states, subjected to stochastic, temporally correlated unitary disturbances.

We model the atmosphere using a Gaussian von K\'arm\'an phase screen with frozen-flow dynamics in which a spatial phase pattern is shifted across the telescope with velocity $v$ by the wind,
\begin{equation}
    \eta(x,t)
    \simeq
    \eta_0(x-vt),
    \label{eq:frozen-flow-main}
\end{equation}
with covariance determined by the Fried parameter and the inner and outer turbulence scales~\cite{Poyneer2009,liu2020,Jia2015}. 

In full imaging the task is to estimate image parameters from an unknown scene. This addresses the realistic situation where the telescope is directed towards an unknown part of the sky and as such there is no prior information on the properties of objects. Below, we consider two representative imaging tasks to numerically evaluate the RFS advantage under specific seeing conditions, assuming one-dimensional pupil for simplicity. First, we consider the standard superresolution task of estimating the angular separation between two incoherent point sources. We then consider tracking moving objects, where the limited observation time makes efficient use of each detected photon particularly important.

\subsection{Binary-source separation}

Without loss of generality, we consider two equally bright incoherent point sources with centroid $\bar\theta$ and separation $\Delta$,
\begin{equation}
    \rho_{\bar\theta,\Delta}
    =
    \frac12
    \ket{\psi_{\bar\theta-\Delta/2}}
    \bra{\psi_{\bar\theta-\Delta/2}}
    +
    \frac12
    \ket{\psi_{\bar\theta+\Delta/2}}
    \bra{\psi_{\bar\theta+\Delta/2}} .
    \label{eq:two_source_dm}
\end{equation}

\begin{figure*}[htbp]
 \centering
 \includegraphics[width=17cm]{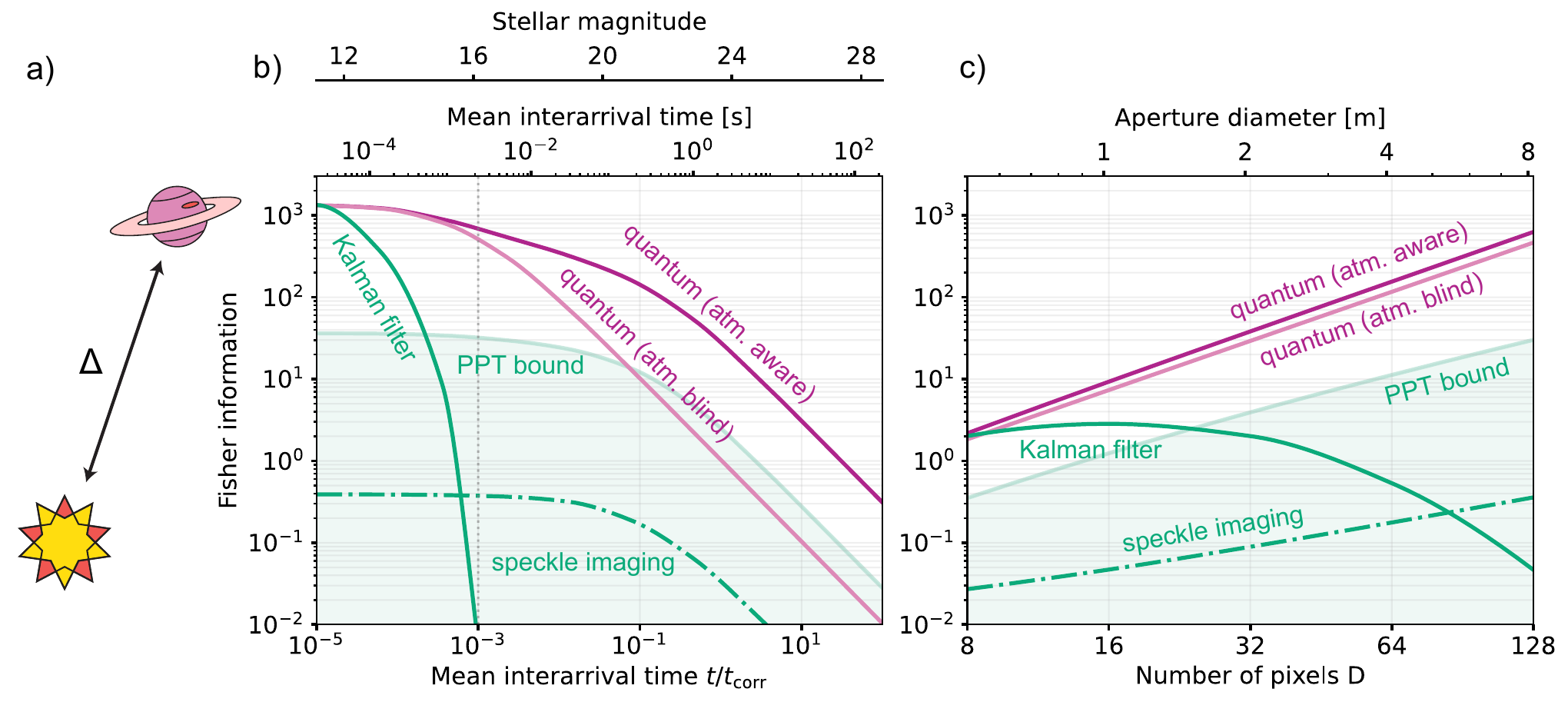}
\caption{\textbf{Fisher information for superresolution under realistic frozen-flow turbulence.} (a) Task schematic: estimating the angular separation $\Delta$ between two point sources. Here, we consider $\Delta=0.1\cdot \lambda/D_{\rm apt}$ (1D aperture of size $D_{\rm apt}$), corresponding to 10\% of the Rayleigh limit. (b) FI as a function of mean photon interarrival time $t/t_{\rm cor}$ at fixed $D=128$. (c) FI as a function of spatial dimension and number of pixels $D$ at fixed $t/t_{\rm cor}=10^{-3}$. Parameters: $\lambda=550\,{\rm nm}$ and $\Delta\lambda=1\,{\rm nm}$. Quantum curves represent naive RFS (atm. blind) and RFS with full knowledge of atmospheric statistics (atm. aware). Classical curves include Kalman filtering (e.g. adaptive optics), speckle imaging (product direct imaging with any allowed classical correlations), and a PPT bound on any nuisance-blind separable measurement, including task-aware ones.}
 \label{fig:atmosphere_scans}
\end{figure*}

The binary-source model is used to turn the information carried by the RFS data into a single scalar Fisher information for illustration; the RFS measurement itself is unchanged and continues to acquire the Fourier observables required for a general image. The Fisher-information calculations use a flat one-dimensional pupil discretized into $D$ spatial modes and the Gaussian von K\'arm\'an frozen-flow covariance described in the Supplementary Information. Atmospheric parameters are chosen to be representative of leading high-quality ground-based sites such as Paranal and Maunakea with an 8-m-class telescope. The simulations are local in source separation, so the plotted information can be interpreted directly as the photon cost of superresolution.

Photon arrivals are modeled as a stationary Poisson process. We denote by $t$ the mean photon interarrival time and by $t_{\rm cor}$ the atmospheric correlation time. Consecutive detected photons are grouped into pairs; their actual temporal separations are therefore random and the RFS Fisher information is averaged over the corresponding differential phase covariance. The classical history-assisted receiver is given the same photon record together with the frozen-flow covariance and wind model, but not the realized phase screen, and uses the Kalman predictor described above.

Figure~\ref{fig:atmosphere_scans} shows the resulting Fisher information. Panel (b) fixes $D=128$, corresponding to a 128-mode optical field (pixels), and varies the mean photon interarrival time $t/t_{\rm cor}$. For $t/t_{\rm cor}\ll1$, consecutive photons see nearly the same phase screen and the RFS information remains close to its common-noise value. As the interarrival time approaches the atmospheric correlation time, the two wavefronts decorrelate, the differential nuisance grows, and the quantum advantage decreases. The history-assisted receiver (Kalman filtering) loses information earlier because it must infer the absolute high-dimensional phase from the finite classical measurement record, whereas RFS depends primarily on the phase difference across a single accepted pair. Classical strategies also include optimistic performance of speckle imaging which is allowed to use any classical correlations, and a PPT bound on any nuisance-blind separable measurement, including task-aware ones.

Figure~\ref{fig:atmosphere_scans}c fixes $t/t_{\rm cor}=10^{-3}$ and varies the number of resolved spatial modes by increasing the physical size of the telescope. Increasing $D$ provides access to finer spatial structure, but it also increases the dimensionality of the atmospheric nuisance that a nuisance-learning full-imaging receiver must track. This produces a resolution--nuisance tradeoff: using too few modes discards useful spatial frequencies, whereas resolving too many modes spreads a finite wavefront-sensing budget over a larger nuisance space. RFS avoids much of this tradeoff because the common component is removed coherently and only the residual differential phase enters the pair measurement. For the parameter range used in the simulations, the resulting photon-sample reduction exceeds three orders of magnitude over a broad experimentally relevant region. The space-optimized quantum implementation at $D=128$ uses compressive photonic routing (Sec.~\ref{sec:hardware}) into two 7-qubit sample registers together with one processing ancilla (totalling 15 qubits) and the depth-9 RFS circuit.

Nuisance-blind strategies show similar to RFS growth since $\Delta$-estimation task does not have an exponential penalty with $D$ (only $\Omega(D)$) (see SI for details and examples of observables that show exponential scaling). We also note that the FI for the small binary source separation in the deep and noisy sub-Rayleigh range acquires $\Omega(\Delta^{2})$ penalty (so-called 'Rayleigh curse'~\cite{oh2021quantum}), which dominates the gap between RFS and classical FI in this regime (see SI for details).

\subsection{Moving-object tracking}

Moving-object tracking provides another application of coherent nuisance suppression, where improved sensitivity to small changes in an optical signal translates into more accurate prediction of its future evolution. We consider a single faint moving point source, such as an asteroid, observed through correlated atmospheric aberrations similar to those in the binary-source example above (see SI for details). A quantum receiver uses a fraction of the photons for absolute-position measurements and coherently compares time-separated photon pairs to estimate their relative displacement. Both the quantum and classical receivers must estimate the unknown position and velocity from the same total photon budget.

\begin{figure}[H]
    \centering
    \includegraphics[width=8.5cm]{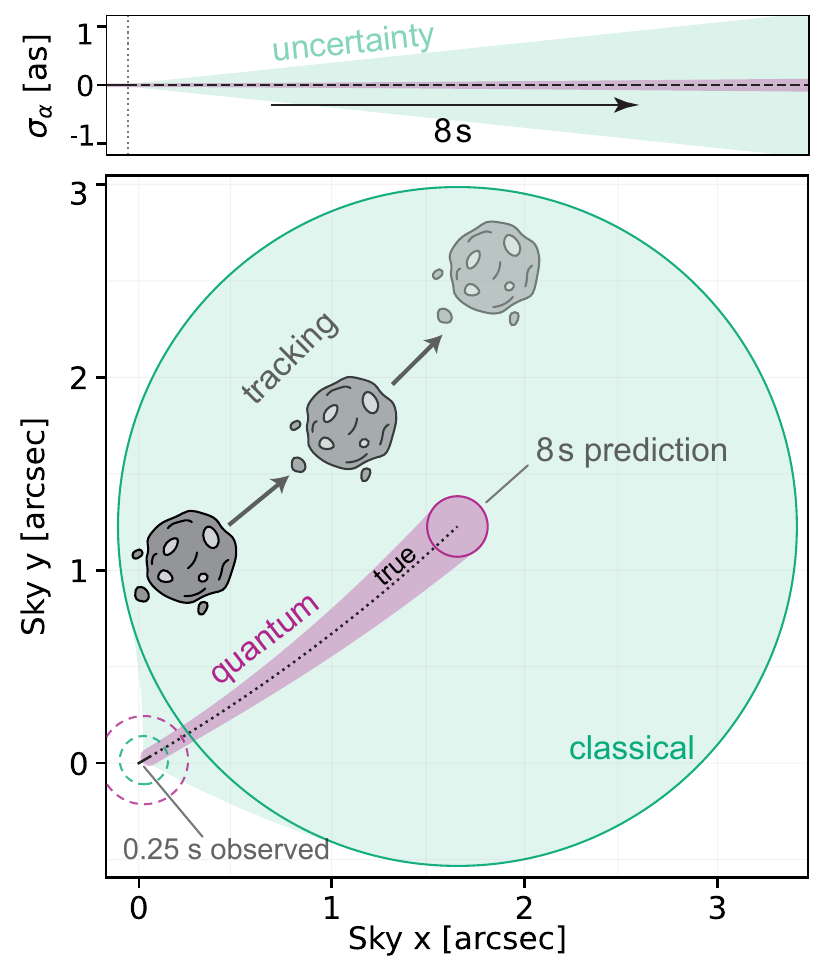}
    \caption{\textbf{Short-arc trajectory estimation and prediction.} A single moving point source is observed for $0.25\,\mathrm{s}$, followed by $8\,\mathrm{s}$ of prediction. The true trajectory is shown by a black solid line during observation and a dashed line during prediction. Dashed ellipses indicate the $5\,\sigma$ uncertainty in the reconstructed centroid position, solid ellipses indicate the $1\,\sigma$ uncertainty in the object position at $8\,\mathrm{s}$. Top panel shows time dependence of position uncertainty along the trajectory. Quantum is shown in purple and classical is shown in green.}
    \label{fig:tracking}
\end{figure}

Figure~\ref{fig:tracking} illustrates a short-observation regime with parameters relevant to faint, fast-moving near-Earth asteroids with the similar acquisition parameters as in Fig.~\ref{fig:atmosphere_scans}b. We consider an $8.2\,\mathrm{m}$ telescope with $D=128$ optical modes, a $40\,\mathrm{ms}$ photon-pair delay and a detected flux of approximately $1600$ photons/s, giving only $\sim400$ photons during the $0.25\,\mathrm{s}$ observation, followed by $8\,\mathrm{s}$ of trajectory prediction. The source has an apparent speed of $250\,\mathrm{mas/s}$, while its position offset and residual velocity are jointly unknown. The effective Fisher-information model predicts an approximately $137$-fold quantum advantage in residual-velocity information, corresponding to a $\sim11.7$-fold reduction in the one-dimensional velocity standard error. In the illustrative two-dimensional model, the predicted $1\,\sigma$ position uncertainty $8\,\mathrm{s}$ after observation is approximately $157\,\mathrm{mas}$ per coordinate for the quantum receiver, compared with $1.76\,\mathrm{arcsec}$ for classical. As a classical benchmark we use direct imaging with any classical two-photon correlations (such as speckle interferometry), both receivers are allowed to know the wind velocity vector.

This improved sensitivity substantially reduces the photon budget required for trajectory prediction with desired precision. For example, for a target $1\,\sigma$ position uncertainty of $100\,\mathrm{mas}$ per coordinate at $8\,\mathrm{s}$ after observation, the illustrative two-dimensional model requires approximately $985$ detected photons for the quantum receiver, compared with $1.24\times10^5$ for classical, during the same $0.25\,\mathrm{s}$ observing window. This corresponds to an approximately $126$-fold reduction in the FI-limited photon requirement, equivalent to a stellar magnitude gain of $5.25$ under equal throughput assumptions. As the time interval for which we wish to determine an object's position increases, the advantage grows (Fig.~\ref{fig:tracking}, top panel). The atmospheric model, numerical likelihood calculation, photon budget, and limitations are detailed in the SI.

The quantum receiver exploits self-referencing by directly measuring relations between photons, coherently suppressing correlated nuisance disturbances without explicitly reconstructing them. This approach is particularly advantageous when relational observables, such as binary-source separation or the velocity of a moving object, carry most of the information of interest. These examples illustrate the central mechanism of quantum learning under correlated nuisance: coherent joint measurements can preserve information about the signal that is strongly obscured in individual samples. The numerical code used to reproduce the full imaging, binary-source separation, and moving-object tracking results are publicly available~\cite{Sirotin2026QuantumCameraCode}.

\section{Hardware implementation and noise}\label{sec:hardware}

In the cavity-based implementation considered here, an incoming photon interacts coherently with the corresponding quantum pixel through a spin-dependent photon--cavity phase gate~\cite{reiserer2014quantum,nguyen2019integrated} (Fig.~\ref{fig:camera_full}). The most direct implementation stores the optical state as a one-hot excitation over \(D\) quantum pixels and performs relational Fourier sampling in this basis (Fig.~\ref{fig:camera_full}). Alternatively, light from each pixel is optically routed, addressing the receiving qubits corresponding to the nonzero bits of its binary index. For example, a photon from pixel $5$, represented by $101$ in binary, interacts with the first and third receiver qubits. Photonic logarithmic routing encodes the spatial information into a compact register of \(\lceil\log_2 D\rceil\) receiving qubits in expense of routing loss.

\begin{figure}[H]
 \centering
 \includegraphics[width=8.5cm]{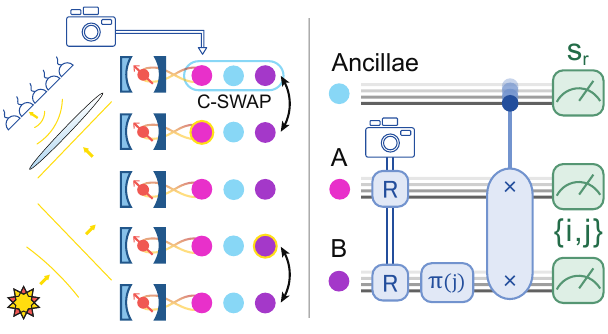}
\caption{\textbf{Simplified RFS circuit for the one-hot implementation with $D=5$.} The gate between registers $\rm A$, $\rm B$ and ancillae is controlled-$\rm SWAP$, '$\pi(j)$' is the permutation operation (also shown with arrows), and '$\rm R$' is a phase rotation.}
 \label{fig:circuits}
\end{figure}

The one-hot implementation of RFS requires approximately $3\cdot D$ qubits (two registers and ancillae) and can be realized with a constant-depth circuit (Fig.~\ref{fig:circuits}). Two photonic states are stored in registers $A$ and $B$. We first apply nearest-neighbor permutations $\pi(j)$ to register $B$, followed by local phase rotations of $0$ or $\pi/2$ on selected modes depending on the permutation pattern. We then perform controlled-SWAP measurements between corresponding qubits of the two registers using independent ancillae, followed by final readout. The two detected excitation positions $j<k$, which may belong to the same register, determine the Fourier indices $m=k-j$ and $n=\pi^{-1}(k)-\pi^{-1}(j)$, while the product of the corresponding ancilla outcomes, $s=s_js_k=\pm1$, gives the exchange parity. Its conditional expectation satisfies $\langle s\rangle_{j,k}\propto\operatorname{Re}[e^{i\chi_{jk}}\Phi_m^*\Phi_n]$, where $\chi_{jk}$ is the known relative phase introduced by the $\pi(j)$-dependent rotations. Outcomes with $n=m$ measure Fourier powers $|\Phi_m|^2$, whereas outcomes with $n=m\pm1$ access neighboring Fourier coherences. Choosing phase settings $0$ and $\pi/2$ allows both real and imaginary quadratures to be reconstructed. The permutation patterns can be randomized and adapted to sample the desired Fourier powers and coherences at the required rates.

The photonic logarithmic routing implementation of RFS requires only \(q=\lceil\log_2 D\rceil\) qubits per sample, but independent ancillae are insufficient to preserve the information needed to identify the excitation positions. A straightforward space-efficient implementation uses a single ancilla that sequentially interacts with the entire register, resulting in a depth-$q$ circuit. Alternatively, distributed cat-state ancillae could enable parallel interactions, potentially reducing the circuit depth to $O(\log q)$, including ancilla-state preparation.

For $D=128$ optical modes, logarithmic photonic routing stores each photon state in seven receiving qubits and uses a $15$-qubit processor with one traveling ancilla, whereas one-hot encoding uses $128$ receiving qubits with total $3\cdot 128$ qubits and parallel local controlled-SWAP measurements. We estimate the sample-complexity overhead of RFS assuming independent depolarizing errors of $1\%$ per receiving qubit, $0.1\%$ per controlled-SWAP gate, and $0.01\%$ per single-qubit gate, with ideal qubit permutations, photon capture, readout, and no routing loss. For photonic logarithmic routing implementation of RFS this error budget increases the required photon-pair number by only $\sim1.26$, and one-hot encoding incurs an overhead of $\sim6.64$. For the examples above, the substantial advantage persists in the realistic scenario and is compatible with the current hardware without requiring fault-tolerant operations~\cite{Chiu2025, evered2026}.

The dominant hardware imperfections arise during photon-to-qubit transduction, suggesting that error detection via parity and consistency checks could significantly improve quantum camera performance. Once the photonic state has been transferred into qubits, storage, processing, and readout can be done either with high-fidelity operations on physical qubits or  protected fault-tolerantly through quantum error correction~\cite{cotler2025noisy}. While the level of quantum processing and error rates required for the quantum camera are compatible with current hardware, the main hardware challenge lies in scaling the quantum pixels performing the photon-to-qubit transduction. Recent experimental demonstrations have implemented the transduction for two pixels~\cite{stas2026entanglement} and the spin-photon entangling operation has been demonstrated with high fidelity with both silicon vacancy centers~\cite{Bhaskar2020} and atoms~\cite{reiserer2014quantum}. Wafer-scale thin-film diamond processing~\cite{Riedel2026} and cavity arrays for parallel atom-photon interfacing~\cite{Shaw2026} offer promising routes to scale.

The spectral bandwidths commonly used in astronomical observations (\(\gtrsim 1\) nm) are typically three to four orders of magnitude broader than those of quantum memories (\(\lesssim 1\) GHz)~\cite{stas2026entanglement}, potentially limiting the achievable advantage of a quantum camera. While narrowband applications, such as high-resolution spectroscopy, can be addressed directly, broadband observations may require frequency multiplexing to fully exploit quantum-enhanced detection~\cite{mokeev2026enhancing}. In this approach, the collected light is divided into multiple spectral channels, each individually frequency-converted to interface with quantum memories. Similar to spatial-mode encoding, \(N\) spectral channels can be encoded using only \(\lceil\log_2 N\rceil\) memory qubits through photonic routing~\cite{khabiboulline2019optical}.

The architecture is not tied to a single physical platform. Cavity-coupled atoms and color centers naturally combine optical interfaces with long-lived memories~\cite{reiserer2014quantum,nguyen2019integrated,stas2026entanglement}, while neutral atom arrays and trapped ions provide high-fidelity processing and long coherence times~\cite{bluvstein2022quantum,ransford202698,shi2026efficient}. Superconducting circuits may be particularly attractive for microwave sensing or when combined with coherent optical-to-microwave transduction~\cite{google2025quantum}. Hybrid architectures are therefore also natural: photon capture, quantum storage, and processing may be implemented in different subsystems optimized for each function~\cite{mokeev2026enhancing,bulgarelli2026quantum,brunet2024quantum}.

\section{Conclusion and Outlook}

We have identified \textit{coherent nuisance suppression} as a general mechanism for quantum computational sensing and imaging: by coherently storing and jointly processing nearby samples before irreversible measurement, a quantum receiver can access relational observables in which common nuisance is suppressed, rather than first learning the nuisance from classical outcomes. For imaging tasks we established rigorous sample-complexity separations against all PPT measurements, and hence against separable and single-copy strategies, even when the signal model and nuisance statistics are known. Relational Fourier sampling realizes this principle in a source-independent way, directly extracting the observables needed to reconstruct an arbitrary source distribution. For finite correlated Gaussian noise, the framework exhibits an exponential-growth regime controlled by the difference between common- and differential-nuisance attenuation exponents, and the same structure appears against an MMSE-optimal Kalman predictor that uses nuisance information from earlier samples.

We apply this framework to ground-based astronomical imaging, where atmospheric turbulence and absorption impose severe limitations, sometimes necessitating space-based observations. By using nearby photons as coherent references for one another, quantum camera can suppress correlated atmospheric aberrations without estimating the instantaneous atmospheric phase or point-spread function. This capability could bring certain imaging tasks within reach of ground-based observatories, while also improving information extraction and suppressing correlated instrumental noise in future space missions~\cite{feinberg2024habitable, feinberg2026habitable}.

Our simulations show that relational Fourier sampling can reduce the required photon number by several orders of magnitude under realistic atmospheric conditions. We demonstrated this advantage for binary-source superresolution and moving-object tracking, while the same underlying measurement provides the Fourier observables required for general image reconstruction. The moving-object setting is particularly relevant to faint and rapidly evolving targets such as near-Earth asteroids and comets, for which the available photon budget and observation time can be intrinsically limited. Related capabilities may also be useful for ground-based imaging and tracking of artificial satellites and other space objects.

The required quantum processing is modest compared with general-purpose fault-tolerant computation: the examples considered here use only tens of qubits and shallow circuits. Recent experiments have demonstrated key ingredients for coherently mapping optical photons onto stationary qubit registers, and related work continues to improve spin--photon interfaces and quantum-assisted optical imaging~\cite{stas2026entanglement,mokeev2026enhancing}. 

More broadly, the mechanism identified here is not specific to astronomy or optical imaging. Correlated stochastic nuisance arises whenever the environment evolves continuously on a timescale comparable to, or longer than, the spacing between successive samples. In such settings, moving the measurement boundary from individual samples to small coherently processed groups can expose nuisance-insensitive relational degrees of freedom that are inaccessible after separate measurement. Identifying other sensing problems with this structure, and constructing the corresponding optimal algorithms, is therefore an important direction for future work. It will also be useful to extend the information-theoretic bounds developed here to continuous-variable sensors and beyond. The quantum camera provides an example of a programmable measurement architecture in which a quantum computer acts not merely as a post-processing accelerator, but as part of the sensing apparatus itself. We expect this viewpoint to substantially broaden the range of sensing tasks in which quantum computation can provide a practical advantage.

\begin{acknowledgments}
We thank Soonwon Choi, Susanne Yelin, Jordan Cotler, Isaac Chuang, William Oliver, Saikat Guha, Viatcheslav Dobrovitski, Brittany McClinton, Francisco Machado, Kelly Smith, Gefen Baranes, Yunkai Wang, Aditi Venkatesh, and Hemant Sharma for valuable discussions and insights related to this work. We gratefully acknowledge support from The AWS Quantum Discovery Fund at the Harvard Quantum Initiative, the NASA SAT program, the National Science Foundation (grant numbers PHY-2608266 and OMA- 2326787) and CUA PFC (PHY -2317134). IK acknowledges support from the DoD NDSEG Fellowship, the Nobile Research Initiative and 
the Air Force Office of Scientific Research under award number FA9550-25-C-B010 in the amount of \$134600.

The authors acknowledge the use of OpenAI's ChatGPT (GPT-6) for assistance with Python code development and debugging, and checking mathematical derivations and proofs. All AI-assisted calculations, code, and scientific arguments were independently reviewed and verified by the authors.
\end{acknowledgments}

\bibliography{sample}

@article{Riedel2026,
  title = {Scalable Photonic Quantum Interconnect Platform},
  author = {Riedel, Daniel and Graziosi, Teodoro and Wang, Zhuoxian and De-Eknamkul, Chawina and Abulnaga, Alex and Dietz, Jonathan and Mucchietto, Andrea and Haas, Michael and Sutula, Madison and Barral, Pierre and Pompili, Matteo and Raha, Mouktik and Robens, Carsten and Ha, Jeonghoon and Sukachev, Denis and Levonian, David and Bhaskar, Mihir and Markham, Matthew and Machielse, Bartholomeus},
  journal = {Phys. Rev. X},
  volume = {16},
  issue = {1},
  pages = {011063},
  numpages = {11},
  year = {2026},
  month = {Mar},
  publisher = {American Physical Society},
  doi = {10.1103/nfrg-zsts},
  url = {https://link.aps.org/doi/10.1103/nfrg-zsts}
}

@article{Shaw2026,
	author = {Shaw, Adam L. and Soper, Anna and Shadmany, Danial and Kumar, Aishwarya and Palm, Lukas and Koh, Da-Yeon and Kaxiras, Vassilios and Taneja, Lavanya and Jaffe, Matt and Schuster, David I. and Simon, Jonathan},
	date = {2026/02/01},
	doi = {10.1038/s41586-025-10035-9},
	id = {Shaw2026},
	isbn = {1476-4687},
	journal = {Nature},
	number = {8101},
	pages = {320--326},
	title = {A cavity-array microscope for parallel single-atom interfacing},
	url = {https://doi.org/10.1038/s41586-025-10035-9},
	volume = {650},
	year = {2026}}

@misc{liu2026,
      title={Quantum-optimized phase closure for robust interferometric imaging}, 
      author={Zhenning Liu and Emil T. Khabiboulline and Lorcan O. Conlon and Brittany McClinton and Aqil Sajjad and Saikat Guha and Jayadev Rajagopal and Alexey V. Gorshkov and Daniel Gottesman and Stephen T. Ridgway},
      year={2026},
      eprint={2610.06786},
      archivePrefix={arXiv},
      primaryClass={quant-ph},
      url={https://arxiv.org/abs/2610.06786}, 
}

@misc{deshler2026,
      title={Attaining Fundamental Limits of Multiparameter Incoherent Optical Imaging Using Joint-Detection Quantum Measurements}, 
      author={Nico Deshler and Aakash Warke and Michael R. Grace and Amit Ashok and Saikat Guha},
      year={2026},
      eprint={2608.19524},
      archivePrefix={arXiv},
      primaryClass={quant-ph},
      url={https://arxiv.org/abs/2608.19524}, 
}

@article{Chiu2025,
	author = {Chiu, Neng-Chun and Trapp, Elias C. and Guo, Jinen and Abobeih, Mohamed H. and Stewart, Luke M. and Hollerith, Simon and Stroganov, Pavel L. and Kalinowski, Marcin and Geim, Alexandra A. and Evered, Simon J. and Li, Sophie H. and Lyu, Xingjian and Peters, Lisa M. and Bluvstein, Dolev and Wang, Tout T. and Greiner, Markus and Vuleti{\'c}, Vladan and Lukin, Mikhail D.},
	date = {2025/10/01},
	doi = {10.1038/s41586-025-09596-6},
	id = {Chiu2025},
	isbn = {1476-4687},
	journal = {Nature},
	number = {8087},
	pages = {1075--1080},
	title = {Continuous operation of a coherent 3,000-qubit system},
	url = {https://doi.org/10.1038/s41586-025-09596-6},
	volume = {646},
	year = {2025}}

@article{Bhaskar2020,
	author = {Bhaskar, M. K. and Riedinger, R. and Machielse, B. and Levonian, D. S. and Nguyen, C. T. and Knall, E. N. and Park, H. and Englund, D. and Lon{\v c}ar, M. and Sukachev, D. D. and Lukin, M. D.},
	date = {2020/04/01},
	doi = {10.1038/s41586-020-2103-5},
	id = {Bhaskar2020},
	isbn = {1476-4687},
	journal = {Nature},
	number = {7801},
	pages = {60--64},
	title = {Experimental demonstration of memory-enhanced quantum communication},
	url = {https://doi.org/10.1038/s41586-020-2103-5},
	volume = {580},
	year = {2020}}

@misc{evered2026,
      title={High-fidelity entangling gates and nonlocal circuits with neutral atoms}, 
      author={Simon J. Evered and Muqing Xu and Sophie H. Li and Alexandra A. Geim and J. Pablo Bonilla Ataides and Marcin Kalinowski and Dolev Bluvstein and Nishad Maskara and Christian Kokail and Markus Greiner and Vladan Vuletić and Mikhail D. Lukin},
      year={2026},
      eprint={2604.25987},
      archivePrefix={arXiv},
      primaryClass={quant-ph},
      url={https://arxiv.org/abs/2604.25987}, 
}

@article{Bates1982,
title = {Astronomical speckle imaging},
journal = {Physics Reports},
volume = {90},
number = {4},
pages = {203-297},
year = {1982},
issn = {0370-1573},
doi = {https://doi.org/10.1016/0370-1573(82)90121-1},
url = {https://www.sciencedirect.com/science/article/pii/0370157382901211},
author = {R.H.T. Bates}
}

@article{Petrov86,
author = {R. Petrov and F. Roddier and C. Aime},
journal = {J. Opt. Soc. Am. A},
number = {5},
pages = {634--644},
publisher = {Optica Publishing Group},
title = {Signal-to-noise ratio in differential speckle interferometry},
volume = {3},
month = {May},
year = {1986},
url = {https://opg.optica.org/josaa/abstract.cfm?URI=josaa-3-5-634},
doi = {10.1364/JOSAA.3.000634},
}

@article{Jia2015,
	author = {Jia, Peng and Cai, Dongmei and Wang, Dong and Basden, Alastair},
	doi = {10.1093/mnras/stu2655},
	eprint = {https://academic.oup.com/mnras/article-pdf/447/4/3467/5697714/stu2655.pdf},
	issn = {0035-8711},
	journal = {Monthly Notices of the Royal Astronomical Society},
	month = {03},
	number = {4},
	pages = {3467-3474},
	title = {Simulation of atmospheric turbulence phase screen for large telescope and optical interferometer},
	url = {https://doi.org/10.1093/mnras/stu2655},
	volume = {447},
	year = {2015}}

@article{liu2020,
	author = {Liu, Xuewen and Morris, Tim and Saunter, Chris and de Cos Juez, Francisco Javier and Gonz{\'a}lez-Guti{\'e}rrez, Carlos and Bardou, Lisa},
	doi = {10.1093/mnras/staa1558},
	eprint = {https://academic.oup.com/mnras/article-pdf/496/1/456/33395007/staa1558.pdf},
	issn = {0035-8711},
	journal = {Monthly Notices of the Royal Astronomical Society},
	month = {07},
	number = {1},
	pages = {456-464},
	title = {Wavefront prediction using artificial neural networks for open-loop adaptive optics},
	url = {https://doi.org/10.1093/mnras/staa1558},
	volume = {496},
	year = {2020}}

@article{Poyneer2009,
author = {Lisa Poyneer and Marcos van Dam and Jean-Pierre V\'{e}ran},
journal = {J. Opt. Soc. Am. A},
number = {4},
pages = {833--846},
publisher = {Optica Publishing Group},
title = {Experimental verification of the frozen flow atmospheric turbulence assumption with use of astronomical adaptive optics telemetry},
volume = {26},
month = {Apr},
year = {2009},
url = {https://opg.optica.org/josaa/abstract.cfm?URI=josaa-26-4-833},
doi = {10.1364/JOSAA.26.000833},
}

@article{Na2017,
	author = {Ji, Na},
	date = {2017/04/01},
	doi = {10.1038/nmeth.4218},
	id = {Ji2017},
	isbn = {1548-7105},
	journal = {Nature Methods},
	number = {4},
	pages = {374--380},
	title = {Adaptive optical fluorescence microscopy},
	url = {https://doi.org/10.1038/nmeth.4218},
	volume = {14},
	year = {2017}}

@article{Bate2025,
  title = {Experimental Distributed Quantum Sensing in a Noisy Environment},
  author = {Bate, J. and Hamann, A. and Canteri, M. and Winkler, A. and Koong, Z. X. and Krutyanskiy, V. and D\"ur, W. and Lanyon, B. P.},
  journal = {Phys. Rev. Lett.},
  volume = {135},
  issue = {22},
  pages = {220801},
  numpages = {8},
  year = {2025},
  month = {Nov},
  publisher = {American Physical Society},
  doi = {10.1103/3hgx-wcdn},
  url = {https://link.aps.org/doi/10.1103/3hgx-wcdn}
}

@article{Hamann2025,
doi = {10.1088/2058-9565/add61b},
url = {https://doi.org/10.1088/2058-9565/add61b},
year = {2025},
month = {may},
publisher = {IOP Publishing},
volume = {10},
number = {3},
pages = {035028},
author = {Hamann, Arne and Aigner, Paul and Dür, Wolfgang and Sekatski, Pavel},
title = {Selective and noise-resilient wave estimation with quantum sensor networks},
journal = {Quantum Science and Technology}
}

@article{Sekatski2020,
  title = {Optimal distributed sensing in noisy environments},
  author = {Sekatski, P. and W\"olk, S. and D\"ur, W.},
  journal = {Phys. Rev. Res.},
  volume = {2},
  issue = {2},
  pages = {023052},
  numpages = {8},
  year = {2020},
  month = {Apr},
  publisher = {American Physical Society},
  doi = {10.1103/PhysRevResearch.2.023052},
  url = {https://link.aps.org/doi/10.1103/PhysRevResearch.2.023052}
}

@article{conlon2023,
	author = {Conlon, Lorc{\'a}n O. and Vogl, Tobias and Marciniak, Christian D. and Pogorelov, Ivan and Yung, Simon K. and Eilenberger, Falk and Berry, Dominic W. and Santana, Fabiana S. and Blatt, Rainer and Monz, Thomas and Lam, Ping Koy and Assad, Syed M.},
	date = {2023/03/01},
	doi = {10.1038/s41567-022-01875-7},
	id = {Conlon2023},
	isbn = {1745-2481},
	journal = {Nature Physics},
	number = {3},
	pages = {351--357},
	title = {Approaching optimal entangling collective measurements on quantum computing platforms},
	url = {https://doi.org/10.1038/s41567-022-01875-7},
	volume = {19},
	year = {2023}}

@article{Huang2021,
  title = {Information-Theoretic Bounds on Quantum Advantage in Machine Learning},
  author = {Huang, Hsin-Yuan and Kueng, Richard and Preskill, John},
  journal = {Phys. Rev. Lett.},
  volume = {126},
  issue = {19},
  pages = {190505},
  numpages = {7},
  year = {2021},
  month = {May},
  publisher = {American Physical Society},
  doi = {10.1103/PhysRevLett.126.190505},
  url = {https://link.aps.org/doi/10.1103/PhysRevLett.126.190505}
}

@article{King2024,
  title = {Exponential Learning Advantages with Conjugate States and Minimal Quantum Memory},
  author = {King, Robbie and Wan, Kianna and McClean, Jarrod R.},
  journal = {PRX Quantum},
  volume = {5},
  issue = {4},
  pages = {040301},
  numpages = {41},
  year = {2024},
  month = {Oct},
  publisher = {American Physical Society},
  doi = {10.1103/PRXQuantum.5.040301},
  url = {https://link.aps.org/doi/10.1103/PRXQuantum.5.040301}
}

@article{Tsang2019,
  title = {Semiparametric estimation for incoherent optical imaging},
  author = {Tsang, Mankei},
  journal = {Phys. Rev. Res.},
  volume = {1},
  issue = {3},
  pages = {033006},
  numpages = {14},
  year = {2019},
  month = {Oct},
  publisher = {American Physical Society},
  doi = {10.1103/PhysRevResearch.1.033006},
  url = {https://link.aps.org/doi/10.1103/PhysRevResearch.1.033006}
}

@article{Booth2014,
	author = {Booth, Martin J},
	date = {2014/04/01},
	doi = {10.1038/lsa.2014.46},
	id = {Booth2014},
	isbn = {2047-7538},
	journal = {Light: Science \& Applications},
	number = {4},
	pages = {e165--e165},
	title = {Adaptive optical microscopy: the ongoing quest for a perfect image},
	url = {https://doi.org/10.1038/lsa.2014.46},
	volume = {3},
	year = {2014}}

@article{Hunziker2018,
	author = {{Hunziker, S.} and {Quanz, S. P.} and {Amara, A.} and {Meyer, M. R.}},
	title = {PCA-based approach for subtracting thermal background emission in high-contrast imaging data},
	DOI= "10.1051/0004-6361/201731428",
	url= "https://doi.org/10.1051/0004-6361/201731428",
	journal = {Astronomy \& Astrophysics},
	year = 2018,
	volume = 611,
	pages = "A23",
}

@article{Rao2024,
	author = {Rao, Changhui and Zhong, Libo and Guo, Youming and Li, Min and Zhang, Lanqiang and Wei, Kai},
	date = {2024/05/01},
	doi = {10.1186/s43074-024-00118-7},
	id = {Rao2024},
	isbn = {2662-1991},
	journal = {PhotoniX},
	number = {1},
	pages = {16},
	title = {Astronomical adaptive optics: a review},
	url = {https://doi.org/10.1186/s43074-024-00118-7},
	volume = {5},
	year = {2024}}

@misc{allen2025,
      title={Quantum Computing Enhanced Sensing}, 
      author={Richard R. Allen and Francisco Machado and Isaac L. Chuang and Hsin-Yuan Huang and Soonwon Choi},
      year={2025},
      eprint={2501.07625},
      archivePrefix={arXiv},
      primaryClass={quant-ph},
      url={https://arxiv.org/abs/2501.07625}, 
}

@misc{dimitroff2026,
      title={Learning Gaussian optical states with quantum computers}, 
      author={Spencer Dimitroff and John Kallaugher and Ashe Miller and Mohan Sarovar},
      year={2026},
      eprint={2605.05325},
      archivePrefix={arXiv},
      primaryClass={quant-ph},
      url={https://arxiv.org/abs/2605.05325}, 
}

@article{Kelly2026,
  title = {Quantum-processing-assisted classical communication},
  author = {Werker Smith, Kelly and Boroson, Don and Guha, Saikat and Borregaard, Johannes},
  journal = {Phys. Rev. Appl.},
  volume = {25},
  issue = {4},
  pages = {044037},
  numpages = {11},
  year = {2026},
  month = {Apr},
  publisher = {American Physical Society},
  doi = {10.1103/wp82-718s},
  url = {https://link.aps.org/doi/10.1103/wp82-718s}
}

@article{Kalman1960,
  author  = {Kalman, Rudolph E.},
  title   = {A New Approach to Linear Filtering and Prediction Problems},
  journal = {Journal of Basic Engineering},
  volume  = {82},
  number  = {1},
  pages   = {35--45},
  year    = {1960}
}

@inproceedings{ChenCotlerHuangLi2021,
  author    = {Chen, Sitan and Cotler, Jordan and Huang, Hsin-Yuan and Li, Jerry},
  title     = {Exponential Separations Between Learning with and without Quantum Memory},
  booktitle = {Proceedings of the 62nd IEEE Symposium on Foundations of Computer Science},
  pages     = {574--585},
  publisher = {IEEE},
  year      = {2021}
}

@article{AharonovCotlerQi2022,
  author  = {Aharonov, Dorit and Cotler, Jordan and Qi, Xiao-Liang},
  title   = {Quantum Algorithmic Measurement},
  journal = {Nature Communications},
  volume  = {13},
  pages   = {887},
  year    = {2022}
}

@article{HuangLearningExperiments2022,
  author  = {Huang, Hsin-Yuan and Broughton, Michael and Cotler, Jordan and Chen, Sitan and Li, Jerry and Mohseni, Masoud and Neven, Hartmut and Babbush, Ryan and Kueng, Richard and Preskill, John and McClean, Jarrod R.},
  title   = {Quantum Advantage in Learning from Experiments},
  journal = {Science},
  volume  = {376},
  number  = {6598},
  pages   = {1182--1186},
  year    = {2022}
}

@article{mokeev2026enhancing,
  title={Enhancing Optical Imaging via Quantum Computation},
  author={Mokeev, Aleksandr and Saif, Babak and Lukin, Mikhail D and Borregaard, Johannes},
  journal={PRX Quantum},
  volume={7},
  number={1},
  pages={010318},
  year={2026},
  publisher={APS}
}

@article{stas2026entanglement,
  title={Entanglement-assisted Non-local Optical Interferometry in a Quantum Network},
  author={Stas, P-J and Wei, Y-C and Sirotin, M and Huan, YQ and Yazlar, U and Abdo Arias, F and Knyazev, E and Baranes, G and Machielse, B and Grandi, S and others},
  journal={Nature},
  volume={651},
  number={8105},
  pages={326--332},
  year={2026},
  publisher={Nature Publishing Group}
}

@article{cotler2025noisy,
  title={Noisy Quantum Learning Theory},
  author={Cotler, Jordan and Gong, Weiyuan and Kannan, Ishaan},
  journal={arXiv preprint arXiv:2512.10929},
  year={2025}
}

@article{kannan2026exponential,
  title={Exponential speedups in fault-tolerant processing of quantum experiments},
  author={Kannan, Ishaan and Putterman, Harald and Cotler, Jordan},
  journal={arXiv preprint arXiv:2605.02057},
  year={2026}
}

@article{cotler2026quantum,
  title={Quantum Advantage for Sensing Properties of Classical Fields},
  author={Cotler, Jordan and Danielson, Daine L and Kannan, Ishaan},
  journal={arXiv preprint arXiv:2602.17591},
  year={2026}
}

@article{khabiboulline2019optical,
  title={Optical interferometry with quantum networks},
  author={Khabiboulline, Emil T and Borregaard, Johannes and De Greve, Kristiaan and Lukin, Mikhail D},
  journal={Physical Review Letters},
  volume={123},
  number={7},
  pages={070504},
  year={2019},
  publisher={APS}
}

@article{wang2026entanglement,
  title={Entanglement advantage in sensing power-law spatiotemporal noise correlations},
  author={Wang, Yu-Xin and Brady, Anthony J and Belliardo, Federico and Gorshkov, Alexey V},
  journal={arXiv preprint arXiv:2603.15742},
  year={2026}
}

@article{li2026exponential,
  title={An exponential sample-complexity advantage for coherent quantum inference},
  author={Li, Zhaoyi and Theil, Elias and Harrow, Aram W and Chuang, Isaac},
  journal={arXiv preprint arXiv:2605.21457},
  year={2026}
}

@article{lidar2014review,
  title={Review of Decoherence-Free Subspaces, Noiseless Subsystems, and Dynamical Decoupling},
  author={Lidar, Daniel A},
  journal={Quantum information and computation for chemistry},
  pages={295--354},
  year={2014},
  publisher={Wiley Online Library}
}

@article{reiserer2014quantum,
  title={A quantum gate between a flying optical photon and a single trapped atom},
  author={Reiserer, Andreas and Kalb, Norbert and Rempe, Gerhard and Ritter, Stephan},
  journal={Nature},
  volume={508},
  number={7495},
  pages={237--240},
  year={2014},
  publisher={Nature Publishing Group UK London}
}

@article{nguyen2019integrated,
  title={An integrated nanophotonic quantum register based on silicon-vacancy spins in diamond},
  author={Nguyen, CT and Sukachev, DD and Bhaskar, MK and Machielse, Bartholomeus and Levonian, DS and Knall, EN and Stroganov, Pavel and Chia, Cleaven and Burek, MJ and Riedinger, Ralf and others},
  journal={Physical Review B},
  volume={100},
  number={16},
  pages={165428},
  year={2019},
  publisher={APS}
}

@article{bluvstein2022quantum,
  title={A quantum processor based on coherent transport of entangled atom arrays},
  author={Bluvstein, Dolev and Levine, Harry and Semeghini, Giulia and Wang, Tout T and Ebadi, Sepehr and Kalinowski, Marcin and Keesling, Alexander and Maskara, Nishad and Pichler, Hannes and Greiner, Markus and others},
  journal={Nature},
  volume={604},
  number={7906},
  pages={451--456},
  year={2022},
  publisher={Nature Publishing Group UK London}
}

@article{ransford202698,
  title={A 98-qubit trapped-ion quantum computer with all-to-all connectivity},
  author={Ransford, Anthony and Allman, MS and Arkinstall, Jake and Campora, JP and Cooper, Samuel F and Delaney, Robert D and Dreiling, Joan M and Estey, Brian and Figgatt, Caroline and Hall, Alex and others},
  journal={Nature},
  volume={655},
  number={8121},
  pages={81--86},
  year={2026},
  publisher={Nature Publishing Group}
}

@article{google2025quantum,
  title={Quantum error correction below the surface code threshold},
  journal={Nature},
  volume={638},
  number={8052},
  pages={920--926},
  year={2025},
  publisher={Nature Publishing Group UK London}
}

@article{shi2026efficient,
  title={Efficient implementation of a quantum algorithm with a trapped ion qudit},
  author={Shi, Xiaoyang and Sinanan-Singh, Jasmine and Burke, Timothy J and Chiaverini, John and Chuang, Isaac L},
  journal={Nature Communications},
  volume={17},
  number={1},
  pages={1911},
  year={2026},
  publisher={Nature Publishing Group UK London}
}

@article{cotler2026noisy,
  title={Noisy quantum learning theory},
  author={Cotler, Jordan and Gong, Weiyuan and Kannan, Ishaan},
  journal={Nature Communications},
  year={2026},
  publisher={Nature Publishing Group UK London}
}

@article{bulgarelli2026quantum,
  title={Quantum computing for astrophysics and cosmology: Results from the INAF activities},
  author={Bulgarelli, Andrea and Burigana, Carlo and Cappelli, Luca and Cardone, Vincenzo Fabrizio and Cavuoti, Stefano and Farsian, Farida and Graziotti, Irene and Macaluso, Antonio and Meneghetti, Massimo and Murante, Giuseppe and others},
  journal={Astronomy and Computing},
  pages={101108},
  year={2026},
  publisher={Elsevier}
}

@article{brunet2024quantum,
  title={Quantum radio astronomy: Data encodings and quantum image processing},
  author={Brunet, Thomas and Tolley, Emma and Corda, Stefano and Ilic, Roman and Broekema, P Chris and Kneib, J-P},
  journal={Astronomy and Computing},
  volume={47},
  pages={100796},
  year={2024},
  publisher={Elsevier}
}

@article{Huggins_2021,
   title={Virtual Distillation for Quantum Error Mitigation},
   volume={11},
   ISSN={2160-3308},
   url={http://dx.doi.org/10.1103/PhysRevX.11.041036},
   DOI={10.1103/physrevx.11.041036},
   number={4},
   journal={Physical Review X},
   publisher={American Physical Society (APS)},
   author={Huggins, William J. and McArdle, Sam and O’Brien, Thomas E. and Lee, Joonho and Rubin, Nicholas C. and Boixo, Sergio and Whaley, K. Birgitta and Babbush, Ryan and McClean, Jarrod R.},
   year={2021},
   month=Nov }

@article{Cotler_2019,
   title={Quantum Virtual Cooling},
   volume={9},
   ISSN={2160-3308},
   url={http://dx.doi.org/10.1103/PhysRevX.9.031013},
   DOI={10.1103/physrevx.9.031013},
   number={3},
   journal={Physical Review X},
   publisher={American Physical Society (APS)},
   author={Cotler, Jordan and Choi, Soonwon and Lukin, Alexander and Gharibyan, Hrant and Grover, Tarun and Tai, M. Eric and Rispoli, Matthew and Schittko, Robert and Preiss, Philipp M. and Kaufman, Adam M. and Greiner, Markus and Pichler, Hannes and Hayden, Patrick},
   year={2019},
   month=July }

@misc{montanaro2017learningstabilizerstatesbell,
      title={Learning stabilizer states by Bell sampling}, 
      author={Ashley Montanaro},
      year={2017},
      eprint={1707.04012},
      archivePrefix={arXiv},
      primaryClass={quant-ph},
      url={https://arxiv.org/abs/1707.04012}, 
}

@misc{montanaro2018surveyquantumpropertytesting,
      title={A Survey of Quantum Property Testing}, 
      author={Ashley Montanaro and Ronald de Wolf},
      year={2018},
      eprint={1310.2035},
      archivePrefix={arXiv},
      primaryClass={quant-ph},
      url={https://arxiv.org/abs/1310.2035}, 
}

@misc{khan2026quantumcomputationalsensing,
      title={Quantum Computational Sensing}, 
      author={Saeed A. Khan and Sridhar Prabhu and Logan G. Wright and Peter L. McMahon},
      year={2026},
      eprint={2507.16918},
      archivePrefix={arXiv},
      primaryClass={quant-ph},
      url={https://arxiv.org/abs/2507.16918}, 
}

@article{babbush2026grand,
  title={Grand challenge of quantum applications},
  author={Babbush, Ryan and King, Robbie and Boixo, Sergio and Huggins, William and Khattar, Tanuj and Low, Guang Hao and McClean, Jarrod R and O’Brien, Thomas and Rubin, Nicholas C},
  journal={PRX Quantum},
  volume={7},
  number={2},
  pages={020101},
  year={2026},
  publisher={APS}
}

@article{huang2026vast,
  title={Vast World of Quantum Advantage},
  author={Huang, Hsin-Yuan and Choi, Soonwon and McClean, Jarrod R and Preskill, John},
  journal={Physical Review X},
  volume={16},
  number={3},
  pages={030501},
  year={2026},
  publisher={APS}
}

@article{heeman2019clinical,
  title={Clinical applications of laser speckle contrast imaging: a review},
  author={Heeman, Wido and Steenbergen, Wiendelt and van Dam, Gooitzen M and Boerma, E Christiaan},
  journal={Journal of biomedical optics},
  volume={24},
  number={8},
  pages={080901--080901},
  year={2019},
  publisher={Society of Photo-Optical Instrumentation Engineers}
}

@article{oh2021quantum,
  title={Quantum limits of superresolution in a noisy environment},
  author={Oh, Changhun and Zhou, Sisi and Wong, Yat and Jiang, Liang},
  journal={Physical Review Letters},
  volume={126},
  number={12},
  pages={120502},
  year={2021},
  publisher={APS}
}

@article{knox1974recovery,
  title={Recovery of images from atmospherically degraded short-exposure photographs},
  author={Knox, Keith T and Thompson, Brian J},
  journal={Astrophysical Journal, vol. 193, Oct. 1, 1974, pt. 2, p. L45-L48.},
  volume={193},
  pages={L45--L48},
  year={1974}
}

@inproceedings{sarovar2023quantum,
  title={Quantum computational imaging and sensing},
  author={Sarovar, Mohan},
  booktitle={Quantum Nanophotonic Materials, Devices, and Systems 2023},
  volume={12657},
  pages={6--9},
  year={2023},
  organization={SPIE}
}

@misc{depradenne2026restrictionsnoncliffordfaulttolerance,
      title={Restrictions on non-Clifford fault tolerance and ruling out beyond-SQL quantum metrology}, 
      author={Constantin Cedillo Vayson de Pradenne and Ishaan Kannan and Harald Putterman and Jordan Cotler},
      year={2026},
      eprint={2607.27342},
      archivePrefix={arXiv},
      primaryClass={quant-ph},
      url={https://arxiv.org/abs/2607.27342}, 
}

@article{weigelt1977modified,
  title={Modified astronomical speckle interferometry “speckle masking”},
  author={Weigelt, GP},
  journal={Optics Communications},
  volume={21},
  number={1},
  pages={55--59},
  year={1977},
  publisher={Elsevier}
}

@inproceedings{feinberg2024habitable,
  title={The Habitable Worlds Observatory engineering view: status, plans, and opportunities},
  author={Feinberg, Lee and Ziemer, John and Ansdell, Megan and Crooke, Julie and Dressing, Courtney and Mennesson, Bertrand and O'Meara, John and Pepper, Joshua and Roberge, Aki},
  booktitle={Space Telescopes and Instrumentation 2024: Optical, Infrared, and Millimeter Wave},
  volume={13092},
  pages={548--559},
  year={2024},
  organization={SPIE}
}

@article{feinberg2026habitable,
  title={Habitable Worlds Observatory’s concept and technology maturation: initial feasibility and trade space exploration},
  author={Feinberg, Lee D and Sitarski, Breann N and McElwain, Michael W and Arney, Giada N and Baker, Caleb and Bolcar, Matthew R and Levine, Marie and Liu, Alice and Mennesson, Bertrand and Roberge, Aki and others},
  journal={Journal of Astronomical Telescopes, Instruments, and Systems},
  volume={12},
  number={4},
  pages={041031--041031},
  year={2026},
  publisher={Society of Photo-Optical Instrumentation Engineers}
}

@misc{Sirotin2026QuantumCameraCode,
  author       = {Maxim Sirotin and others},
  title        = {Quantum Camera: Simulation and Figure Reproduction Code},
  year         = {2026},
  howpublished = {\url{https://github.com/MaximSirotin/Quantum_Camera}}
}

\clearpage

\end{document}


\maketitle
\tableofcontents
\clearpage

\section{Scope of the comparisons}
\label{sec:si-imaging-advantage}\label{sec:si-scope}
In the main text, quantum advantage in imaging refers to the reduced number of detected source photons needed to resolve scientific parameters of a scene when the same correlated nuisance affects nearby optical samples. An image contains many possible science directions, and a receiver optimized for one known parameter need not acquire all of them efficiently. We therefore distinguish a task-specific directional separation from a {worst-direction imaging advantage}, defined by the least informative identifiable science direction of a fixed acquisition receiver. The definition is local and refers to a specified source model, nuisance law, photon-pair input state, and measurement-access class. It does not assert an advantage for every science parameter or a global minimax guarantee for reconstructed images. As a classical benchmark we use universal classical receiver (direct imaging with any allowed classical correlations), PPT receiver (any separable measurement, including task-aware ones) and Kalman-filtering receiver which learns the nuisance, see definitions and optimality conditions below.

We compare the number of detected photons required for a fixed local precision when consecutive samples experience unknown correlated nuisance. The relevant resource is the number of input photon pairs (we also call them samples or copies). The advantage is an Fisher-Information-limited sample-complexity ratio for a specified task, nuisance model, and measurement class.

Three kinds of statements are distinguished. First, the binary-source theorems in Secs.~\ref{sec:si-binary-twirl}--\ref{sec:si-finite-gaussian} compare a coherent measurement with {all PPT} measurements of the same two-photon state. Second, the Fourier-power result of Sec.~\ref{sec:si-universal} concerns acquisition of many unknown image parameters with {one fixed receiver}; it does not imply an $O(1/D)$ penalty for every single known observable. Third, numerical classical receivers---direct imaging with classical correlations and a linear--Gaussian Kalman predictor---are {explicit benchmarks}, not universal optimizations over all classical histories.

\begin{figure*}[htbp]
 \centering
 \includegraphics[width=17cm]{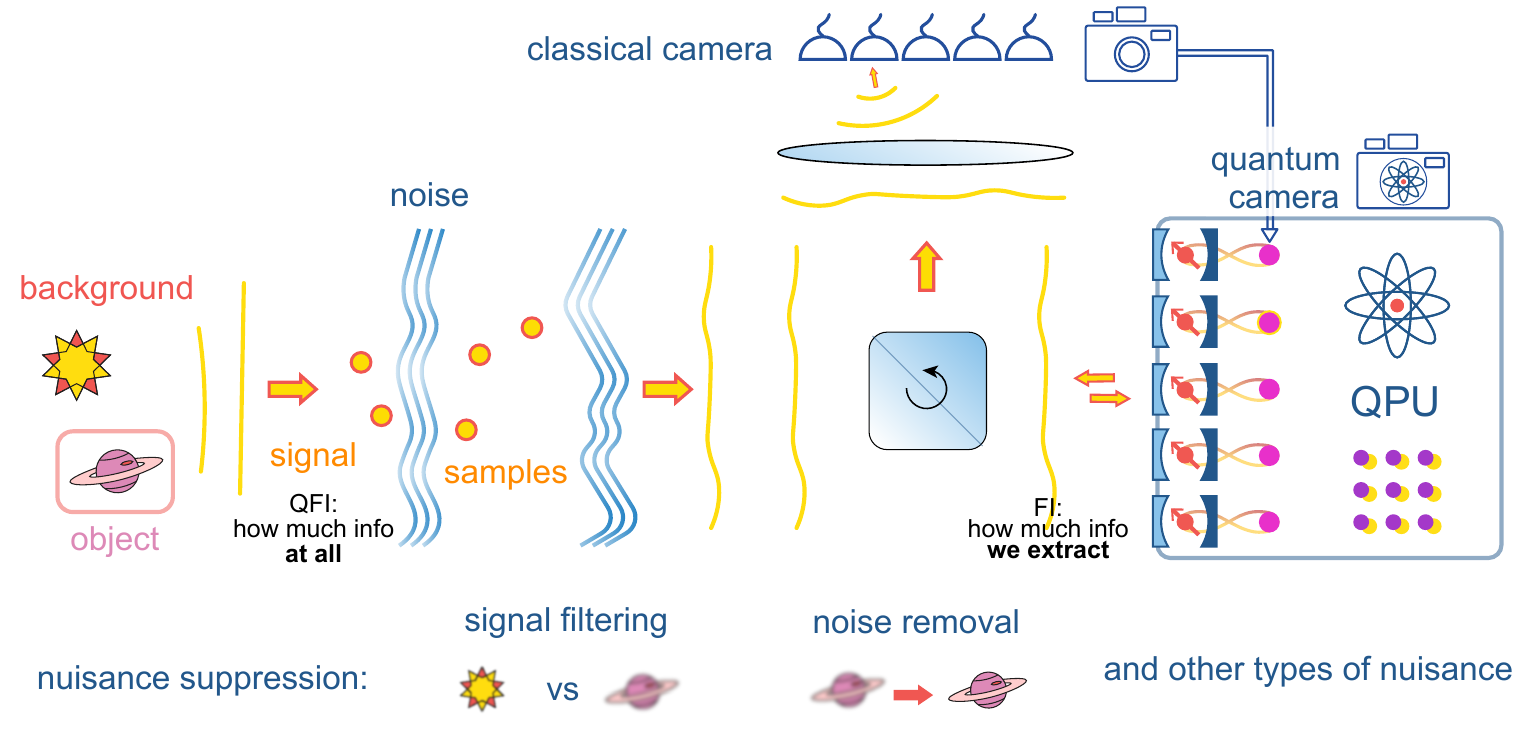}
 \caption{\textbf{Quantum camera concept and coherence nuisance suppression examples.}}
 \label{fig:concept}
\end{figure*}

\section{Quantum camera, setting and definitions}
\label{sec:si-setting}\label{app:definitions}\label{app:setting}

The purpose of this Supplementary Information is to identify when correlated nuisance makes separate measurements statistically suboptimal, and to quantify the information retained by coherent relational measurements. We begin with equal-brightness binary separation, for which the science parameter and the measurement advantage can both be stated explicitly. We first solve the complete common-phase-twirl experiment, then turn on finite common Gaussian noise, and finally allow the two phase screens (additive diagonal noise) to differ. The later sections extend the measurement statements to general source observables and discuss classical-history assistance. Following the access-model viewpoint of quantum-learning theory, each comparison specifies the input state, the allowed measurement class, and the information supplied to the receiver. Most pair-state converse bounds use the positive-partial-transpose (PPT) condition, a necessary separability criterion; they therefore bound the corresponding separable and single-copy measurement classes, including within-pair classical feedforward. They do not automatically bound history-assisted estimation with temporally correlated nuisance. The following definitions fix notation and resource conventions.

\textbf{Optical modes, pupil, baselines, and coherences.}
The entrance telescope pupil is discretized into $D$ orthogonal spatial modes, with one-photon Hilbert space $\mathcal H_{\rm opt}\simeq\mathbb C^D$ and pixel basis $\{\ket j\}_{j=0}^{D-1}$. We use ordinary non-wrapping indices for a finite physical pupil; cyclic indexing, $\mathbb Z_D=\{0,\ldots,D-1\}$, is used only where an explicitly periodic contrast model is specified. The optical dimension $D$ is a numerical and hardware resolution parameter and should not be identified with the number of physical atmospheric degrees of freedom. A {pupil baseline} is a pair of pupil modes $(j,k)$, or, for a translation-invariant one-dimensional pupil, their relative offset
\begin{equation}
    m=k-j,\qquad 0\le j<k<D,
    \label{eq:app_baseline_definition}
\end{equation}
When needed, $r_m$ denotes the corresponding physical separation in the pupil plane. In this Appendix ``baseline'' means a separation between pupil modes of one telescope, not a separation between distinct telescopes.

For a one-photon density operator $\rho$, the matrix element
\begin{equation}
    c_{jk}=\bra j\rho\ket k
    \label{eq:app_mutual_coherence}
\end{equation}
is the spatial first-order coherence between pupil modes $j$ and $k$. For the shift-invariant source family used below, all matrix elements with the same offset are equal up to the common normalization. We define the normalized complex visibility
\begin{equation}
    \Phi_m
    :=
    D\,\bra{j+m}\rho\ket j ,
    \qquad
    \Phi_0=1 ,
    \label{eq:app_visibility_definition}
\end{equation}
which is independent of $j$ for a uniform pupil wherever $0\le j,j+m<D$. Equivalently, if $P(\theta)$ is the normalized angular brightness distribution,
\begin{equation}
    \Phi_m
    =
    \int d\theta\,
    P(\theta)
    e^{2\pi i m\theta/D}.
    \label{eq:app_visibility_fourier}
\end{equation}
Thus the optical coherences of the pupil state are the Fourier coefficients, or {visibilities}, of the source brightness distribution. For $r$ point sources with normalized brightnesses $p_a$, Eq.~\eqref{eq:app_visibility_fourier} becomes $\Phi_m=\sum_{a=1}^r p_a e^{2\pi i m\theta_a/D}$. The quantities $|\Phi_m|^2$ will be called {visibility powers}, while
\begin{equation}
    C_m:=\Phi_m^*\Phi_{m+1}
    \label{eq:app_visibility_link}
\end{equation}
will be called {neighboring visibility links}. The set $\{|\Phi_m|^2,C_m\}$ determines the complex visibility sequence whenever the retained visibility graph is connected and the relevant amplitudes do not vanish; zeros can be bridged by using longer links $\Phi_m^*\Phi_{m+\ell}$.\\

\textbf{Quantum camera, quantum telescope, transduction, and heralding.}
A {quantum camera} is an optical receiver that coherently transfers the spatial quantum state carried by an incident photon into a matter register while producing a classical herald that certifies successful capture without revealing the stored spatial label. Candidate matter memories include cavity-coupled rubidium (Rb) atoms or solid-state emitters such as silicon-vacancy (SiV) centers. A spin-dependent cavity reflection phase can implement a controlled-$Z$ (CZ) interaction between the optical mode and the local matter qubit. Related optical-to-memory and quantum-computation-assisted imaging architectures are discussed in Refs.~\cite{stas2026entanglement,mokeev2026enhancing}; implementation-dependent loss and error assumptions are detailed in Sec.~\ref{sec:si-hardware}.

By {transduction} we mean coherent transfer of a quantum state from the optical degree of freedom to the matter-memory degree of freedom. Abstractly, after the local photon--memory interaction, the successful branch can be represented as
\begin{equation}
    \sum_{j=0}^{D-1}\alpha_j\ket j_{\rm opt}\ket{0}_{\rm mem}
    \longrightarrow
    \sum_{j=0}^{D-1}\alpha_j\ket j_{\rm opt}\ket j_{\rm mem},
    \label{eq:app_preherald_transduction}
\end{equation}
where $\ket j_{\rm mem}$ denotes the logical memory state associated with pupil mode $j$. The optical photon is then measured in a basis mutually unbiased with respect to the pupil basis. We use the $D$-dimensional discrete Fourier transform
\begin{equation}
    F_D\ket j
    =
    \frac{1}{\sqrt D}
    \sum_{\ell=0}^{D-1}
    e^{2\pi i j\ell/D}
    \ket\ell ,
    \qquad
    \ket{\widetilde\ell}
    =
    F_D^\dagger\ket\ell
    =
    \frac{1}{\sqrt D}
    \sum_{j=0}^{D-1}
    e^{-2\pi i j\ell/D}\ket j .
    \label{eq:app_optical_fourier_basis}
\end{equation}
For $D=2$, $F_2$ is the Hadamard transform and $\{\ket{\widetilde0},\ket{\widetilde1}\}$ is the usual $X$ or $\{\ket+,\ket-\}$ basis. Experimentally, a Fourier-transforming optical stage followed by a spatially resolving photodetector or charge-coupled-device (CCD) camera can implement the corresponding erasure measurement; for more than two discrete modes an appropriate Fourier mode sorter or interferometric network may be used.

Conditioned on Fourier outcome $\ell$, Eq.~\eqref{eq:app_preherald_transduction} leaves the memory in
\begin{equation}
    \ket{\psi_{\rm mem}^{(\ell)}}
    =
    Z_\ell^\dagger
    \sum_{j=0}^{D-1}\alpha_j\ket j_{\rm mem},
    \qquad
    Z_\ell
    =
    \sum_{j=0}^{D-1}
    e^{2\pi i j\ell/D}
    \ket j\!\bra j .
    \label{eq:app_herald_byproduct}
\end{equation}
The outcome $\ell$ therefore produces only a known diagonal byproduct $Z_\ell^\dagger$, which can be corrected physically or tracked as a known classical phase-frame record. Because $|\braket{\widetilde\ell|j}|^2=1/D$ for every $j,\ell$, the heralding detector does not reveal the pupil mode $j$ and hence does not measure the astronomical spatial information. In this sense {heralding} means recording that the photon was successfully transduced, together with the known Fourier-basis byproduct label, while leaving the source state available for later coherent processing.

A {quantum telescope} is a conventional collecting telescope equipped with such a quantum camera and connected to a quantum processor capable of preserving several heralded photon states and performing coherent multi-copy measurements before final destructive readout.

\textbf{Logical compression and stored copies.}
The most direct physical memory representation can be one-hot, with $D$ spatial memory elements and one logical excitation or phase record associated with the occupied pupil mode. Because the one-photon sector is only $D$ dimensional, it may be coherently compressed into
\begin{equation}
    q=\left\lceil\log_2D\right\rceil
    \label{eq:app_q_definition}
\end{equation}
qubits through an isometry
\begin{equation}
    \ket{1_j}_{\rm one\mbox{-}hot}
    \longrightarrow
    \ket j_{\rm addr}.
    \label{eq:app_logical_compression}
\end{equation}

\textbf{POVMs and receiver classes.}
A positive-operator-valued measure (POVM) is a collection of positive operators $\{M_y\}$ satisfying $M_y\succeq0$ and $\sum_yM_y=I$; for state $\rho$, outcome $y$ occurs with probability $p_y=\Tr(M_y\rho)$. A {joint} two-copy POVM is an unrestricted POVM on $\mathcal H_{\rm opt}^{\otimes2}$, or on the corresponding two stored matter registers, and its effects may be entangled across the two (or more) copies. A {product} measurement measures the two copies independently. A {local operation with classical communication} (LOCC) may additionally adapt the measurement of one copy to classical outcomes obtained from the other. For the temporal ordering relevant here, a no-quantum-memory receiver implements a one-way-LOCC POVM,
\begin{equation}
    M_{a,b}
    =
    A_a\otimes B_{b|a},
    \label{eq:app_oneway_locc}
\end{equation}
where the second one-photon POVM may depend on the first classical result $a$, but no quantum system carrying information about the first photon is preserved until the second arrives.

A POVM is {separable} (SEP) if every effect can be written
\begin{equation}
    M_y
    =
    \sum_\lambda
    A_{y,\lambda}\otimes B_{y,\lambda},
    \qquad
    A_{y,\lambda}\succeq0,\quad
    B_{y,\lambda}\succeq0.
    \label{eq:app_sep_definition}
\end{equation}
A positive operator has {positive partial transpose} (PPT) if $M_y^\Gamma\succeq0$, where $\Gamma$ denotes transposition on one of the two copies in a fixed basis. We denote the class of POVMs whose effects are PPT operators as ${\rm PPT}$. The measurement classes obey
\begin{equation}
    {\rm one\mbox{-}way\ LOCC}
    \subseteq
    {\rm LOCC}
    \subseteq
    {\rm SEP}
    \subseteq
    {\rm PPT}
    \subseteq
    {\rm joint\ POVM}.
    \label{eq:app_measurement_hierarchy}
\end{equation}
An upper bound proved for all PPT measurements automatically applies to separable and no-quantum-memory receivers. Direct imaging, one-photon spatial-mode demultiplexing (SPADE), adaptive one-photon mode measurements, randomized informationally complete measurements like classical shadows, and arbitrary classical correlations of their outcomes all remain inside the no-quantum-memory/separable side of this comparison.\\

\textbf{SWAP, exchange parity, and measurement nonlocality.}
The two-copy SWAP operator is
\begin{equation}
    S\ket{j,k}
    =
    \ket{k,j},
    \qquad
    P_{\rm sym}
    =
    \frac{I+S}{2},
    \qquad
    P_{\rm asym}
    =
    \frac{I-S}{2}.
    \label{eq:app_swap_definition}
\end{equation}
A global SWAP test measures only the symmetric/antisymmetric parity and satisfies $\langle S\rangle_{\rho\otimes\rho}=\Tr\rho^2$, so it directly estimates purity. A {block-resolved SWAP} additionally resolves the relative pupil baseline. For a nonzero offset $m$, define the two-dimensional exchange block
\begin{equation}
    \mathcal H_{j,m}
    =
    \operatorname{span}
    \{\ket{j,j+m},\ket{j+m,j}\}
    \label{eq:app_exchange_block_definition}
\end{equation}
and its symmetric/antisymmetric states
\begin{equation}
    \ket{\pm_{j,m}}
    =
    \frac{\ket{j,j+m}\pm\ket{j+m,j}}{\sqrt2}.
    \label{eq:app_exchange_states_definition}
\end{equation}
The corresponding block-resolved projectors are
\begin{equation}
    \Pi_{m,\pm}
    =
    \sum_{j\in\mathcal J_m}
    \ket{\pm_{j,m}}\!\bra{\pm_{j,m}},
    \label{eq:app_brs_definition}
\end{equation}
where $\mathcal J_m$ is chosen so that each unordered exchange block is counted once; the canonical convention is specified explicitly in Sec.~\ref{app:u1d_proof}.

By {measurement nonlocality} or {data hiding} we mean that information present in a multipartite state is accessible strictly more efficiently to some joint POVM than to any LOCC/separable measurement. 

\textbf{Nuisance, common-mode rejection, and twirling.}
The {scientific parameters} $\bm\vartheta$ are the source properties to be inferred, such as position, separation, brightnesses, velocity, or an image. A {nuisance parameter} is a variable that affects the measured state but is not itself the scientific target. The principal nuisance here is the atmospheric phase vector $\bm\eta_t\in\mathbb R^{d_\eta}$, where $d_\eta$ is the effective number of physical atmospheric degrees of freedom and need not equal $D$. A two-copy relational measurement is {common-mode invariant} with respect to a nuisance group if its statistics are unchanged when the same nuisance transformation acts on both copies. The SWAP operator, for example, obeys
\begin{equation}
    [S,U\otimes U]=0
    \label{eq:app_swap_common_mode}
\end{equation}
for every one-copy unitary $U$. If the two photons instead experience $U_t$ and $U_{t+\tau}$, the residual error depends on their {relative} or {differential nuisance}, represented later by $\delta\bm\eta_\tau=\bm\eta_{t+\tau}-\bm\eta_t$ and its covariance. The symbol $d_{\rm eff}$ denotes the effective nuisance dimension relevant to that differential quantity; typically $d_{\rm eff}\ll d_\eta$ for sufficiently short temporal separation.

A {twirl} is the average of a unitary group action on an input operator with respect to a specified probability measure on the group. For example, on a two-copy state,
\begin{equation}
    \mathcal T_{\mathcal G}(X)
    =
    \int_{\mathcal G}
    (U_g\otimes U_g)\,
    X\,
    (U_g^\dagger\otimes U_g^\dagger)\,
    dg,
    \label{eq:app_twirl_definition}
\end{equation}
where $dg$ is the normalized invariant group measure. The $U(1)^D$ phase twirl used below independently randomizes the phase of every pupil mode, whereas the full Haar $U(D)$ twirl averages over all $D$-dimensional unitaries. These are exact theorem models of unlearned nuisance; the realistic atmosphere is introduced later as a correlated Gaussian process rather than a full twirl.\\

\textbf{Relational Fourier sampling.}
Relational Fourier sampling (RFS) is the family of joint two-copy measurements introduced in this work to estimate source Fourier information without first reconstructing the instantaneous common phase screen. Its elementary operation is exchange-parity measurement resolved by relative pupil offset. For the translation-invariant state of Eq.~\eqref{eq:app_visibility_definition}, the {plain block-resolved SWAP} measures the projectors $\Pi_{m,\pm}$ in Eq.~\eqref{eq:app_brs_definition}. After conditioning on the offset block and factoring out its known occurrence probability, the parity contrast is
\begin{equation}
    \frac{p_{m,+}-p_{m,-}}{p_{m,+}+p_{m,-}}
    =
    |\Phi_m|^2 .
    \label{eq:app_plain_rfs_signal}
\end{equation}
Thus plain RFS estimates the visibility powers $\{|\Phi_m|^2\}$. These are exactly the source quantities that survive the full common $U(1)^D$ phase twirl considered in Sec.~\ref{app:u1d_proof}.

The {rotated block-resolved SWAP} supplements the same relative-coordinate decomposition with a coherent two-mode rotation between neighboring offset sectors $m$ and $m+1$ before exchange-parity readout. Writing this controllable rotation as $R_{m,m+1}(\chi)$ we get basis vectors
\begin{equation}
\ket{\phi_{j,\pm}^{(\chi)}}
=
\frac{1}{\sqrt{2}}
\left(
\ket{\pm_{j,m}}
+
e^{i\chi}\ket{\pm_{j,m+1}}
\right),
\end{equation}
the displayed vectors specify selected effects within the matched subspaces,
\begin{equation}
\Pi_{m,\pm}^{(\chi)}
=
\sum_{j\in\mathcal J_m}
\ket{\phi_{j,\pm}^{(\chi)}}
\bra{\phi_{j,\pm}^{(\chi)}}.
\end{equation}
The orthogonal partners $(\ket{\pm_{j,m}}-e^{i\chi}\ket{\pm_{j,m+1}})/\sqrt2$ and the untouched or unmatched sectors are also measured so that the collection of effects is complete. The indicated sums are therefore selected effects, not by themselves a normalized POVM; matching is restricted to pairs of physical, non-overlapping exchange blocks, and the registered outcome includes the measurement setting. Absorbing the known block-occurrence normalization into the definition of the reported contrast, the resulting parity signal is
\begin{equation}
    \mathcal R_{m,\chi}
    =
    \operatorname{Re}
    \left[
      e^{-i\chi}
      \Phi_m^*\Phi_{m+1}
    \right].
    \label{eq:app_rotated_rfs_signal}
\end{equation}
The settings $\chi=0$ and $\chi=\pi/2$ recover respectively the real and imaginary parts of $C_m=\Phi_m^*\Phi_{m+1}$, up to the fixed sign convention of the implemented rotation. The combined RFS data are therefore
\begin{equation}
    \mathcal D_{\rm RFS}
    =
    \left\{
      |\Phi_m|^2,\,
      \Phi_m^*\Phi_{m+1}
    \right\}_{m}.
    \label{eq:app_rfs_dataset}
\end{equation}
When the visibility-link graph is connected, these quantities determine the complex visibility sequence and hence, by an inverse Fourier transform, the source brightness distribution up to the usual global/discrete ambiguities of the chosen finite model. The rotated links are {not} invariants of the full $U(1)^D$ twirl; they require the spatial correlation structure treated later. This distinction is why the exact full-twirl theorem is stated for the plain block-resolved SWAP.\\

\textbf{Fisher information and sample complexity}

For a scalar parameter $\vartheta$ encoded in a family of states
$\rho_\vartheta$, a POVM $M=\{E_y\}_y$ produces outcome probability distribution
\begin{equation}
    p_y(\vartheta)
    =
    \Tr(E_y\rho_\vartheta).
\end{equation}
The corresponding classical Fisher information (FI) is
\begin{equation}
    F_M(\vartheta)
    =
    \sum_y
    \frac{
        [\partial_\vartheta p_y(\vartheta)]^2
    }{
        p_y(\vartheta)
    }
    =
    \sum_y
    \frac{
        [\Tr(E_y\dot{\rho}_\vartheta)]^2
    }{
        \Tr(E_y\rho_\vartheta)
    },
    \qquad
    \dot{\rho}_\vartheta
    :=
    \partial_\vartheta\rho_\vartheta .
    \label{eq:app_fi_definition}
\end{equation}
The symmetric logarithmic derivative (SLD) $L_\vartheta$ is defined by
\begin{equation}
    \dot{\rho}_\vartheta
    =
    \frac12
    \left(
        L_\vartheta\rho_\vartheta
        +
        \rho_\vartheta L_\vartheta
    \right).
\end{equation}
Measurement in the eigenbasis of the SLD maximizes scalar Fisher information over all POVMs; the maximum is the quantum Fisher information (QFI)
\begin{equation}
    H(\vartheta)
    =
    \Tr\!\left(
        \rho_\vartheta L_\vartheta^2
    \right)
    =
    \sup_M F_M(\vartheta).
    \label{eq:app_qfi_definition}
\end{equation}
Thus $H$ denotes the unrestricted QFI, while $F_M$ denotes the FI of a
specified measurement. In particular, we define
\begin{equation}
    F_{\rm PPT}^{\star}(\vartheta)
    :=
    \sup_{M\in{\rm PPT}}F_M(\vartheta),
\end{equation}
for the optimal FI attainable by PPT measurements, and use $F_J$ for the FI
of the particular achievable joint measurement considered below. In general, $F_J\le H$ and $F_{\rm PPT}^{\star}\le H$, but no ordering between $F_J$ and $F_{\rm PPT}^{\star}$ follows: a coherent joint measurement can lie outside the PPT class. The relation $F_J\le F_{\rm PPT}^{\star}$ applies only when the selected joint measurement is itself PPT.

For a multiparameter family, the corresponding quantities are the classical
FI matrix $F_M$ and the SLD QFI matrix $H$, with
\begin{equation}
    F_M \preceq H .
\end{equation}
This matrix inequality does not by itself imply simultaneous attainability of
all QFI entries. Whenever simultaneous attainment is used below, the required
compatibility follows from the existence of the common eigenbasis specified
there.

Throughout, we convert local Fisher information into sample complexity using
the convention
\begin{equation}
    N_M(\epsilon;\delta)
    :=
    \frac{1}{
        \epsilon^2 F_M(\delta)
    },
    \qquad
    N_{\rm PPT}^{\star}(\epsilon;\delta)
    :=
    \frac{1}{
        \epsilon^2 F_{\rm PPT}^{\star}(\delta)
    },
    \qquad
    N_J(\epsilon;\delta)
    :=
    \frac{1}{
        \epsilon^2 F_J(\delta)
    }.
    \label{eq:si-sample-convention}
\end{equation}
Here $\delta$ denotes the parameter being estimated and $\epsilon$ is the
target local root-mean-square error in $\delta$. Consequently,
\begin{equation}
    \frac{N_{\rm PPT}^{\star}}{N_J}
    =
    \frac{F_J}{F_{\rm PPT}^{\star}} .
\end{equation}
For the reference information scale $V$ used below, we similarly write
\begin{equation}
    N_0
    :=
    \frac{1}{\epsilon^2 V}.
\end{equation}

Unless stated otherwise, all Fisher informations are normalized {per
input photon pair}, so rejected measurement branches are included in the
resource accounting. Accordingly, $N_M$ counts input photon pairs rather
than only successful or accepted outcomes. If every photon belongs to one
pair, the corresponding photon number is $2N_M$.

The relation $N_M\simeq 1/(\epsilon^2F_M)$ is the local Cramér--Rao
sample-complexity convention used throughout our comparisons; it should not
be interpreted as a separate fixed-confidence minimax statement at a
singular point. In particular, statements ``at coincidence'' refer to the
limit $\delta\downarrow0$ from positive separation. A smooth
reparametrization, for example from $\delta$ to $z$, transforms both the FI
and the target precision consistently, so the sample-complexity ratios used
below are invariant under such a change of parameter.\\

\textbf{Binary source and separation}

Let $\mathcal H\simeq\mathbb C^D$ be the retained one-photon mode space. The two distinguishable photon arrivals occupy $\mathcal H_A\otimes\mathcal H_B$. For the flat one-dimensional pupil, let $x_j=aj$, $j=0,\ldots,D-1$, and define
\begin{equation}
 |\psi_\theta\rangle=\frac1{\sqrt D}\sum_j e^{i\theta x_j}|j\rangle,
 \qquad G=\sum_jx_j|j\rangle\langle j|.
 \label{eq:si-point-source}
\end{equation}
Here $\theta$ is an angular spatial frequency: a physical small sky angle $\theta_{\rm sky}$ corresponds to $\theta=(2\pi/\lambda_{\rm opt})\theta_{\rm sky}$. An arbitrary normalized intensity distribution $P_{\boldsymbol\vartheta}(\theta)$ gives
\begin{equation}
 \rho_{\boldsymbol\vartheta}=\int P_{\boldsymbol\vartheta}(\theta)|\psi_\theta\rangle\langle\psi_\theta|\,d\theta,
 \quad \Phi_m=\int P_{\boldsymbol\vartheta}(\theta)e^{iam\theta}\,d\theta,
 \quad [\rho_{\boldsymbol\vartheta}]_{jk}=\Phi_{j-k}/D.
 \label{eq:si-general-source}
\end{equation}
Thus $\Phi_0=1$ and $\Phi_{-m}=\Phi_m^*$. The equal-brightness binary source, with known centroid set to zero and full separation $\delta=|\Delta|\ge0$, is
\begin{equation}
 \rho_\delta=\frac12|\psi_{-\delta/2}\rangle\langle\psi_{-\delta/2}|+
             \frac12|\psi_{+\delta/2}\rangle\langle\psi_{+\delta/2}|,
 \qquad
 \alpha_m(\delta)=|\Phi_m|^2=\cos^2\!\left(\frac{\delta r_m}{2}\right),
 \quad r_m=am.
 \label{eq:si-binary}
\end{equation}
We write a dot for a derivative with respect to $\delta$; in particular,
\begin{equation}
 \dot\alpha_m=-\frac{r_m}{2}\sin(\delta r_m),\qquad
 V:=\Var_{\psi_0}(G)=\frac{a^2(D^2-1)}{12}.
 \label{eq:si-binary-derivative}
\end{equation}
Let $L=(D-1)a$ be the span of the pupil sampling centers and put
\begin{equation}
 z=L\delta,\qquad \kappa_D=V/L^2=\frac{D+1}{12(D-1)},\qquad D\ge2.
 \label{eq:si-normalized-separation}
\end{equation}

\section{Quantum advantage regimes in sensing and imaging}
\label{sec:si-overview}

Two photons may encounter a strongly distorted yet nearly common wavefront. A coherent exchange measurement accesses a relative observable whose differential atmospheric phases are weak, whereas a classical receiver may lose access to the same contrast. Increasing correlated nuisance can therefore enlarge a {measurement-access} separation while preserving relational information; noise does not increase the unrestricted QFI. All comparisons are local and use the same pair-sample convention of Eq.~\eqref{eq:si-sample-convention}.

For a binary separation $\delta=|\Delta|$, put $z=L\delta$. Write $\Xi_{\rm C}$ for an attenuation exponent of a reference-sensitive classical contribution and $\Xi_{\rm J}$ for the corresponding differential-phase attenuation of the relational measurement. Their difference may create a finite parameter window with an advantage proportional to $\exp(\Xi_{\rm C}-\Xi_{\rm J})$, provided the residual PPT information floors and the task-dependent prefactors are also controlled. This does {not} imply an asymptotically unbounded advantage for a fixed physical atmosphere with finite outer scale.

\begin{table}[t]
\centering\small
\begin{tabularx}{\linewidth}{@{}p{0.35\linewidth}X@{}}
\toprule
Regime & Qualified conclusion \\
\midrule
Known wavefront; negligible nuisance & Task-specific local measurements may already attain the relevant QFI; no generic separation is implied.\\
Complete common $U(1)^D$ twirl & RFS reaches the retained QFI; PPT bounds give dimension and small-separation penalties, of order $\max\{D,z^{-2}\}$ in the stated regime.\\
Finite common Gaussian noise & The coherent exchange dark sector remains protected under symmetric-support assumptions; its one-paragraph proof is in Sec.~\ref{sec:si-finite-gaussian}.\\
Partially correlated Gaussian noise & The full-state PPT theorem gives an explicit finite window with sample-FI gain proportional to $e^{\Xi_{\rm C}-\Xi_{\rm Q}}$ when its twirled floor is sufficiently small (Sec.~\ref{sec:si-finite-gaussian}).\\
Differential-noise-dominated regime & A filled exchange-dark sector removes the protected small-$z$ scaling, although other advantages may remain.\\
Reference learned from prior photons & The Kalman benchmark concerns a specified linear--Gaussian sensor and has a distinct history and information budget with the scaling regime $e^{\Xi_{\rm Kalman}-\Xi_{\rm Q}}$ (Sec.~\ref{sec:si-history}).\\
\bottomrule
\end{tabularx}
\caption{A compact guide to the physical limits.}
\label{tab:si-overview-regimes}
\end{table}
We refer to the effect of rapid FI decay when the binary source separation is small and the measurement is noisy as dark-sector effect also known as Rayleigh curse in superresolution imaging. This effect is connected to the noisy SWAP-test loosing asymptotic advantage: false antisymmetric events are indistinguishable from the source properties, so the dark counts background limits the achievable FI for a small separation.

The binary PPT dimension certificate and the dark-sector certificate apply in complementary ranges of $z$ and nuisance strength; the maximum of valid lower bounds may be used, but a bound saturating does not establish an upper limit on the true advantage. The finite outer scale, the finite pupil, and the differential phase variance determine the end of the exponential window. The full mathematical statements and crossover conditions are given in Secs.~\ref{sec:si-binary-twirl}--\ref{sec:si-finite-gaussian}; the comparison with classical history is in Sec.~\ref{sec:si-history}.

\section{Binary separation after a complete common phase twirl}
\label{sec:si-binary-twirl}\label{app:u1d_proof}

The simplest complete example already contains two effects: a dimension-dependent PPT information loss and an additional small-separation dark-space loss. Neither requires an unknown science task.

\subsection{Exchange blocks and an optimal joint measurement}

Let $S|j,k\rangle=|k,j\rangle$, $P_\pm=(I\pm S)/2$, and $\Pi_{\rm d}=\sum_j|j,j\rangle\langle j,j|$. For the finite-pupil offset $m$, let
\begin{equation}
 \begin{split}
 \Pi_m&=\sum_{j=0}^{D-1-m}(|j,j+m\rangle\langle j,j+m|+|j+m,j\rangle\langle j+m,j|),\\
 X_m&=\sum_{j=0}^{D-1-m}(|j,j+m\rangle\langle j+m,j|+|j+m,j\rangle\langle j,j+m|),\\
 \Pi_{m,\pm}&=(\Pi_m\pm X_m)/2.
 \end{split}
 \label{eq:si-exchange-ops}
\end{equation}
The complete common twirl draws the $D$ phases independently and uniformly within a pair, using the same screen on both photons. A matrix element $|j,k\rangle\langle l,n|$ survives precisely when $\{j,k\}=\{l,n\}$ as multisets. Therefore
\begin{equation}
 \Omega_\delta^{\rm tw}=\mathcal T(\rho_\delta^{\otimes2})
 =\frac1{D^2}\left(I+\sum_{m=1}^{D-1}\alpha_m(\delta)X_m\right).
 \label{eq:si-twirl-binary-state}
\end{equation}
The relational Fourier measurement resolves the offset and exchange parity, not the ordering of the two pupil labels.

\begin{theorem}[Binary RFS optimality in the twirled experiment]
\label{thm:si-binary-rfs}
For Eq.~\eqref{eq:si-twirl-binary-state}, the source-independent measurement $\{\Pi_{\rm d},\Pi_{m,+},\Pi_{m,-}\}_m$ attains the QFI at every regular separation. Its probabilities and information are
\begin{equation}
 p_{\rm d}=D^{-1},\quad p_{m,\pm}=\frac{n_m}{D^2}(1\pm\alpha_m),\quad
 F_{\rm RFS}^{\rm tw}=H_\delta^{\rm tw}=
 \frac2{D^2}\sum_mn_m\frac{\dot\alpha_m^2}{1-\alpha_m^2}.
 \label{eq:si-twirl-rfs-fi}
\end{equation}
Moreover $F_{\rm RFS}^{\rm tw}=V+O(\delta^2)$ as $\delta\downarrow0$ for fixed pupil.
\end{theorem}
\begin{proof}
Each unordered pair spans a block $D^{-2}\left(\begin{smallmatrix}1&\alpha_m\\\alpha_m&1\end{smallmatrix}\right)$ with parameter-independent eigenvectors $(|j,k\rangle\pm|k,j\rangle)/\sqrt2$. Their eigenvalues give Eq.~\eqref{eq:si-twirl-rfs-fi}. Since $\alpha_m=1-\delta^2r_m^2/4+O(\delta^4)$, the contribution of offset $m$ tends to $n_mr_m^2/D^2$. Finally $D^{-2}\sum_{j<k}(x_j-x_k)^2=V$. Equivalently, away from removable singularities the exact contribution is $2n_mr_m^2\alpha_m/[D^2(1+\alpha_m)]$.
\end{proof}
The limit $V$ is per pair. The clean, reference-known binary can have pair information $2V$; optimality here refers to the twirled state, not to the information removed by twirling.

\subsection{Two elementary PPT tools}

\begin{lemma}[Partial-transpose QFI certificate]
\label{lem:si-ptqfi}
For a differentiable state family $\rho_\theta$ that is PPT in a neighborhood of the operating point,
\begin{equation}
 F_{\rm PPT}^\star(\rho_\theta,\dot\rho_\theta)
 \le\min\{H(\rho_\theta,\dot\rho_\theta),H(\rho_\theta^\Gamma,\dot\rho_\theta^\Gamma)\}.
 \label{eq:si-ptqfi}
\end{equation}
\end{lemma}
\begin{proof}
For a PPT POVM, $\{E_y^\Gamma\}$ is also a POVM. Its probabilities and derivatives on $\rho^\Gamma$ equal those of $\{E_y\}$ on $\rho$, because $\Tr(E_y\rho)=\Tr(E_y^\Gamma\rho^\Gamma)$. The unrestricted QFI of $\rho^\Gamma$ bounds that FI. This is an algebraic relaxation, not a physical implementation of transpose.
\end{proof}

\begin{lemma}[PPT symmetric weight and a Hilbert--Schmidt FI bound]
\label{lem:si-ppt-background}
Every positive PPT effect satisfies
\begin{equation}
 \Tr(P_+E)\ge\tfrac12\Tr E.
 \label{eq:si-ppt-half}
\end{equation}
If a state $\rho$ obeys $\Tr(E\rho)\ge b\Tr E$ for all such effects, with $b>0$, then any Hermitian tangent $Y$ satisfies
\begin{equation}
 \sup_{M\in\mathrm{PPT}}\sum_y\frac{[\Tr(E_yY)]^2}{\Tr(E_y\rho)}
 \le \frac{\Tr(Y^2)}b.
 \label{eq:si-hs-bound}
\end{equation}
\end{lemma}
\begin{proof}
The identity $S^\Gamma=D|\Phi_D\rangle\langle\Phi_D|\succeq0$ implies $\Tr(SE)=\Tr(S^\Gamma E^\Gamma)\ge0$, proving Eq.~\eqref{eq:si-ppt-half}. Cauchy--Schwarz in the positive functional $Z\mapsto\Tr(EZ)$ gives $[\Tr(EY)]^2\le\Tr E\,\Tr(EY^2)$. Divide by $b\Tr E$ and sum, using $\sum_yE_y=I$. The proof also covers continuously labelled POVMs by integration with respect to their trace measure.
\end{proof}
This second bound retains $\Tr(Y^2)$ rather than replacing it by $D^2\|Y\|_\infty^2$, and can therefore improve the dimension-dependent coefficient.

\subsection{A binary-specific dimension bound and a dark-space refinement}

\begin{theorem}[Two complementary all-PPT binary bounds]
\label{thm:si-twirl-ppt}
For the flat-pupil binary after complete common phase twirling, at every regular $\delta>0$,
\begin{equation}
 F_{\rm PPT}^{\star,\rm tw}(\delta)\le B_{\rm tw}(\delta):=
 \min\left\{\frac{2V}{D},\ Q_D(\delta)\right\},\qquad
 Q_D(\delta):=\frac4{D^2}\sum_{j<k}\dot\alpha_{|j-k|}(\delta)^2.
 \label{eq:si-Btw}
\end{equation}
In particular,
\begin{equation}
 Q_D(\delta)\le \frac{\delta^2}{D^2}\sum_{j<k}(x_j-x_k)^4
              =Vc_Dz^2,
 \qquad
 \frac{F_{\rm RFS}^{\rm tw}}{F_{\rm PPT}^{\star,\rm tw}}
 \ge\frac{F_{\rm RFS}^{\rm tw}/V}{\min\{2/D,c_Dz^2\}}.
 \label{eq:si-twirl-ratio}
\end{equation}
Thus the local overhead is at least $\max\{D/2,(c_Dz^2)^{-1}\}[1+O(z^2)]$. The two bounds are combined by a minimum of information ceilings, not by multiplying their advantage factors.
\end{theorem}
\begin{proof}
Partial transpose maps exchange coherences to the diagonal sector. Defining $[\sigma_\delta]_{jk}=D^{-1}\alpha_{|j-k|}$, with $\alpha_0=1$, gives
\begin{equation}
 (\Omega_\delta^{\rm tw})^\Gamma
 =\left(1-\frac1D\right)\tau_{\rm off}\oplus\frac1D\sigma_\delta,
 \qquad
 \sigma_\delta=\frac12|\psi_0\rangle\langle\psi_0|
  +\frac14|\psi_\delta\rangle\langle\psi_\delta|
  +\frac14|\psi_{-\delta}\rangle\langle\psi_{-\delta}|.
 \label{eq:si-shadow}
\end{equation}
The first block and both block weights are parameter independent. Convexity of QFI and $H(|\psi_{\pm\delta}\rangle)=4V$ imply $H(\sigma_\delta)\le2V$. Lemma~\ref{lem:si-ptqfi} proves $F_{\rm PPT}\le2V/D$.

For the second bound, $0\le\alpha_m\le1$ implies $\Omega_\delta^{\rm tw}\succeq P_+/D^2$ for every $\delta$. Lemma~\ref{lem:si-ppt-background} therefore gives $b=1/(2D^2)$. Orthogonality of the exchange blocks yields
\begin{equation}
 \Tr[(\dot\Omega_\delta^{\rm tw})^2]
 =\frac2{D^4}\sum_{j<k}\dot\alpha_{|j-k|}^2.
\end{equation}
Equation~\eqref{eq:si-hs-bound} proves $F_{\rm PPT}\le Q_D$. Finally $|\sin(\delta r)|\le\delta|r|$ gives the displayed fourth-moment bound.
\end{proof}

The dark-space mechanism is visible directly at coincidence:
\begin{equation}
 \Omega_0^{\rm tw}=\frac1{D^2}\sum_j|jj\rangle\langle jj|
 +\frac2{D^2}\sum_{j<k}|jk;+\rangle\langle jk;+|.
 \label{eq:si-twirl-zero}
\end{equation}
Its support is exactly the symmetric subspace. An antisymmetric joint outcome is dark, but Eq.~\eqref{eq:si-ppt-half} forces every nonzero PPT effect to have symmetric weight and therefore background probability. The binary changes that background only at order $\delta^2$, giving the quadratic PPT FI bound. This proves a separation dependence in the same physical model that already exhibits the dimension bound.


\section{Partially correlated Gaussian noise: exponential window}
\label{sec:si-finite-gaussian}

The complete phase twirl gives a useful limiting case, but real wavefronts are neither uniformly random nor exactly identical for successive photons. Here we retain finite Gaussian phase correlations and compare RFS with {every PPT measurement of the full, unmodified two-photon state}. The main result is an explicit finite window in which a common-noise attenuation exponent can be much larger than the differential-noise exponent relevant to RFS.

\subsection{Common and differential phase noise}

Let the phase vectors of the two photons be jointly zero-mean Gaussian, with covariance $\Sigma$. In the uniform-pupil binary model,
\begin{equation}
 \begin{aligned}
 \Sigma&=\begin{pmatrix}C_A&R\\R^T&C_B\end{pmatrix},\qquad
 [\Omega_\delta(\Sigma)]_{jk;ln}
 =\frac{c_{jl}(\delta)c_{kn}(\delta)}{D^2}e^{-h^T\Sigma h/2},\\
 h&=\binom{e_j-e_l}{e_k-e_n},\qquad
 c_{jl}(\delta)=\cos\frac{\delta(x_j-x_l)}2.
 \end{aligned}
 \label{eq:si-gaussian-full}
\end{equation}
Introduce common and differential phases $c=(\phi^A+\phi^B)/2$ and $d=\phi^A-\phi^B$. Their covariances and Gaussian regression give
\begin{equation}
 \begin{split}
 C_c&=(C_A+C_B+R+R^T)/4,\qquad
 C_d=C_A+C_B-R-R^T,\\
 B&=(C_A-R+R^T-C_B)/2,\qquad
 K=BC_d^+,\qquad C_*=C_c-BC_d^+B^T.
 \end{split}
 \label{eq:si-regression-cov}
\end{equation}
Here $c=c_*+Kd$ with $c_*$ independent of $d$ and covariance $C_*$. Thus the physical pair state is obtained from a state with only the common screen $c_*$ by adding parameter-independent correlated local phase rotations. These rotations cannot increase the optimum PPT information:
\begin{equation}
 F_{\rm PPT}^\star\bigl(\Omega_\delta(\Sigma)\bigr)
 \le F_{\rm PPT}^\star\bigl(\Omega_\delta^{\rm c}(C_*)\bigr).
 \label{eq:si-channel-reduction}
\end{equation}
This reduction also covers nonsymmetric cross-covariances $R$, as arise for displaced frozen-flow screens.

\subsection{A finite-Gaussian ceiling for all PPT measurements}

An unordered pair of pupil modes $b=(j,k)$, $j\le k$, carries the common-phase charge $q_b=e_j+e_k$. For the independent common component define
\begin{equation}
 \mathsf K_{bc}=\exp\!\left[-\tfrac12(q_b-q_c)^TC_*(q_b-q_c)\right],\quad
 a_*=\lambda_{\min}(\mathsf K),\quad
 \kappa_*=\tfrac12\min_{b\ne c}(q_b-q_c)^TC_*(q_b-q_c).
 \label{eq:si-a-star}
\end{equation}
The number $a_*>0$ controls how much of the fully twirled state's positive background remains under finite Gaussian noise. One sufficient, though often conservative, estimate is
\begin{equation}
 a_*\ge\left[1-\left(\frac{D(D+1)}2-1\right)e^{-\kappa_*}\right]_+.
 \label{eq:si-Gershgorin}
\end{equation}
Sharper estimates may use $\lambda_{\min}(\mathsf K)$ or its individual row sums.

Write $\Omega_\delta^{\rm c}(C_*)=\Omega_\delta^{\rm tw}+R_\delta$. The remainder $R_\delta$ consists of matrix elements connecting {different} charge sectors. The amount of separation information in this remainder is bounded by
\begin{equation}
 \begin{split}
 T_*(\delta)&=2D^2\Tr[(\dot R_\delta)^2],\\
 [\dot R_\delta]_{jk;ln}
 &=\frac{\partial_\delta(c_{jl}c_{kn})}{D^2}
 e^{-\frac12q^TC_*q}\,\mathbf 1_{q\ne0},
 \qquad q=e_j+e_k-e_l-e_n,\\
 T_*(\delta)&\le C_D(\delta)e^{-\Xi_{\rm C}},\qquad
 \Xi_{\rm C}=2\kappa_*,\qquad
 C_D(\delta)=2D^2\Tr\!\left[\bigl(\partial_\delta\rho_\delta^{\otimes2}\bigr)^2\right].
 \end{split}
 \label{eq:si-T-star}
\end{equation}
All dependence on common-phase noise in the last line is in the exponential; $C_D$ is calculated from the clean binary state and is explicit.

\begin{theorem}[Finite-Gaussian all-PPT bound]
\label{thm:si-finite-PPT}
For the full Gaussian two-photon state in Eq.~\eqref{eq:si-gaussian-full}, with $a_*>0$, any PPT measurement satisfies
\begin{equation}
  {F_{\rm PPT}^\star\le B_\Sigma:=\min\left\{
 2V,\ \frac{Q_D+T_*}{a_*},\
 \frac{\bigl(\sqrt{B_{\rm tw}}+\sqrt{T_*}\bigr)^2}{a_*}
 \right\}.}
 \label{eq:si-full-finite-bound}
\end{equation}
Here $B_{\rm tw}=\min\{2V/D,Q_D\}$ and $Q_D$ are the explicit binary complete-twirl ceilings of Eq.~\eqref{eq:si-Btw}.
\end{theorem}
\begin{proof}
The Gaussian common-phase channel obeys $\Omega_\delta^{\rm c}\succeq a_*\Omega_\delta^{\rm tw}\succeq a_*P_+/D^2$. Thus any PPT effect has a nonzero outcome-probability floor, by Lemma~\ref{lem:si-ppt-background}. The within-charge tangent is the complete-twirl binary problem of Sec.~\ref{sec:si-binary-twirl}; the different-charge tangent has squared norm $T_*/(2D^2)$. Applying the Hilbert--Schmidt FI bound gives $(Q_D+T_*)/a_*$. Applying the triangle inequality to the two sets of score contributions while retaining the sharper twirled ceiling gives $(\sqrt{B_{\rm tw}}+\sqrt{T_*})^2/a_*$. Finally, the clean-pair QFI bound is $2V$. Equation~\eqref{eq:si-channel-reduction} transfers these ceilings to the original partially correlated state.
\end{proof}

\subsection{Relational FI and the exponential window}

RFS removes the common screen from its exchange contrast; what remains is the {difference} between the two screens. Put
\begin{equation}
 \kappa_{jk}^{\rm d}=\tfrac12(e_j-e_k)^TC_d(e_j-e_k),\qquad
 \gamma_{jk}=e^{-\kappa_{jk}^{\rm d}},\qquad
 \alpha_{jk}=\cos^2\!\frac{\delta(x_j-x_k)}2.
 \label{eq:si-diff-kappa}
\end{equation}
The achievable edge-resolved Fisher information, per photon pair, is
\begin{equation}
 {F_{\rm RFS}(\delta;\Sigma)=\frac2{D^2}\sum_{j<k}
 \frac{\gamma_{jk}^2\dot\alpha_{jk}^{\,2}}
 {1-\gamma_{jk}^2\alpha_{jk}^2}.}
 \label{eq:si-hJ}
\end{equation}

\paragraph{Perfectly common noise and the dark sector.}
For $C_d=0$, the antisymmetric exchange outcome remains dark at zero binary separation, even if the common Gaussian screen is strong; Eq.~\eqref{eq:si-hJ} then has the nonzero limit $F_{\rm RFS}\to V$. The corresponding PPT small-separation information is $O(\delta^2)$, provided $\Omega_0^{\rm c}\succeq\lambda_+P_+$ with $\lambda_+>0$. Indeed, writing $X_{\rm c}=\frac12\partial_\delta^2\Omega_\delta^{\rm c}|_{\delta=0}$, the same PPT-background lemma gives
\begin{equation}
 \limsup_{\delta\downarrow0}\frac{F_{\rm PPT}^{\star,\rm c}(\delta)}{\delta^2}
 \le\frac{8\Tr(X_{\rm c}^2)}{\lambda_+}.
 \label{eq:si-common-dark-bound}
\end{equation}
Nondegenerate relative Gaussian phases ensure the required symmetric support; an exactly known wavefront or pure piston does not. When all informative differential variances are positive, the exchange-dark outcome fills in and $F_{\rm RFS}=\Theta(\delta^2)$ locally. Thus the protected small-separation limit and the Gaussian attenuation advantage below are related but distinct effects.

\paragraph{Main result: an explicit exponential sample-advantage window.}
Define the conservative attenuation exponents and the clean geometrical coefficient
\begin{equation}
 \Xi_{\rm C}=2\kappa_*,\qquad
 \Xi_{\rm Q}=2\max_{j<k:\,\dot\alpha_{jk}\ne0}\kappa_{jk}^{\rm d},\qquad
 A_D(\delta)=\frac2{D^2}\sum_{j<k}\dot\alpha_{jk}^{\,2}=\frac{Q_D}{2}.
 \label{eq:si-exponential-definitions}
\end{equation}
Equation~\eqref{eq:si-hJ} implies $F_{\rm RFS}\ge A_D e^{-\Xi_{\rm Q}}$. Combining this with the full-state PPT theorem and $T_*\le C_De^{-\Xi_{\rm C}}$ yields
\begin{equation}
  {\begin{aligned}
 \frac{N_{\rm PPT}}{N_{\rm RFS}}
 &=\frac{F_{\rm RFS}}{F_{\rm PPT}^\star}
 \ \ge\ \frac{a_*F_{\rm RFS}}
 {\bigl[\sqrt{B_{\rm tw}}+\sqrt{C_D}\,e^{-\Xi_{\rm C}/2}\bigr]^2}\\[2pt]
 &\ge\frac{a_* A_D}{C_D}\,
 \frac{e^{\Xi_{\rm C}-\Xi_{\rm Q}}}
 {\bigl[1+\sqrt{\varepsilon_D}\bigr]^2},\qquad
 \varepsilon_D=\frac{B_{\rm tw}}{C_D}e^{\Xi_{\rm C}}.
 \end{aligned}}
 \label{eq:si-exponential-floor-summary}
\end{equation}
The factor $\varepsilon_D$ measures the completely twirled classical information floor relative to the attenuated different-charge contribution. In the window
\begin{equation}
  {B_{\rm tw}\ \lesssim\ C_De^{-\Xi_{\rm C}}}
 \quad\Longleftrightarrow\quad
 \Xi_{\rm C}\ \lesssim\ \log\frac{C_D}{B_{\rm tw}},
 \label{eq:si-exponential-window-condition}
\end{equation}
we have $[1+\sqrt{\varepsilon_D}]^2\le4$ whenever $\varepsilon_D\le1$, and therefore
\begin{equation}
  {\frac{N_{\rm PPT}}{N_{\rm RFS}}
 \ge \frac{a_*A_D}{4C_D}\,
 e^{\Xi_{\rm C}-\Xi_{\rm Q}}.}
 \label{eq:si-exponential-window-gain}
\end{equation}
This is the {exponential attenuation window}: the classical charge-sensitive information falls as $e^{-\Xi_{\rm C}}$ while relational information needs only survive $e^{-\Xi_{\rm Q}}$. The window can be broad at finite $D$: even a small $C_De^{-\Xi_{\rm C}}$ can remain much larger than a still smaller $B_{\rm tw}$. For the observables with $\Xi_{\rm C}-\Xi_{\rm Q} \propto D$, the scaling becomes exponential in dimension. Once the twirled floor dominates, this argument saturates.

\subsection{Observables and covariance families with extensive exponents}

An exponent gap growing with $D$ is possible but requires specifying what grows physically. Two complementary constructions make this explicit.

\paragraph{Sparse corner-coherence image at fixed pixel pitch.}
Take a nonnegative periodic angular brightness profile with a single high-frequency harmonic,
\begin{equation}
 P_v(\theta)=\frac{1+2v\cos[(D-1)\theta]}{2\pi},\qquad
 |v|\le\frac12,\qquad
 \rho_v=\frac{I}{D}+\frac{v}{D}
 (|0\rangle\langle D-1|+|D-1\rangle\langle0|).
 \label{eq:si-corner-image}
\end{equation}
Its unknown Fourier visibility $\Phi_{D-1}=v$ occupies only the two opposite corners of the one-photon density matrix. The parameter-dependent {different-charge} two-copy coherences involve charges $\pm(e_0-e_{D-1})$ or $\pm2(e_0-e_{D-1})$. Consequently their smallest Gaussian charge variance is
\begin{equation}
 \Xi_{\rm C}^{\rm corner}
 =\Var(\phi_{D-1}-\phi_0)
 =D_\phi[(D-1)a].
 \label{eq:si-corner-exponent}
\end{equation}
For a fixed-pitch Gaussian random walk this is $(D-1)s^2=\Theta(D)$; in the Kolmogorov range it can grow as $D^{5/3}$ until the atmospheric outer scale becomes important. This is a {task-restricted} exponent, because the unknown $v$ does not change the other off-diagonal coherences. It may sharpen the remainder bound for this observable, but it cannot simply replace the unrestricted $\kappa_*$ of Eq.~\eqref{eq:si-a-star}.

\paragraph{A scaled covariance with an extensive global minimum.}
Let the common screen be a Gaussian random walk across the pupil,
\begin{equation}
 \phi_0=0,\qquad \phi_j=\sum_{r=1}^{j}\xi_r,\qquad
 \xi_r\overset{\rm iid}{\sim}\mathcal N(0,s_D^2).
 \label{eq:si-scaled-brownian}
\end{equation}
For distinct unordered-pair charges $q_b,q_c$, their phase difference contains at least one independent increment with a nonzero integer coefficient. The minimum variance is therefore exactly $s_D^2$, attained when one of the pair's indices moves by one pixel:
\begin{equation}
  {\Xi_{\rm C}=2\kappa_*=s_D^2.}
 \label{eq:si-extensive-global-charge}
\end{equation}
Choosing $s_D^2=cD$ gives a genuinely global $\Xi_{\rm C}=cD$. Equivalently, for the synthetic spatial law $D_\phi(r)=g|r|$, increasing the pitch as $a(D)\propto D$ produces this family. If the two photons experience identical screens, $\Xi_{\rm Q}=0$, and the global Gram matrix is strictly positive. A physical von K\'arm\'an atmosphere has a finite outer scale, so this scaling is a finite-range construction rather than a fixed-conditions asymptotic prediction.

\section{Atmospheric model}
\label{sec:si-atmosphere}
We use Gaussian phase statistics with frozen-flow translation to connect the abstract covariance $\Sigma$ to a telescope. For a stationary phase covariance $C(r)$ and structure function $D_\phi(r)=2[C(0)-C(r)]$, the physical two-dimensional von K\'arm\'an model sampled along a one-dimensional pupil has
\begin{equation}
 C(r)=C_0\frac{2^{1-\nu}}{\Gamma(\nu)}(k_0|r|)^\nu K_\nu(k_0|r|),\quad
 \nu=\frac56,\quad k_0=\frac{2\pi}{L_0},\quad C(0)=C_0.
 \label{eq:si-vK}
\end{equation}
The amplitude $C_0$ is fixed by the Fried parameter through the small-baseline Kolmogorov structure function. A one-dimensional spectrum with the same spectral exponent would instead give $\nu=4/3$, and should not be substituted for the physical two-dimensional slice merely by matching one reference point. In the reported FI scans the values are $r_0=0.157921\,\mathrm m$, $L_0=22\,\mathrm m$, pupil pitch $a=0.0640625\,\mathrm m$, and wind $v=10\,\mathrm{m/s}$.

With $x_j=aj$ and a frozen displacement $u=v\tau$ between photon arrivals,
\begin{equation}
 (C_A)_{jk}=(C_B)_{jk}=C(x_j-x_k),\qquad
 R_{jk}=C(x_j-x_k+u).
 \label{eq:si-atmos-cov}
\end{equation}
The exchange-visibility attenuation depends only on the variance of the {difference} between baseline phase errors:
\begin{equation}
 \gamma_m=e^{-K_2(am,u)},\quad
 K_2(r,u)=D_\phi(r)+D_\phi(u)-\tfrac12[D_\phi(r+u)+D_\phi(r-u)].
 \label{eq:si-K2}
\end{equation}
Inserting $\gamma_m$ into the physical-pair likelihood gives
\begin{equation}
 F_{\rm RFS}(\delta,u)=\frac2{D^2}\sum_{m=1}^{D-1}(D-m)
 \frac{\gamma_m^2\dot\alpha_m^2}{1-\gamma_m^2\alpha_m^2}.
 \label{eq:si-atmos-rfs}
\end{equation}
For small separation and weak differential phase variance, each term is approximately
\begin{equation}
 F_m\simeq\frac{(D-m)r_m^2}{D^2}
 \frac{\delta^2r_m^2}{4K_2(r_m,u)+\delta^2r_m^2}.
 \label{eq:si-atmos-window}
\end{equation}
Thus a nonzero differential error introduces a well-defined crossover in the binary FI; replacing the whole result by a clean-limit QFI multiplied by one global visibility would miss it.

For numerical quantum-information comparisons we also use an {auxiliary} total-index dephasing $j+k=l+n$; it produces blocks of dimension at most $D$ whose non-normalized density matrices and parameter derivatives are diagonalized to calculate QFI. The PPT mixed-SLD Schur-complement solver similarly acts on the explicitly stated {total-index-twirled} state. These diagnostic quantities should not be confused with the full-state PPT theorem of Sec.~\ref{sec:si-finite-gaussian}; in particular, dephasing a state before optimizing a classical receiver can underestimate classical information on the original state. Its status is identified with each numerical curve below.

\section{Binary source simulations}
\label{sec:si-benchmarks}\label{sec:si-binary-numerics}
This section specifies the two simulations behind the binary-source panels in the main text: FI as a function of mean time between photons, and FI as the pupil dimension is increased. The source is the {equal-brightness binary} of Eq.~\eqref{eq:si-binary}; the measured parameter is its {full} angular separation $\delta$, not the half-separation. The algorithm is unchanged across the scan: the binary family is used only to evaluate its scalar FI.

\subsection{Likelihoods and what each curve means}
For each fixed lag $u$ we form the exact Gaussian-averaged two-copy state in Eq.~\eqref{eq:si-gaussian-full}. Quantum RFS information comes from Eq.~\eqref{eq:si-atmos-rfs}. The full-state universal classical receiver records {both} mode-sorter results, not independent nuisance-averaged marginals:
\begin{equation}
 p_{pq}^{\rm FT}(\delta,u)=\Tr[(|f_p\rangle\langle f_p|\otimes|f_q\rangle\langle f_q|)\Omega_\delta(u)],\quad
 F_{\rm FT}=\sum_{p,q}\frac{(\partial_\delta p_{pq}^{\rm FT})^2}{p_{pq}^{\rm FT}},
 \quad |f_p\rangle=D^{-1/2}\sum_je^{2\pi ipj/D}|j\rangle.
 \label{eq:si-product-likelihood}
\end{equation}
A speckle-type difference statistic, when shown separately, is obtained by coarse-graining $(p,q)$; its FI cannot exceed the complete direct imaging with correlations record. The QFI curve denoted {power-spectrum QFI} is calculated on the total-index-dephased state described in Sec.~\ref{sec:si-atmosphere}, not the full two-copy QFI. The PPT curve from the sparse numerical mixed-SLD solver is an upper certificate on PPT FI of that {dephased state}.

The Kalman curve uses the causal linear--Gaussian reference-sensing model of Sec.~\ref{sec:si-history}. Specifically, the implementation uses a modewise steady-state, fixed-{mean}-cadence one-photon FI model, multiplied by two for nominal per-pair comparison.

\subsection{Time and dimension scans}
For Poisson photon arrivals at mean interval $\bar t$, retain the observed time lag $\tau$ and integrate the {conditional FI}, not the density matrix evaluated at the mean lag:
\begin{equation}
 F_{\rm tagged}(\bar t)=\int_0^\infty\frac{e^{-\tau/\bar t}}{\bar t}
 F_{\rm fixed}(v\tau)\,d\tau.
 \label{eq:si-poisson-FI}
\end{equation}
The released code calculates the conditional FI on a finite logarithmic grid in physical displacement $u$ and analytically integrates a {piecewise-linear interpolation}, with constant endpoint extension. The resulting PPT interpolant is an exploratory numerical curve, not a rigorously certified continuous-lag PPT bound. The fixed-mean-cadence Kalman model is evaluated at $u=v\bar t$ rather than Poisson averaged. In the atmosphere panels the convenient time coordinate $\bar t/(L_0/v)=v\bar t/L_0$ uses the outer-scale advection time as a reference; it should not be confused with a universal coherence time for every baseline.

For the time scan the nominal optical dimension is $D=128$, with physical aperture $Da=8.2\,\mathrm m$. The main-text binary schematic uses $\delta_{\rm sky}=0.1\lambda/D_{\rm apt}$; in the finite-pupil code this corresponds to a phase step $\delta a=2\pi(0.1)a/D_{\rm apt}$ (approximately $0.2\pi/D$ when $D_{\rm apt}=Da$). For the dimension scan, $a$, the physical atmosphere, and the mean displacement ratio $v\bar t/L_0=10^{-3}$ are held fixed while the number of pupil modes is varied. Thus {the physical aperture changes with $D$}. Two useful source conventions are supported in the code: a fixed angular separation (fixed spatial phase step $\delta a$) and a fixed fraction of the nominal Rayleigh scale ($\delta a\propto D^{-1}$). These produce different scaling curves and must not be combined; the published figure and metadata must select one explicitly.

\begin{table}[t]
\centering\small
\begin{tabularx}{\linewidth}{@{}p{0.34\linewidth}X@{}}
\toprule
Quantity & Scan convention and interpretation \\
\midrule
Atmospheric model & $\nu=5/6$, $r_0=0.157921$ m, $L_0=22$ m, wind $10$ m/s, $a=0.0640625$ m \\
Time panel & $D=128$; $u/L_0=v\tau/L_0$ sampled and exponentially weighted for time tags \\
Dimension panel & $D$ varied at fixed pixel pitch and frozen-flow law, mean $\bar u/L_0=10^{-3}$ \\
RFS & RFS FI on the full pair state, equal to the power-spectrum QFI in atmosphere-aware case \\
Speckle imaging & FI of the direct imaging with all classical correlations of the full pair state \\
PPT bound & Upper bound on the PPT receivers, including task-aware ones \\
Kalman & History-assisted linear--Gaussian Kalman filtering performance \\
\bottomrule
\end{tabularx}
\caption{Conventions for the binary-source FI scans. The code repository records the exact separation convention, grid endpoints, regularization, and numerical settings for each figure.}
\label{tab:si-binary-simulations}
\end{table}

The plots use local FI, so inverse-FI sample ratios describe a Cram\'er--Rao photon budget, not finite-photon estimator success. The scan demonstrates the physical tradeoff: RFS retains information when within-pair differential noise is weak, while a reference-learning receiver must infer the instantaneous wavefront using its granted measurement history. For the {known binary task}, however, the nuisance-blind all-PPT theorem does not generically yield an exponential-in-$D$ separation; the strong extensive-exponent statement requires a separately specified image observable or universal learning requirement. At fixed finite $u$ the binary RFS FI ultimately develops the quadratic-in-separation Rayleigh penalty of Eq.~\eqref{eq:si-hJ} as $\delta\to0$.

The public Python package contains the full-state Fourier likelihood, fixed-lag quadrature, Poisson interpolation, power-block QFI, PPT solver, and the dimension scanner, including reproducibility diagnostics. Numerically reported certificates should be checked for PSD, normalization, eigensolver cutoffs, interpolation range, and residual error before they are interpreted as rigorous bounds.

\section{General image observables and the RFS circuit}
\label{sec:si-general}
Binary separation demonstrates the information mechanism for one announced task. General imaging instead calls for a fixed receiver that records enough {different} spatial Fourier quantities to reconstruct a scene without choosing its model in advance. Only the following simple measurement facts are needed here.

\subsection{Power-spectrum acquisition}
\label{sec:si-universal}
For an arbitrary vector of {retained} powers $\boldsymbol\alpha=(|\Phi_m|^2)_m$, the fully common-twirled state is $\Omega_{\boldsymbol\alpha}^{\rm tw}=D^{-2}(I+\sum_m\alpha_mX_m)$. It has a parameter-independent exchange eigenbasis. Therefore the same, scene-independent block-resolved SWAP measures all powers with the QFI matrix
\begin{equation}
 H_{mn}=\delta_{mn}\frac{2(D-m)}{D^2(1-\alpha_m^2)}.
 \label{eq:si-power-H}
\end{equation}
\begin{theorem}[Optimal measurement of the retained powers]
\label{thm:si-power-optimal}
On this commuting contrast family, RFS attains Eq.~\eqref{eq:si-power-H} simultaneously. The result also holds for any identifiable source parameters obtained by a smooth pullback $\boldsymbol\alpha(\boldsymbol\vartheta)$.
\end{theorem}
\begin{proof}
The probabilities $p_{m,\pm}=(D-m)(1\pm\alpha_m)/D^2$ are diagonal in a fixed basis for every parameter. Differentiating them gives the displayed QFI, which is thus attained by the observed probabilities.
\end{proof}

For one fixed POVM required to cover an entire bank of $K=\lfloor D/2\rfloor$ independent periodic-model powers, the normalized mean information fraction is
\begin{equation}
 \mathcal G(M)=K^{-1}\Tr(H^{-1/2}F_MH^{-1/2}).
\end{equation}
For the {periodic contrast model}, PPT positivity and exchange-block Cauchy--Schwarz give $\mathcal G(M)\le\min\{1,(1+a_0)/K\}$ with $a_0=\max_m|\alpha_m|<1$. The key restriction is that the exchange term of every PPT effect is bounded both by the local positive exchange weight and by its diagonal-sector weight, so that one measurement cannot be highly sensitive to every offset independently. For completeness, in the periodic model a PPT effect may be symmetrized without changing its FI and written as $E_y=c_y\Pi_{\rm d}+\sum_m(a_{m,y}\Pi_m+b_{m,y}X_m)$. Positivity and partial-transpose positivity imply $|b_{m,y}|\le a_{m,y}$ and $|b_{m,y}|\le c_y$. With $t_m=n_m/D$ and $A_y=\sum_m t_m(a_{m,y}+\alpha_mb_{m,y})$, one has $p_y=(c_y+2A_y)/D$. Cauchy--Schwarz yields $(1-\alpha_m^2)b_{m,y}^2\le c_y(1+a_0)(a_{m,y}+\alpha_mb_{m,y})$, and hence $K\mathcal G\le2(1+a_0)\sum_yc_yA_y/(c_y+2A_y)\le1+a_0$. This proves the stated periodic bound without imposing a per-direction restriction. For the finite nonperiodic pupil, the corresponding occurrence-weighted argument uses the degree-two offset graph and gives $\mathcal G_w\le 2(1+a_0)/(D-1)$. These are {full-bank} information budgets, not scalar bounds for every known Fourier mode. Indeed a sorter optimized for a selected single contrast can retain a nonvanishing fraction of its QFI. The binary-specific all-PPT bound is separately proved in Sec.~\ref{sec:si-binary-twirl}.

\subsection{Phase links, inversion, and physical readout}
\label{sec:si-links}\label{sec:si-circuit-observables}
The complex image also requires the neighboring links $\beta_m=\Phi_m^*\Phi_{m+1}$. These are {erased} by the complete common $U(1)^D$ twirl; they can survive a correlated finite Gaussian screen. For a selected calibrated two-dimensional link block with fixed occurrence weight $w$ and off-diagonal signal $g\,\mathrm{Re}(e^{-i\chi}\beta_m)$, measuring its two exchange parities is scalar-QFI optimal for that {block}. This does not establish simultaneous full-state multiparameter QFI optimality for every Gaussian source or imply that every circuit realization has identical FI.

The actual one-hot permutation-based circuit gives the direct operational connection. After a known permutation $\pi$ and phase rotation of register $B$, the detected sites $j<k$ and product ancilla parity $s=s_js_k$ define
\begin{equation}
 m=k-j,\quad n=\pi^{-1}(k)-\pi^{-1}(j),\quad
 \mathbb E[s\mid j,k]=\eta_{jk}\mathrm{Re}[e^{i\chi_{jk}}\Gamma_{jk}^{(\pi)}\Phi_m^*\Phi_n],
 \label{eq:si-circ-contrast}
\end{equation}
where $\eta_{jk}=2\sqrt{W_1W_2}/(W_1+W_2)$ is the known pupil-weight factor ($\eta_{jk}=1$ for a flat pupil) and $\Gamma_{jk}^{(\pi)}$ is the calibrated Gaussian phase average. To see this, undo the programmed $B$ permutation in the two-ordering exchange vector; its two source amplitudes are $\langle k|\rho|j\rangle\langle u|\rho|v\rangle$ with $u=\pi^{-1}(j)$ and $v=\pi^{-1}(k)$, and their interference gives Eq.~\eqref{eq:si-circ-contrast}. Two excitations registered in the same output register remain valid events because the local SWAPs conserve the {combined} occupation of each site. With $\pi=I$, $n=m$ and the parity reads $|\Phi_m|^2$ under perfectly common nuisance; an adjacent transposition can make $n=m\pm1$, and two phase choices $0,\pi/2$ give both link quadratures after the calibrated attenuation factor is included. The frequencies with which these settings occur are included in total FI.

When all required powers and nonzero connected links are retained, fixing one visibility phase and applying $\Phi_{m+1}=\beta_m/\Phi_m^*$ reconstructs the complex Fourier sequence. A final inverse Fourier transform gives the source-intensity representation. Missing or very small coefficients can disconnect the graph or amplify errors; additional links or regularization are then needed. For image coherences supported on a long physical pupil baseline $r$, the reference-sensitive Gaussian exponent includes a term of the order of the phase structure function $D_\phi(r)$; in a Kolmogorov range this grows approximately as $(r/r_0)^{5/3}$ before the finite outer scale cuts it off. If the corresponding differential-phase exponent stays much smaller for the chosen pair delay, its particular FI ratio may therefore grow much more rapidly with increasing pupil size than the {known binary-separation} ratio. This is a conditional, observable-specific finite-range mechanism, not a claim that every visibility has such an advantage.

The cat and Einstein-cross demonstrations in the code repository implement the periodic-grid version of this chaining and inversion. They illustrate recovery from stipulated observables, not an end-to-end finite-photon hardware validation.

\section{Kalman filtering}
\label{sec:si-history}\label{app:kalman_realistic}

The preceding theorems concern a specified pair-state law. We now allow earlier photons to predict the phase screen of the current pair. This is the nuisance-learning branch of the comparison. The Kalman receiver is retained as the MMSE-optimal causal estimator of the granted linear-Gaussian observation model, with a universal task-coverage condition and a trace-information budget. Those assumptions are stated here once rather than interleaved with the pair-state proofs.

\subsection{Conditional states and information accounting}

Let $H_{<t}$ denote the entire earlier classical record, including measurement settings. An adaptive experiment has
\begin{equation}
 F_\delta(H_{1:n})=\sum_{k=1}^n\mathbb E\bigl[F_\delta(Y_k\mid H_{<k})\bigr].
 \label{eq:si-history-chain}
\end{equation}
Within-pair feedforward is already included in a PPT optimization on the correct conditional pair state. In contrast, information from earlier correlated pairs changes the prior and cannot be ignored. If pairs have independently refreshed nuisance, their conditional state is the same prior-averaged pair state and the uniform single-pair bound simply adds. If history predicts the screen, its conditional law is substituted before applying any per-pair theorem.

\subsection{The linear-Gaussian observation model}

Let the physical nuisance vector have dimension $d_\eta$ and dynamics
\begin{equation}
 \eta_{k+1}=A_k\eta_k+\xi_k,\qquad
 y_k=H_k\eta_k+n_k,\qquad
 \xi_k\sim\mathcal N(0,Q_k),\quad n_k\sim\mathcal N(0,R_k).
 \label{eq:si-Kalman-model}
\end{equation}
The prior is Gaussian, the noises are independent with specified covariances, and the granted nuisance observations have this likelihood. Matrices may be selected from the past record and are then known when conditioning. The observation information is $J_k=H_k^TR_k^{-1}H_k$, subject to the chosen photon-information budget $\Tr J_k\le B_k$. The budget and any division of photons between phase sensing and science acquisition are included in the resource convention.

\begin{proposition}[Kalman MMSE optimality and the covariance recursion]
\label{prop:si-kalman}
For Eq.~\eqref{eq:si-Kalman-model}, the causal conditional mean is the minimum-mean-square-error estimator. Its covariance satisfies
\begin{equation}
 P_k^-=A_{k-1}P_{k-1}^+A_{k-1}^T+Q_{k-1},\qquad
 P_k^+=\bigl[(P_k^-)^{-1}+J_k\bigr]^{-1},
 \label{eq:si-Riccati}
\end{equation}
with the usual support-restricted interpretation for singular covariances.
\end{proposition}
\begin{proof}
Gaussian conditioning of a quadratic observation likelihood gives the posterior precision $(P_k^-)^{-1}+H_k^TR_k^{-1}H_k$. Propagation through the linear dynamics gives the prediction covariance. For any estimator $\widehat\eta(H)$, the conditional-mean orthogonality identity writes its MSE as the posterior variance plus $\mathbb E\|\widehat\eta-\mathbb E(\eta\mid H)\|^2$, proving MMSE optimality.
\end{proof}
This is the standard Kalman setting~\cite{Kalman1960}; predictive Fourier wavefront control provides a directly relevant atmospheric realization of this viewpoint~\cite{Poyneer2007}. MMSE optimality is for this likelihood. A trace constraint on the information supplied by arbitrary photon measurements does not itself prove that every such measurement is statistically dominated by the granted Gaussian sensor. The Kalman and Kalman-conditioned science curves therefore retain their benchmark interpretation.

Suppose the pupil phase is $G_\eta\eta$ and $P$ is the predicted residual covariance at the first photon. Before using either current-photon result, propagation to the second arrival gives the conditional phase-state covariance, in nuisance coordinates,
\begin{equation}
 \Sigma_K=\begin{pmatrix}P&PA_\tau^T\\A_\tau P&A_\tau PA_\tau^T+Q_\tau\end{pmatrix},
 \qquad
 P_{\rm diff}=(A_\tau-I)P(A_\tau-I)^T+Q_\tau.
 \label{eq:si-Kalman-pair}
\end{equation}
Apply $G_\eta$ to both blocks to obtain the pupil covariance. The known posterior mean is included in the measurement or removed by correction. Replacing this matrix by a scalar multiple of the prior covariance is a model, not the full Kalman prediction. When the conditional Gaussian law is independent of $\delta$ for the granted phase-observation model, Theorem~\ref{thm:si-finite-PPT} can be applied to each conditional pair. For a general photon-generated posterior that depends on $\delta$, its derivative is part of the conditional likelihood and must not be dropped.

\subsection{Why temporal correlation controls the benefit of history}

Kalman filtering remains MMSE optimal even when history is almost useless. To separate these notions, take a stationary scalar process with $A_\tau=a_\tau$ and prior variance $\sigma_\eta^2$. Then $Q_\tau=(1-a_\tau^2)\sigma_\eta^2$ and
\begin{equation}
 P^-_{k+1}=a_\tau^2P_k^++(1-a_\tau^2)\sigma_\eta^2.
 \label{eq:si-correlation-predict}
\end{equation}
As $a_\tau\to0$, the prediction reverts to the prior regardless of how well the previous screen was measured. As $a_\tau\to1$, past information can persist, provided the modes are observed well enough. In a weak-photon regime, useful history-assisted reference recovery therefore requires both temporal correlation across the available record and a sufficient observation budget in the relevant nuisance modes. This is the interpretation of the correlation-dependent Kalman improvement, not a statement that its MMSE theorem holds only at high correlation.

\subsection{Trace-information budget and universal task coverage}

For a static isotropic Gaussian nuisance prior $\eta\sim\mathcal N(0,\sigma_\eta^2I_{d_\eta})$ with total expected observation FI $\Tr\mathbb E J\le nB$, the Bayesian information inequality gives
\begin{equation}
 \mathbb E\Tr\Cov(\eta\mid H)
 \ge\frac{d_\eta\sigma_\eta^2}{1+nB\sigma_\eta^2/d_\eta}.
 \label{eq:si-trace-budget}
\end{equation}
This follows from the matrix Bayesian Cram\'er--Rao inequality and $\Tr M^{-1}\ge d_\eta^2/\Tr M$; it also follows directly for a fixed Gaussian observation channel. It is used as a reference-window information budget, while the evolving posterior is propagated by Eq.~\eqref{eq:si-Riccati}. It is not a pointwise lower bound on every posterior covariance or on every optical charge variance.

For a universal receiver, let $\{v_\alpha,w_\alpha\}$ describe the nuisance charges and fixed nonnegative normalized weights required by the acquisition task. A weighted coverage condition
\begin{equation}
 \sum_\alpha w_\alpha v_\alpha v_\alpha^T\succeq c_{\rm cov}I_{d_\eta}
 \label{eq:si-coverage}
\end{equation}
implies $\sum_\alpha w_\alpha v_\alpha^TPv_\alpha\ge c_{\rm cov}\Tr P$. This is where task independence can matter: a receiver told only one desired scalar combination may instead direct its phase-sensing resources toward that combination. Universality is not a hypothesis of the Kalman recursion itself and is not needed for the earlier binary PPT theorems.

The actual Gaussian amplitude attenuation is $\exp[-v_\alpha^TPv_\alpha/2]$. Define the specified-weight FI-level survival diagnostic and exponent
\begin{equation}
 \Lambda(P)=\sum_\alpha w_\alpha e^{-v_\alpha^TPv_\alpha},\qquad
 \Xi(P)=-\log\Lambda(P).
 \label{eq:si-Lambda}
\end{equation}
The likelihood denominators must be controlled before identifying this diagnostic with FI. In particular, Jensen gives
\begin{equation}
 \Xi(P)\le\sum_\alpha w_\alpha v_\alpha^TPv_\alpha,
 \label{eq:si-Jensen}
\end{equation}
so a large mean residual variance alone does not prove exponentially small survival.

\begin{lemma}[A sufficient extensive-exponent condition]
\label{lem:si-tail}
If for constants $b_1,b_2>0$ the total weight of directions with $v_\alpha^TPv_\alpha<b_1d_\eta$ is at most $e^{-b_2d_\eta}$, then
\begin{equation}
 \Lambda(P)\le2e^{-\min(b_1,b_2)d_\eta},\qquad
 \Xi(P)\ge\min(b_1,b_2)d_\eta-\log2.
 \label{eq:si-tail-bound}
\end{equation}
\end{lemma}
\begin{proof}
Split the weighted sum into the small-variance set and its complement. Bound the first by its total weight and each exponential in the second by $e^{-b_1d_\eta}$.
\end{proof}
The assumption concerns physical charge couplings, not mode counting. For example, independent pupil phases with covariance $\sigma^2I$ and a single-copy charge $v=e_j-e_k$ give $v^TPv=2\sigma^2$, independent of $D$. Also, if all charges have bounded Euclidean norm, $c_{\rm cov}$ in Eq.~\eqref{eq:si-coverage} cannot remain positive independently of $d_\eta$. An exponent proportional to a nuisance dimension therefore requires an explicitly extensive coupling/variance regime such as Lemma~\ref{lem:si-tail}, not merely a large matrix dimension.

\subsection{Reference recovery, relational information, and the exponential window}

Let $\Xi_C=\Xi(P_K)$ describe the reference-sensitive benchmark after its optimal Kalman prediction, and let $\Xi_Q=\Xi(P_{\rm diff})$ describe the selected relational charges. When the stated science likelihood and task assumptions give information of the form $F_C^{\rm ref}\asymp V e^{-\Xi_C}$ and $F_J\asymp V e^{-\Xi_Q}$, the reference-dominated comparison has
\begin{equation}
 \frac{N_C}{N_J}\asymp e^{\Xi_C-\Xi_Q}.
 \label{eq:si-history-exponential}
\end{equation}
If the exponents separately have extensive forms $\Xi_C=\kappa_Cd_\eta+O(1)$ and $\Xi_Q=\kappa_Jd_{\rm eff}+O(1)$, this is the shorthand $\exp[\kappa_Cd_\eta-\kappa_Jd_{\rm eff}]$ or $\Xi_C-\Xi_Q \propto D$.

\section{Moving-object tracking}
\label{sec:si_tracking}
The numerical tracking experiment differs from the static binary-separation theorem: its parameter is a residual temporal displacement, its classical comparator is an explicit classical (direct imaging with correlations) receiver, and the quantum information used for the forecast is an idealized registered-exchange model. The local FI calculations below specify the exact meaning of the prediction advantage.

\subsection{Moving-object tracking with $D=128$ and an $8.2\,\mathrm m$ aperture}
\label{sec:si_tracking_D128}

\paragraph{Tracking task and observation model.}
We consider a faint moving point source observed through a one-dimensional, uniformly sampled telescope pupil containing $D=128$ optical modes of pitch $a=0.0640625\,\mathrm{m}$, corresponding to a participating pupil length $Da=8.2\,\mathrm{m}$. The wavelength is $\lambda=550\,\mathrm{nm}$. For one angular coordinate, we write the apparent position as a known nominal ephemeris plus an unknown residual position and velocity,
\begin{equation}
 \theta(t)=\theta_{\mathrm{eph}}(t)+\theta_0+v\left(t-\frac{T}{2}\right),
 \qquad 0\le t\le T+H,
 \label{eq:track128_motion}
\end{equation}
where $T=0.25\,\mathrm{s}$ is the observation duration, $H$ is the prediction horizon measured from the {end} of the observation, and $(\theta_0,v)$ are both unknown to either receiver. The nominal ephemeris is assumed known, but its residual offset and linear drift are estimated from the same finite photon budget. No unknown residual acceleration is fitted. We use a detected photon rate $R=1600\,\mathrm{s}^{-1}$, giving an expected $N_{\mathrm{ph}}=RT=400$ detected photons, or $N_{\mathrm{pair}}=RT/2=200$ disjoint nominal photon pairs. The comparison assumes ideal photon capture and pairing, without an additional loss penalty.

It is convenient to work in the dimensionless pupil phase-gradient coordinate $\varphi=ka\theta$, with $k=2\pi/\lambda$ and $\theta$ in radians. The conversion back to milliarcseconds is
\begin{equation}
 \kappa=\frac{\lambda}{2\pi a}\frac{180\times3600\times10^3}{\pi}
 =281.8409\,\mathrm{mas},\qquad \theta_{\mathrm{mas}}=\kappa\varphi.
 \label{eq:track128_kappa}
\end{equation}
For a photon detected at time $t$, the conditional normalized pupil state is
\begin{equation}
 |\psi_t\rangle=\frac{1}{\sqrt D}\sum_{j=0}^{D-1}
  \exp\!\left\{i\left[j\varphi(t)+\chi_j(t)\right]\right\}|j\rangle,
 \label{eq:track128_photon}
\end{equation}
where $\chi_j(t)$ is the unknown atmospheric phase at pupil site $j$. The temporal pair separation is $\tau=0.040\,\mathrm{s}$, and the angular parameter inferred by the temporal exchange measurement is the residual phase displacement $\delta=\dot\varphi\tau$ after subtraction of the nominal ephemeris motion.

\paragraph{Atmospheric model and classical receiver likelihood.}
We use a single-layer frozen-flow Gaussian phase-screen model with Fried parameter $r_0=0.177\,\mathrm{m}$, outer scale $L_0=22\,\mathrm{m}$, and wind speed $w=10\,\mathrm{m/s}$. A periodic one-dimensional Gaussian phase slice is generated on 2048 spatial samples with spectral density proportional to $(f^2+L_0^{-2})^{-4/3}$, with its amplitude normalized to the one-pitch phase-structure function
\begin{equation}
 \mathbb E\!\left[\big(\chi(x+a)-\chi(x)\big)^2\right]
 =6.88\left(\frac{a}{r_0}\right)^{5/3}=1.26466\;\mathrm{rad}^2.
 \label{eq:track128_structure}
\end{equation}
The second screen is a translation of the first by $w\tau=0.40\,\mathrm{m}$, equivalent to $s=w\tau/a=6.2439$ pupil pitches; a fractional-pixel shift is realized by Fourier interpolation. This is a calibrated {one-dimensional slice model}, not a full two-dimensional von K\'arm\'an phase-screen simulation. The assumed translation is known exactly and no screen evolution beyond frozen flow is included.

For comparison, each photon is detected separately in the Fourier plane; the complete correlated pair of focal-plane outcomes is retained by the classical estimator. Using $M=256$ zero-padded Fourier outcomes per photon, the conditional one-photon likelihood for outcome $u$ is
\begin{equation}
 p_{\mathrm{FT}}(u\mid\varphi,\boldsymbol\chi)
 =\frac{1}{DM}\left|\sum_{j=0}^{D-1}
  e^{i[j\varphi+\chi_j-2\pi uj/M]}\right|^2,
 \qquad u=0,\ldots,M-1.
 \label{eq:track128_oneclick}
\end{equation}
For the two photon arrival times, parameterize the source phase gradients by $\varphi_{1,2}=\varphi_c\mp\delta/2$. The nuisance-averaged joint classical (direct imaging with correlations) likelihood is
\begin{equation}
 p_{uv}(\varphi_c,\delta)=\mathbb E_{\chi}\!\left[
 p_{\mathrm{FT}}(u\mid\varphi_c-\delta/2,\boldsymbol\chi_t)
 p_{\mathrm{FT}}(v\mid\varphi_c+\delta/2,\boldsymbol\chi_{t+\tau})\right].
 \label{eq:track128_pair_likelihood}
\end{equation}
We estimate this likelihood and its analytic phase derivatives at zero residual motion using $10^5$ independent Gaussian-screen realizations. The two independent halves of the Monte Carlo ensemble are cross-multiplied in the FI numerator to reduce the positive plug-in bias from sampling noise. The resulting per-pair Fisher matrix, in the $(\varphi_c,\delta)$ coordinates, is
\begin{equation}
 \mathcal F_{\mathrm{FT}}^{\mathrm{pair}}=
 \sum_{u,v}\frac{1}{p_{uv}}
 \begin{pmatrix}
 (\partial_{\varphi_c}p_{uv})^2 &
 (\partial_{\varphi_c}p_{uv})(\partial_{\delta}p_{uv})\\
 (\partial_{\varphi_c}p_{uv})(\partial_{\delta}p_{uv}) &
 (\partial_{\delta}p_{uv})^2
 \end{pmatrix}
 \simeq
 \begin{pmatrix}
 1.26576194&0.00135629\\
 0.00135629&7.67921033
 \end{pmatrix}.
 \label{eq:track128_classical_pair_FI}
\end{equation}
Here the numerical coefficients are the cross-validated finite-Monte-Carlo estimates, with the full two-click correlations included. The small cross term is reported but omitted from the centered-time diagonal tracking model below. This comparator is the specified classical (direct imaging with correlations), not a bound on all separable or classical measurements.

\paragraph{Quantum temporal exchange and finite pupil overlap.}
We compare photons separated by $\tau$ after a known wind-aligned spatial translation. In the ideal frozen-flow limit, $\chi_{j+s}(t+\tau)=\chi_j(t)$ and the common atmospheric phases cancel in the corresponding shifted-exchange interference. The remaining pair phase is proportional to $(j-\ell)\delta$, independently of the unknown absolute source phase gradient. For a uniform $D$-site pupil, the ideal two-photon relative-displacement FI scale is
\begin{equation}
 F_{\delta}^{\mathrm{ideal,pair}}
 =2\operatorname{Var}(j)=\frac{D^2-1}{6}
 =2730.5\qquad (D=128).
 \label{eq:track128_ideal_delta_FI}
\end{equation}
We allocate $q_Y=0.65$ of the nominal pairs to temporal Y-quadrature exchange, $q_a=0.30$ to an ordinary FT absolute-position anchor, and leave $0.05$ uninformative. The effective model includes a registered exchange visibility $\gamma$, usable-pair efficiency $\eta$, and a geometrical pupil-overlap model $f=1-w\tau/(Da)$. Thus the effective relative-displacement FI {per nominal incident pair}, including the temporal-branch allocation, is
\begin{equation}
 F_{\delta,Q}
 =q_Y\eta\gamma^2 f^4\frac{D^2-1}{6},
 \qquad f=1-\frac{0.40}{8.2}=0.9512195.
 \label{eq:track128_quantum_delta_FI}
\end{equation}
For the idealized choices $\eta=\gamma=1$, this gives $F_{\delta,Q}=1453.04324$. The factor $f^4$ is a modeling approximation for the reduced spatial overlap and available baselines, not an exact FI calculation of a spatially shifted hardware circuit. The quantum FI is therefore an idealized exchange-receiver benchmark rather than an independent simulation of the complete noisy quantum POVM.

\paragraph{Joint position--velocity Fisher information.}
Let $t_c=(t_1+t_2)/2$ be the pair midpoint. Its locally relevant parameters obey
\begin{equation}
 \varphi_c=\varphi_0+\dot\varphi\left(t_c-\frac T2\right),
 \qquad \delta=\dot\varphi\tau,
 \qquad
 J(t_c)=\frac{\partial(\varphi_c,\delta)}
 {\partial(\varphi_0,\dot\varphi)}
 =\begin{pmatrix}1&t_c-T/2\\0&\tau\end{pmatrix}.
 \label{eq:track128_jacobian}
\end{equation}
We model pair midpoint times as uniform over $[\tau/2,T-\tau/2]$, so their mean is $T/2$ and their variance is
\begin{equation}
 V_t=\mathbb E\!\left[(t_c-T/2)^2\right]
 =\frac{(T-\tau)^2}{12}=0.003675\,\mathrm{s}^2.
 \label{eq:track128_time_variance}
\end{equation}
Summing the pair information over the observing window gives the following centered, diagonal approximations for the residual midpoint position and velocity in phase-gradient units:
\begin{align}
 \mathcal I_Q&=N
 \begin{pmatrix} A_Q&0\\0&B_Q\end{pmatrix},
 &A_Q&=q_aF_{c,\mathrm{FT}},
 &B_Q&=q_aF_{c,\mathrm{FT}}V_t+F_{\delta,Q}\tau^2,\nonumber\\
 \mathcal I_{\mathrm{FT}}&=N
 \begin{pmatrix} A_{\mathrm{FT}}&0\\0&B_{\mathrm{FT}}\end{pmatrix},
 &A_{\mathrm{FT}}&=F_{c,\mathrm{FT}},
 &B_{\mathrm{FT}}&=F_{c,\mathrm{FT}}V_t+F_{\delta,\mathrm{FT}}\tau^2,
 \label{eq:track128_joint_FI}
\end{align}
where $F_{c,\mathrm{FT}}=1.26576194$ and $F_{\delta,\mathrm{FT}}=7.67921033$. Both receivers receive the same total number of photons. In the illustrative two-coordinate display we independently assign half the $N_{\mathrm{pair}}=200$ available nominal pairs to each coordinate, giving $N=100$ pairs per coordinate for either receiver. The per-pair coefficients and resulting numerical matrices are
\begin{align}
 (A_Q,B_Q)&=(0.37972858,\,2.32626469\,\mathrm{s}^2),&
 \mathcal I_Q&=\operatorname{diag}(37.972858,\,232.626469),\nonumber\\
 (A_{\mathrm{FT}},B_{\mathrm{FT}})&=(1.26576194,\,0.01693841\,\mathrm{s}^2),&
 \mathcal I_{\mathrm{FT}}&=\operatorname{diag}(126.576194,\,1.693841).
 \label{eq:track128_joint_FI_values}
\end{align}
In this model the velocity-FI enhancement is therefore
\begin{equation}
 \mathcal A_v=\frac{B_Q}{B_{\mathrm{FT}}}=137.34.
 \label{eq:track128_vel_gain}
\end{equation}
The classical velocity information includes that obtained by following the absolute position throughout the observation ($F_{c,\mathrm{FT}}V_t$), not merely by taking a noisy pairwise difference.

\paragraph{Forecast covariance, sample complexity, and displayed trajectories.}
In angular units, we take the Cram\'er--Rao covariance to be $C_M=\kappa^2\mathcal I_M^{-1}$ for receiver $M=Q$ or $\mathrm{FT}$. The midpoint uncertainties per coordinate are
\begin{align}
 \sigma_{\theta_0,Q}&=45.737\,\mathrm{mas},&
 \sigma_{\theta_0,\mathrm{FT}}&=25.051\,\mathrm{mas},\nonumber\\
 \sigma_{v,Q}&=18.479\,\mathrm{mas/s},&
 \sigma_{v,\mathrm{FT}}&=216.555\,\mathrm{mas/s}.
 \label{eq:track128_marginal_uncertainties}
\end{align}
For a prediction at time $t_\star=T+H$, write $h=t_\star-T/2=H+T/2$. Because the position--velocity covariance vanishes in the centered symmetric approximation, the one-coordinate forecast variance is
\begin{equation}
 \sigma_M^2(H)=\kappa^2\left(\frac{1}{NA_M}
 +\frac{(H+T/2)^2}{NB_M}\right)
 =\frac{2\kappa^2}{\zeta N_{\mathrm{ph}}}
 \left(\frac1{A_M}+\frac{(H+T/2)^2}{B_M}\right),
 \qquad \zeta=\tfrac12,
 \label{eq:track128_forecast_variance}
\end{equation}
where the final form uses $N=\zeta N_{\mathrm{ph}}/2$. At a fixed target forecast standard deviation $\epsilon$, the required detected photon number and the equal-precision photon-number advantage are
\begin{align}
 N_{\mathrm{ph},M}^{\mathrm{req}}(\epsilon,H)
 &=\frac{2\kappa^2}{\zeta\epsilon^2}
 \left(\frac{1}{A_M}+\frac{(H+T/2)^2}{B_M}\right),\nonumber\\
 \mathcal A_{\mathrm{pred}}(H)
 &=\frac{N_{\mathrm{ph},\mathrm{FT}}^{\mathrm{req}}}
 {N_{\mathrm{ph},Q}^{\mathrm{req}}}
 =\frac{A_{\mathrm{FT}}^{-1}+(H+T/2)^2B_{\mathrm{FT}}^{-1}}
 {A_Q^{-1}+(H+T/2)^2B_Q^{-1}}
 =\frac{\sigma_{\mathrm{FT}}^2(H)}{\sigma_Q^2(H)}.
 \label{eq:track128_photon_advantage}
\end{align}
For our $400$-photon reference observation, the numerical results are
\begin{equation}
\begin{array}{c|cc|c|cc}
 H & \sigma_Q\ (\mathrm{mas}) & \sigma_{\mathrm{FT}}\ (\mathrm{mas})
 & \mathcal A_{\mathrm{pred}} &
 N_{\mathrm{ph},Q}^{\mathrm{req}}(100\,\mathrm{mas}) &
 N_{\mathrm{ph},\mathrm{FT}}^{\mathrm{req}}(100\,\mathrm{mas})\\ \hline
 4\,\mathrm{s} & 88.89 & 893.64 & 101.06 &316&31\,944\\
 8\,\mathrm{s} &156.95 &1759.69 &125.70&985&123\,860
\end{array}
\label{eq:track128_forecast_table}
\end{equation}
The required photon counts are FI-limited expectation values rounded to whole photons; they do not include telescope throughput or a practical photon-memory bandwidth limit. For long horizons, $\mathcal A_{\mathrm{pred}}(H)\rightarrow\mathcal A_v$, so reducing source brightness does not by itself change the local-FI advantage.

For visualization, we prescribe a known nominal circular arc with speed $v_{\mathrm{sky}}=250\,\mathrm{mas/s}$, initial heading $\beta=30^\circ$, and turn $\Omega T=0.4^\circ$ during the observation:
\begin{equation}
 \boldsymbol\theta_{\mathrm{eph}}(t)
 =\frac{v_{\mathrm{sky}}}{\Omega}
 \begin{pmatrix}
 \sin(\beta+\Omega t)-\sin\beta\\
 \cos\beta-\cos(\beta+\Omega t)
 \end{pmatrix}.
 \label{eq:track128_arc}
\end{equation}
For each coordinate and receiver, we draw fitted residual parameters once from $\mathcal N(\boldsymbol\beta_{\mathrm{true}},C_M)$ and propagate them through Eq.~\eqref{eq:track128_motion}. This yields time-correlated fitted trajectories, {not} trajectories reconstructed from simulated photon clicks. The true residual position and velocity are zero in the displayed example.

The uncertainty bands are centered on the nominal trajectory and follow $\sigma_M(t)$. The observation circles use a radius of $5\sigma$, and the forecast circles use $1\sigma$, where $\sigma$ is the Cartesian marginal standard deviation. In two dimensions a circular $1\sigma$ region contains approximately $39.3\%$, not $68\%$, of an isotropic Gaussian distribution. We may suppress the stochastic fitted curves in the plots without changing these bands. The curves use 160 observation time samples and 320 prediction samples.

\paragraph{Scope of the simulation.}
These numbers describe a local, efficient-estimator Fisher-information {model}, not a finite-count simulation of photonic detection or of a specific noisy quantum circuit. They assume a known nominal ephemeris, independent disjoint photon pairs, exact wind registration of a single frozen atmospheric layer, unit usable-pair efficiency, and unit registered visibility over $40\,\mathrm{ms}$. The $f^4$ overlap penalty is approximate, and the half-budget allocation to each sky coordinate is illustrative rather than a demonstrated optimal two-dimensional POVM; in particular, a classical camera may extract information about both sky coordinates from the same photon. We neglect unmodeled residual acceleration, any phase ambiguities or non-Gaussian finite-sample estimator effects, and possible improvements from other classical receivers. The reported $\sim10^2$ forecast advantage is therefore a conditional comparison against the specified universal classical receiver likelihood under these idealized assumptions, not a universal separable-measurement lower bound.

\section{Quantum-camera resources and hardware errors}
\label{sec:si-hardware}

The processing and noise estimates quoted in the main text concern a hypothetical quantum-camera architecture, not an experimental demonstration of end-to-end capture and coherent storage of astronomical photons. We distinguish the optical receiving interface, the logical representation of the stored spatial mode, the joint processing circuit, and a postselected error-detection layer. Their costs need not scale in the same way with the optical dimension $D$.

\subsection{Receiving modes, qubits, and circuit depth}

A spatially resolved input occupies $D$ optical pupil modes. In a one-hot memory, one captured photon is represented by a single excitation distributed among $D$ receiving qubits. Two stored photons therefore involve $2D$ receiving qubits before including any ancilla. If a coherent logical compression or spatial-to-address routing is available, each photon may instead be represented in $q=\lceil\log_2D\rceil$ address qubits, so that two stored photons and a parity ancilla use
\begin{equation}
 n_{\rm process}=2\lceil\log_2D\rceil+1.
 \label{eq:si-hardware-qubits}
\end{equation}
For $D=128$, this gives $q=7$ and $n_{\rm process}=15$. This is a count of processing qubits in the compressed representation. It does not include the $D$ optical access channels, the switching or compression hardware, additional memory ancillas, or a parallel one-hot transduction interface. Coherent compression and routing must preserve the optical superposition; a classical readout of the occupied pupil mode would change the statistical experiment.

An elementary controlled block-SWAP may be synthesized from address-register interactions conditioned on an ancilla. For the illustrative logarithmic implementation, seven sequential bitwise controlled-exchange layers plus preparation and readout can be counted as a nominal depth-nine circuit {in that abstract layer model}. The actual physical depth depends on the elementary gate set, available connectivity, decomposition of controlled-SWAP or CCZ operations, error-detection cycles, and the cost of implementing a rotated block and a controlled translation. In particular, depth nine excludes transduction, coherent compression, long-range routing, and memory waiting time; it is not a verified full-camera depth for both plain and rotated RFS. Every full-image setting also consumes its own allocated photon pairs, as discussed in Sec.~\ref{sec:si-links}.

\subsection{Heralding, code-space rejection, and the scope of error detection}
Heralding or an excitation-number check can convert {some} failures into known erasures, but cannot generically flag phase errors within the one-hot code space. A correct information budget includes the acceptance probability and the science-parameter likelihood of the accepted branch; we assume the rejected branch supplies no retained science information and that acceptance is parameter independent. Independent depolarization on every nominally empty receiving atom, assumed in the following one-hot example, is a deliberately conservative architecture-specific model and is not equivalent to having a transduction fault only on populated modes.

\subsection{Depolarizing-error overhead for rotated relational Fourier sampling}
\label{sec:si-rotated-hardware-depolarizing}

\paragraph{Scope and sample-complexity convention.}
We estimate the additional {incident photon pairs} needed to measure a calibrated rotated-RFS quadrature using a logarithmically routed register or a one-hot memory. The compared quantity is the local, small-contrast measurement Fisher-information (FI) overhead, $\mathcal O_{\rm rot}=N_{\rm err}/N_{\rm ideal}=F_{\rm ideal}/F_{\rm err}$. We consider only the rotated-RFS measurement branch, without direct-FT anchoring photons, plain-RFS branches, or a full tracking-estimator FI matrix. The ideal comparison uses the same source, optical dimension, pupil geometry, permutation schedule, and quadrature allocation. All quoted FI values are per {incoming photon pair}, before any postselection. The estimates assume negligible optical loss, perfect programmed mode permutations and phase settings, perfect final measurement records, and no additional memory decoherence. In particular, no physical errors or optical losses are assigned to the predetermined spatial permutation of register $B$, and no controlled-translation gates are counted.

\paragraph{Measured rotated-RFS observable.}
Two stored photon states occupy registers $A$ and $B$. A predetermined, coherent permutation $\pi$ and phase pattern $\{\beta_r\}$ are applied to register $B$ before the exchange measurement. If the measured unordered output sites are $j<k$, the input-basis exchange states are
\begin{equation}
 \ket{\widetilde\pm_{jk}}
 =\frac{\ket{j,\pi^{-1}(k)}\pm\ket{k,\pi^{-1}(j)}}{\sqrt 2},
 \quad m=k-j,\quad n=\pi^{-1}(k)-\pi^{-1}(j).
 \label{eq:si-depol-rotbasis}
\end{equation}
For uniform pupil illumination and ideal correlated atmospheric phases, the conditional parity contrast is $\langle s\rangle_{jk}=\operatorname{Re}[e^{i(\beta_k-\beta_j)}\Phi_m^*\Phi_n]$, where $s=\pm1$ is the global exchange parity. Nonuniform illumination introduces a calculable block-visibility factor, while incomplete atmospheric correlations introduce additional source-dependent attenuation. We use the familiar local quadrature regime in which the ideal contrast is near zero and the FI is proportional to the square of its derivative. Under a purely multiplicative contrast reduction $V$, the FI-retention model is therefore $R=V^2$, and $\mathcal O_{\rm rot}=1/R$. These expressions are not universal for arbitrary finite contrasts, source mixtures, or Fourier links.

\paragraph{Depolarizing channels and gate budgets.}
We use independent stochastic Pauli errors, with probabilities $p_r=10^{-2}$ per receiving qubit, $p_3=10^{-3}$ per elementary three-qubit controlled-SWAP, and $p_1=10^{-4}$ per budgeted one-qubit gate. Specifically, $\mathcal D_p^{(1)}(\rho)=(1-p)\rho+(p/3)\sum_{P\in\{X,Y,Z\}}P\rho P$, and $\mathcal D_p^{(3)}(\rho)=(1-p)\rho+(p/63)\sum_{P\in\mathcal P_3\setminus\{I^{\otimes3}\}}P\rho P$. We write $\lambda_r=1-4p_r/3$, $\lambda_1=1-4p_1/3$, and $g_3=1-64p_3/63$ for the corresponding Pauli-observable attenuation factors. In the routed architecture, each photon occupies $q=\log_2D$ receiving address qubits, and a single traveling ancilla sequentially implements $q$ controlled-SWAP gates, with two additional single-qubit ancilla preparation/readout rotations. The permutation and its phases are incorporated in the ideal optical routing stage. In one-hot encoding, two registers of $D$ qubits undergo $D$ parallel local controlled-SWAP measurements, each with its own ancilla; in the postprocessing, only the two ancilla signs corresponding to the detected occupied sites are used. We charge one ancilla-basis rotation at each of those two informative sites. Independent faults on all $2D$ one-hot receiving qubits, including nominally empty optical modes, are assumed; this is the consequential pessimistic feature of the model.

\paragraph{Logarithmically routed receiver.}
For a uniformly weighted phase-gradient signal near the calibrated zero-contrast quadrature, the receiving-qubit depolarizing channel attenuates the two-photon exchange response in a pair of binary address states $j,k$ as $\lambda_r^{2h(j,k)}$, where $h(j,k)$ denotes their binary Hamming distance. Summing the squared response with the local $Y$-quadrature tracking weight $(j-k)^2$ yields
\begin{equation}
 R_{r,\rm route}
 =\frac{\sum_{j,k=0}^{D-1}(j-k)^2\lambda_r^{4h(j,k)}}{\sum_{j,k=0}^{D-1}(j-k)^2}
 =r\left(\frac{1+r}{2}\right)^{q-1},\qquad r=\lambda_r^4.
 \label{eq:si-depol-Rrroute}
\end{equation}
We represent an unheralded three-qubit gate fault on the traveling-ancilla circuit as a multiplicative attenuation $g_3$ of the useful ancilla-parity contrast, and each of the two single-qubit rotations as a factor $\lambda_1$. With $q$ controlled-SWAP operations and no translation gates, the resulting modeled FI retention and overhead are
\begin{equation}
 R_{\rm route}=R_{r,\rm route}\,g_3^{2q}\lambda_1^4,
 \qquad \mathcal O_{\rm route}=R_{\rm route}^{-1}.
 \label{eq:si-depol-Rroute}
\end{equation}
This is exact for the stated scalar-attenuation calculation, but it is not a full Pauli-error propagation result for an arbitrary routed imaging state.

\paragraph{One-hot population errors and two-excitation postselection.}
For one-hot encoding, the dominant additional effect is that depolarizing $X/Y$ components can change the excitation number in nominally empty modes. The ideal two-photon exchange block has one excitation at each of two distinct spatial sites, $j\ne k$. After local controlled-SWAP interactions, we read out both data registers and retain an event only when {exactly two total excitations at distinct spatial sites} are detected, regardless of which output register contains them. Thus two excitations appearing in the same register are accepted, but three or more excitations, fewer than two excitations, and two excitations at one physical site are rejected. This postselection is performed after the local ancilla measurements, so it does not reveal the premeasurement register assignment.

Write $b=2p_r/3$, $t=1-2b$, and $g_3=1-64p_3/63$. For one pupil site, let $u_{ab}$ denote the probability that the combined transduction and three-qubit gate error flips the logical occupations of the data qubits $A_r$ and $B_r$ according to $(a,b)\in\{0,1\}^2$. Uniform three-qubit Pauli faults give
\begin{equation}
 \begin{aligned}
 u_{00}&=\tfrac14(1+2g_3t+g_3t^2),\\
 u_{10}=u_{01}&=\tfrac14(1-g_3t^2),\\
 u_{11}&=\tfrac14(1-2g_3t+g_3t^2).
 \end{aligned}
 \label{eq:si-depol-uab}
\end{equation}
An initially empty site ends with zero, one, or two excitations with respective probabilities $(u_{00},2u_{10},u_{11})$. An initially singly occupied site ends with zero, one, or two excitations with probabilities $(u_{10},u_{00}+u_{11},u_{10})$. Consequently, the probability to accept exactly two excitations at distinct sites is
\begin{equation}
 \begin{split}
 P_{2,\rm distinct}={}&(u_{00}+u_{11})^2u_{00}^{D-2}\\
 &+4(D-2)(u_{00}+u_{11})u_{10}^2u_{00}^{D-3}\\
 &+4\binom{D-2}{2}u_{10}^4u_{00}^{D-4}.
 \end{split}
 \label{eq:si-depol-paccept}
\end{equation}
The first term represents retaining both original sites with no extra excitations, the second a replacement of one original site by a spurious site, and the third replacement of both sites by spurious sites. The probability of {no occupation-changing fault anywhere} is
\begin{equation}
 P_{\rm clean}=u_{00}^{D}.
 \label{eq:si-depol-pclean}
\end{equation}
Events contributing to $P_{2,\rm distinct}-P_{\rm clean}$ can be accepted even though their occupied positions no longer correspond to the original pair. For a conservative, readily reproducible model, we assume that this accepted corrupted component carries no first-order rotated quadrature signal and has unbiased parity. A full joint likelihood using the complete occupation-and-ancilla record could recover additional information.

\paragraph{One-hot conditional contrast and FI.}
Conditioned on no occupation changes, the receiving-channel phase attenuation associated with a contributing four-qubit exchange coherence is $z^4$, with $z=\lambda_r/(1-2p_r/3)$. We assign factors $g_3^2$ and $\lambda_1^2$ to the conditional contrast from the two informative controlled-SWAP sites and their two ancillary readout rotations. This defines the explicit effective contrast and conditional FI-retention approximation
\begin{equation}
 V_{\rm clean}=z^4g_3^2\lambda_1^2,
 \qquad R_{\rm coh}=V_{\rm clean}^2=z^8g_3^4\lambda_1^4.
 \label{eq:si-depol-rcoh}
\end{equation}
In the zero-contrast regime, a mixture of an informative component of probability $P_{\rm clean}$ with an uninformative accepted background of total probability $P_{2,\rm distinct}-P_{\rm clean}$ reduces the accepted-pair contrast by $P_{\rm clean}/P_{2,\rm distinct}$. Multiplying the FI per accepted event by its acceptance probability gives the {per incoming pair} approximation
\begin{equation}
 R_{\rm hot}
 =P_{2,\rm distinct}\left(\frac{P_{\rm clean}}{P_{2,\rm distinct}}\right)^2R_{\rm coh}
 =R_{\rm coh}\frac{P_{\rm clean}^2}{P_{2,\rm distinct}},
 \qquad \mathcal O_{\rm hot}=R_{\rm hot}^{-1}.
 \label{eq:si-depol-Rhot}
\end{equation}
Equation~\eqref{eq:si-depol-Rhot} is a {signal-plus-random-background FI model}, not an exact FI of the entire noisy POVM. It neglects the source-dependent likelihoods of erroneous occupied-site labels, their possible residual parity information, and microscopic fault propagation through the controlled-SWAP gates. In particular, it should not be generalized to near-dark superresolution observables without explicitly recalculating their likelihoods.

\paragraph{Results.}
Substituting $p_r=0.01$, $p_3=0.001$, and $p_1=0.0001$ gives Table~\ref{tab:si-depol-rotated-overhead}. The one-hot acceptance probability becomes small at large $D$ because all nominally empty receiving atoms are subject to independent occupation-changing faults. In contrast, only $O(\log D)$ address qubits contribute such errors in the routed model. These are conditional architectural estimates under the specified channels, and not a measured or hardware-independent claim about either implementation.

\begin{table}[t]
\centering
\caption{Illustrative local rotated-RFS FI retention and incident-photon-pair sample overhead under depolarizing errors, with ideal permutations and no optical loss. One-hot outcomes are accepted only when two total excitations appear at distinct spatial sites. ``FI retained'' includes the postselection penalty and is normalized per incoming photon pair.}
\label{tab:si-depol-rotated-overhead}
\begingroup
\small
\setlength{\tabcolsep}{5pt}
\begin{tabular}{@{}r l r r r@{}}
\toprule
$D$ & Receiver & $P_{\rm acc}$ & FI retained & $N_{\rm err}/N_{\rm ideal}$ \\
\midrule
128 & Routed, shared ancilla & $1$ & $0.79663$ & $1.255$ \\
128 & One-hot, local ancillas & $0.167854$ & $0.150647$ & $6.638$ \\
512 & Routed, shared ancilla & $1$ & $0.75246$ & $1.329$ \\
512 & One-hot, local ancillas & $0.00079125$ & $0.00061579$ & $1.624\times10^3$ \\
\bottomrule
\end{tabular}
\endgroup
\end{table}

\paragraph{Physical limitations and possible improvements.}
The large one-hot overhead is driven mainly by assuming $p_r=1\%$ independent full depolarization on {every} receiving atom. It does not apply unchanged if spin--photon errors occur predominantly on photon-occupied optical branches, are heralded as erasures, or can be recognized from additional syndrome information. Moreover, using only the two selected local ancilla signs avoids the exponential penalty that would arise from multiplying noisy ancilla outcomes at all $D$ sites. Conversely, if local ancilla or data faults generate substantial extra correlations, the simple scalar processing factors used here can fail. We exclude errors in the fixed spatial permutation and quadrature phase pattern; if those phases are implemented by additional noisy gates on many one-hot atoms, their contribution must be budgeted separately. We also exclude readout confusion, leakage, finite memory lifetime, imperfect atmospheric registration, and photon-loss selection bias. Finally, the FI for each finite-contrast rotated link and a general mixed source should be calculated from its full measured likelihood rather than from the present small-contrast model.

\subsection{Relation to the scientific advantage}

The preceding theorems compare POVMs acting on ideal received pairs with a specified nuisance distribution. Hardware loss changes the resource convention: if $\eta_Q$ of incident pairs yield a usable joint outcome, an otherwise unchanged quantum advantage in {incident} photon number becomes approximately $\eta_Q(F_Q/F_{\rm cl})$, with visibility-dependent FI corrections evaluated above. A fair astronomical comparison also includes the respective classical detection throughput and any spatially or spectrally selective quantum capture. Neither a small processing-qubit count nor a modest assumed processing-gate overhead alone demonstrates a net source-photon advantage. The ideal tracking example in Sec.~\ref{sec:si_tracking} explicitly sets pair efficiency to unity and therefore illustrates an optimistic limiting regime; its hardware feasibility should be tested by varying acceptance and residual differential visibility together.